\documentclass[11pt]{article}

\usepackage[letterpaper, margin=1in]{geometry}
\usepackage{setspace}
\usepackage{indentfirst}

\usepackage[T1]{fontenc}
\usepackage[utf8]{inputenc}

\usepackage{microtype}

\usepackage{graphicx}
\usepackage[labelfont=bf,labelsep=period]{caption}
\usepackage{subcaption}
\usepackage{booktabs}
\usepackage{float}
\usepackage{array}
\usepackage{threeparttable}
\usepackage[titletoc,toc,title]{appendix}

\usepackage{amsmath,amssymb}
\usepackage{dsfont}
\usepackage[colorlinks=true, linkcolor=blue, citecolor=blue, urlcolor=blue]{hyperref}
\usepackage{xurl}

\usepackage[authoryear]{natbib}

\title{\bf\Large\onehalfspacing Directional AI Advice: Experimental Evidence from Healthcare\thanks{
This experiment was preregistered at AEA RCT Registry (AEARCTR-0015851), and approved by Institutional Review Board of Guanghua School of Management, Peking University (\#2025-14).
We are grateful for helpful comments from Naijia Guo, Jin Li, Katrine L{\o}ken, Bentley MacLeod, Sultan Mehmood, Juan Pantano, Uta Sch{\"o}nberg, Yanhui Wu, Liang Zhong, Xiaodong Zhu, and seminar participants at the University of Hong Kong.
}}

\author{
    Yuyu Chen\thanks{Guanghua School of Management, Peking University. Email: \href{mailto:chenyuyu@gsm.pku.edu.cn}{chenyuyu@gsm.pku.edu.cn}} \quad
    Hongbin Li\thanks{Stanford University. Email: \href{mailto:hongbinli@stanford.edu}{hongbinli@stanford.edu}} \quad
    Lingsheng Meng\thanks{Stanford University. Email: \href{mailto:lmeng@stanford.edu}{lmeng@stanford.edu}} \quad
    Xinyao Qiu\thanks{Faculty of Business and Economics, The University of Hong Kong. Email: \href{mailto:xinyaoq@hku.hk}{xinyaoq@hku.hk}} \quad
    Qingxu Yang\thanks{School of Economics, Fudan University. Email: \href{mailto:qxyang@fudan.edu.cn}{qxyang@fudan.edu.cn}}
}

\date{\today}
\begin{document}

\maketitle

\begin{abstract}
\noindent
Generative AI is fast becoming the first place people turn for expert advice. The advice it provides can be directional rather than neutral, shaped in part by the choices of its designers and regulators. When clients consult AI before meeting an expert, they carry this directional advice into a relationship that once rested on the expert's judgment alone. We study its consequences in healthcare through a large-scale preregistered field experiment at a Chinese hospital, where we randomize patients' access to an AI chatbot before their outpatient visit. Examination of the conversation logs shows that the chatbot routinely cautions against the use of medications, especially Traditional Chinese Medicine and antibiotics, while issuing clean recommendations for diagnostic testing, consistent with the liability-driven guardrails encoded in AI training. This directionality propagates into clinical practice. Prescription rates decline among treated patients while diagnostic testing increases, and these effects are more pronounced among physicians who are receptive to patient input and those with more intensive prescribing styles. Beyond shifting healthcare utilization, survey results show that AI access reduces patient compliance and satisfaction, shifting the balance of authority between patients and physicians.

\vspace{1em}
\end{abstract}

\newpage

\section{Introduction}

Generative artificial intelligence (AI) is becoming the first place many people turn before they turn to experts. Unlike traditional information technologies such as newspapers and search engines, generative AI goes beyond providing easier access to information. It also interprets, exercises judgment, and often recommends a course of action. These recommendations may not be neutral. AI's responses reflect the priorities of those who build and govern these systems. For example, to limit legal exposure, developers install guardrails that steer their models away from liability-prone recommendations and toward more cautious alternatives. This directionality matters because AI now reaches into decisions people once delegated entirely to experts, in domains such as healthcare, law, and finance. As more people consult AI before turning to an expert, these design choices can propagate into real decisions at scale.

This paper examines whether and how directional advice from AI shifts the decisions experts and clients reach, in the context of healthcare. Healthcare is a canonical credence-good market characterized by severe informational asymmetries and substantial expert discretion \citep{arrow1963uncertainty,darby1973free,mcguire2000physician}, and it has now become a common use case for consumer-facing generative AI tools.\footnote{Health-related queries are among the three most common use cases of ChatGPT (\url{https://openai.com/index/introducing-gpt-5/}).} Its high stakes have drawn attention from both policymakers and developers, with developers implementing cautious guardrails to protect patients from the risks of acting on AI medical advice and to limit legal exposure. When patients consult AI before seeing a physician, the information they receive narrows the asymmetry that underlies physician discretion. But that information is not neutral. It also carries the priorities of AI developers, which may diverge from those of the health system.

To test how AI advice propagates into practice, we conduct a preregistered, month-long field experiment that provides the first randomized evidence on how patients' access to generative AI reshapes clinical decisions and the patient-physician relationship. Our randomized controlled trial takes place in a large public hospital in China and involves over 10,000 outpatient visits. Patients are randomly assigned to receive, or not receive, access to an LLM-based chatbot prior to their appointments. The chatbot invites patients to describe their symptoms and health concerns in detail before the visit and provides free access to AI-generated, personalized guidance and general medical information. We implement a two-layer randomization design. We first randomize physicians into an \textit{Exposed} group or an \textit{Unexposed} group. Among patients scheduling appointments with \textit{Exposed} physicians, half are randomly assigned to receive access to the chatbot (\textit{Treatment}) and half to receive no access (\textit{Control}). Patients scheduling appointments with \textit{Unexposed} physicians never receive chatbot access. This design allows us to compare clinical outcomes between patients with and without access to the chatbot who see the same physician, while also testing for within-physician spillovers in clinical practice affecting patients without access.

We first examine take-up of the chatbot and the content of the advice it provides. Linking usage logs to hospital administrative records, we find that 17 percent of patients offered access used it before their visit, with higher take-up among younger, male, and employed patients and first-time visitors. Turning to the content of these interactions, we find a sharp asymmetry in the advice the chatbot provides. When discussing medications, the chatbot almost always attaches cautions, warning about side effects, contraindications, and the risks of self-medication. Such cautions appear in 69.7\% of general medication mentions, 90.8\% of Traditional Chinese Medicine (TCM) mentions, and 87.6\% of antibiotic mentions, while clean recommendations are rare, at 3.8\% for TCM and 7.8\% for antibiotics. Diagnostic testing is treated very differently, with the chatbot issuing clean recommendations 94.5\% of the time. This asymmetry tracks the liability concerns of LLM developers. Recommending a specific medication carries substantial legal risk if it leads to adverse health outcomes, whereas encouraging diagnostic testing does not, and the cost of unnecessary tests falls on patients and the health system rather than the developer. The advice patients carry into the consultation is therefore directional, steering away from medication and toward diagnostic testing.

This directionality propagates into clinical decisions. Within-physician comparisons show that chatbot access reduces the likelihood of receiving any prescription by 4.6 percentage points relative to a baseline of 87\%, driven partly by declines in TCM and antibiotic prescribing. At the same time, chatbot access increases the probability that a visit includes diagnostic testing by 2.7 percentage points from a baseline of 23\%. The direction of each treatment effect mirrors the direction of the AI's advice. These shifts lower medication expenditures and modestly raise diagnostic spending, leaving total healthcare expenditures unchanged. We find no evidence of worse short-term outcomes; if anything, treated patients are 1.2 percentage points less likely to revisit the same hospital within two weeks, though this estimate is only marginally significant and may reflect care-seeking behavior rather than improvements in health.

Our main results report intent-to-treat effects based on randomized access to the chatbot, regardless of actual usage. We additionally estimate treatment-on-the-treated effects by instrumenting actual chatbot usage with randomized assignment. Chatbot usage reduces the likelihood of receiving any prescription by 27 percentage points, increases diagnostic testing by 16 percentage points, and lowers two-week revisits by 7.0 percentage points. These effects are roughly five times larger than the corresponding intent-to-treat estimates, reflecting the partial take-up rate of 17\%, and should be interpreted as local average treatment effects for patients who choose to use the chatbot when access is provided.

The effects are concentrated among the physicians for whom the intervention has the greatest scope to alter decisions. Because the chatbot operates by changing patients' information and the requests they make, its influence should depend on a physician's receptiveness to patient input and on the extent of baseline prescribing that AI-assisted patients might constrain. We find that the reduction in prescribing and the increase in testing are concentrated among physicians who self-report being open to patient input, and largely absent among physicians who emphasize strict authority. The reductions are also larger among physicians who prescribe most intensively at baseline, whether measured by their stated attitudes toward these medications or by their actual pre-experiment prescribing.

We further explore whether the effects spill over or persist, either to other patients or beyond the period of the experiment. One possibility is that physicians carry the effect. If repeated interactions with AI-prepared patients led physicians to adjust their practice style, we would expect within-physician spillovers to patients without chatbot access during the experiment, as well as persistence in physician practice after the experiment ends. We find neither. Among patients without chatbot access, clinical outcomes do not differ between those who see \textit{Exposed} physicians and those who see \textit{Unexposed} physicians. We also find no persistence of effects: differences in clinical practice across physician groups disappear immediately after the experiment ends. This null result is economically informative. It suggests that generative AI does not automatically reshape physician practices, at least at the exposure rate in our experiment.

A second possibility is that patients carry the effect. While the intervention produces no lasting change in physician practice, its effects on patients may be more persistent. Following the same patients in the months after the experiment, we find that the increase in diagnostic testing remains, at 1.6 percentage points, while the reduction in prescribing fades. The evidence is suggestive, but it indicates that a single exposure to the chatbot may shift patient behavior in ways that outlast direct access. Taken together, these findings indicate that the chatbot's effects operate through patients' own use and do not diffuse to other patients through changes in physician practice. From a policy perspective, this implies that unequal access to AI tools may translate directly into unequal benefits.

Finally, to understand how patient access to the chatbot affects patient behavior and the physician--patient relationship, we combine administrative medical records with survey data from patients and physicians. We find that access to the chatbot does not materially change observable patient behavior, including appointment attendance, medication purchases, or completion of diagnostic tests. However, survey evidence points to a divergence in perceptions of the consultation experience, despite several individual estimates being imprecise given the smaller survey samples. Patients with chatbot access report lower overall satisfaction, perceive communication with physicians as less smooth, and express lower intended compliance with medical advice. Physicians who interact with AI-prepared patients also report lower patient compliance, but at the same time perceive these patients as easier to communicate with, with clearer symptom descriptions and a better understanding of their condition. The reduction in both intended and observed compliance is consistent with \cite{finkelstein2022taste} and suggests that generative AI may shift informational authority within clinical encounters, reshaping how patients view physician expertise and potentially weakening adherence.

The welfare implications are difficult to assess. The intervention reduces prescribing in categories where overuse is well-documented and increases diagnostic testing, a pattern consistent with welfare improvement if overtreatment is the relevant margin. However, we cannot rule out that some reduced prescriptions represent clinically appropriate care, or that the added testing generates costs without matching benefits. The reduction in compliance is similarly ambiguous. Lower adherence can spare patients unnecessary treatment but can also lead them to decline beneficial care. We do not attempt to sign the net welfare effect. We emphasize instead that the direction of these changes reflects defensive guardrails encoded in AI training, and that the objectives AI developers optimize for need not align with those of the health system.

This study makes multiple contributions to the economics literature. First, and most centrally, we provide the first large-scale evidence on a question that has received little attention: how do the norms and constraints encoded in AI training propagate into real-world decisions at scale? LLMs exhibit systematic behavioral patterns shaped by their design, including tendencies that vary with model architecture, training data, and scale \citep[e.g.,][]{filippas2024large,bowen2025measuring,garg2025artificial,bini2025behavioral,cheung2025large,conlon2026ai,waight2026state}. In high-stakes domains such as healthcare, these patterns are not incidental. Major AI developers have explicitly encoded defensive, liability-driven guardrails into their models, steering them away from specific treatment recommendations and toward caution and diagnostic clarification.\footnote{\cite{chan2026optimal} provides a theoretical framework to study medical liability when artificial intelligence acts as a doctor.}  When such models are deployed as consumer-facing information tools, these encoded constraints enter real-life decisions at scale.\footnote{\cite{luettgau2025people} shows that a large majority of users follow personal advice from AI chatbots, even in high-stakes contexts.} Our findings suggest this propagation mechanism is quantitatively important, with the potential to shift aggregate outcomes such as population-level healthcare utilization.

More broadly, we contribute to the rapidly growing economics of AI literature. Much of the existing work studies generative AI as a production technology that augments worker productivity or substitutes for labor in specific tasks \citep[e.g.,][]{dell2023navigating,peng2023impact,noy2023experimental,cui2024effects,brynjolfsson2025generative,dillon2025shifting,reimers2026ai}. Viewed through the lens of expert--client relationships, a smaller stream examines AI tools that assist experts themselves, including physicians \citep{li2024ai,abaluck2026does,he2026helping}, financial analysts \citep{bertomeu2025impact}, and legal workers \citep{choi2024lawyering}. Much less is known about generative AI as a tool for consumers who face informational disadvantages relative to the experts they rely on. The limited existing work includes \cite{spatharioti2023comparing}, who use online experiments to examine how LLM-based tools help consumers search for and compare products, \cite{ash2025ballotbot}, who randomize voters' access to AI-powered chatbots that answer questions about ballot initiatives, and \cite{shah_levy_ai_justice}, who document that generative AI has substantially increased self-representation in U.S. federal civil courts. By providing patients with access to an LLM-based chatbot prior to outpatient visits, our study offers large-scale randomized evidence on how generative AI can empower informationally disadvantaged clients by reducing their reliance on expert authority and reshaping the dynamics of the expert--client relationship.\footnote{Recent randomized trials in clinical settings have tested integrating LLM tools into hospital workflows. \citet{wan2024outpatient} find that an LLM chatbot assisting nurses at outpatient reception improves patient satisfaction and communication efficiency, and \citet{tao2026llm} show that a pre-consultation chatbot performing clinical history-taking, preliminary diagnoses, and test ordering reduces specialist consultation time and improves coordination. Both studies deploy AI to augment provider functions and find improved patient satisfaction. In contrast, our study deploys AI as a patient-side information tool and examines its effects on both clinical decisions and the broader dynamics of the physician--patient relationship.}

Next, we contribute to a literature on the role of patient information in healthcare. Existing studies typically examine information asymmetries in healthcare by exploiting whether patients are physicians themselves \citep{grytten2011expert,johnson2016physicians,frakes2021great,chen2025physicians} or have family members in the medical profession \citep{artmann2022doctors,chen2022roots,finkelstein2022taste} as sources of variation in patients' information, but the evidence from this literature is mixed. Some studies find that better-informed patients experience less overtreatment \citep{johnson2016physicians}, while others report only limited effects \citep{frakes2021great,artmann2022doctors}. By leveraging LLM-based technology, our study contributes to this body of literature by randomly providing a subset of patients with access to personalized, easy-to-understand, but potentially directional advice on their health conditions.\footnote{Related literature studies other types of health information, including second opinions \citep{mimra2016second}, patient decision aids that provide standardized materials about specific treatment options \citep[surveyed by][]{stacey2017decision,stacey2024decision}, provider report cards and online ratings that disclose information about healthcare provider quality \citep[e.g.,][]{dranove2003more,kolstad2013information,angerer2026value}, and educational campaigns on health risks and the importance of diagnostic testing \citep[e.g.,][]{dupas2011teenagers,lopez2024patients}.} More broadly, the project also relates to studies of health-related information and education programs in developing country settings, as surveyed in \cite{dupas2011health}.

Going beyond clinical decisions, we examine how access to information reshapes patient--physician interactions. Similar to \cite{currie2011patient}, who show that knowledgeable simulated patients constrain physicians' propensity to overprescribe antibiotics but experience less courteous interactions, we find that patients' access to LLM tools reduces prescribing while worsening self-reported experiences. At the same time, patients with access to LLM tools report lower compliance with physician recommendations, consistent with observational evidence that patients with family members in the medical profession exhibit reduced adherence \citep{finkelstein2022taste}. These findings highlight the importance of looking beyond treatment and testing decisions to understand how information technologies reshape relational dynamics in clinical care.

Finally, our study contributes to a broad literature on credence goods markets. These markets are economically significant and pervasive. From real estate \citep{levitt2008market} and financial advice \citep{mullainathan2012market} to pharmacy \citep{bronnenberg2015pharmacists}, automobile and computer repair \citep{schneider2012agency,busse2017repairing,rasch2018drives,kerschbamer2023credence}, and taxi services \citep{balafoutas2013drives,balafoutas2017second}, consumers routinely delegate decisions to better-informed experts whose recommendations they cannot easily verify. A common thread across these settings is that informational asymmetries between experts and clients create scope for overtreatment, fraud, or misallocation. Consistent with this, \cite{balafoutas2013drives}, \cite{busse2017repairing}, and \cite{kerschbamer2023credence} show that reducing informational asymmetries can discipline expert behavior. Our study extends this literature by showing that generative AI can serve as a scalable consumer-side information technology, shifting expert--client relationships in one of the most consequential credence goods markets: outpatient healthcare.

\section{Institutional Background}
\label{sec:background}
\subsection{Generative AI in Healthcare: Adoption and Guardrails}
Generative AI has rapidly become a major channel through which people access health information. Health-related queries are among the most common use cases of large language models: OpenAI reports that approximately 230 million people discuss their health with ChatGPT each week,\footnote{\url{https://openai.com/index/introducing-chatgpt-health/}.} and surveys indicate that roughly one in six American adults consults an AI chatbot for health information at least once a month \citep{kff2024poll}. This growth has occurred largely ahead of formal regulatory frameworks, creating a setting in which AI systems interact with patients at massive scale without the oversight structures that govern licensed medical professionals.\footnote{The regulatory debate over AI's role in healthcare is actively evolving. California's Assembly Bill 489, signed into law in October 2025, directly addresses the boundary between AI and licensed medical practice: it prohibits developers and deployers of AI systems from using any terms, phrases, or design elements that imply the system is providing care from a licensed healthcare professional, and authorizes state licensing boards to investigate and penalize violations. \url{https://leginfo.legislature.ca.gov/faces/billNavClient.xhtml?bill_id=202520260AB489}.}

In response to safety concerns and legal exposure, major AI developers have encoded defensive guardrails into their models for health-related interactions. The liability motive is stated explicitly. OpenAI's published guidelines list protecting the company from legal and reputational harm as one of the objectives of model behavior.\footnote{\url{https://model-spec.openai.com/2025-04-11.html}. We cite the version in effect during our experimental period.} On medical matters, the guidelines instruct the models to provide information but not definitive advice, to include disclaimers directing users to licensed professionals, and, for mental health, to refrain from providing diagnoses or recommending specific treatments or medications. Anthropic has similarly stated that Claude is designed to include contextual disclaimers, acknowledge uncertainty, and direct users to healthcare professionals for personalized guidance.\footnote{\url{https://www.anthropic.com/news/healthcare-life-sciences}.} Audit studies confirm that these stated policies manifest in model behavior: disclaimers and referral advice scale with the clinical urgency of the query, differ across developers, and shift in form over time \citep[e.g.,][]{reis2026disclaimers,sharma2025longitudinal}.

Because general-purpose AI models are not subject to the pre-market review or professional oversight that governs medical practice, these guardrails function as self-imposed substitutes for formal regulation, and their content reflects the objectives of developers, which need not coincide with those of the health system. The restraint is also uneven. The guardrails restrict the actions most likely to create liability for the developer, recommending treatments or medications and providing diagnoses, while directing users toward licensed professionals and imposing no comparable restraint on encouraging further evaluation. When patients consult these tools, this defensive and potentially directional logic reaches them at scale, and may reshape their healthcare decisions.

\subsection{Healthcare Market and Institutional Context in China}
Healthcare expenditures have risen rapidly worldwide. Economic research points to both demand-side and supply-side mechanisms as drivers of high healthcare spending. On the demand side, beyond the classic moral hazard problem,\footnote{China has near-universal health insurance coverage through two main public schemes: the Urban Employee Basic Medical Insurance (UEBMI), which covers formal-sector workers, and the Urban and Rural Resident Basic Medical Insurance (URRBMI), which covers rural residents and non-working urban populations.} patients often lack the medical knowledge needed to evaluate the necessity of care and may request treatments that are not clinically indicated. Physicians may comply with such requests, contributing to overutilization \citep[e.g.,][]{kravitz2005influence,kotwani2010factors,linder2014time,finkelstein2016sources,lopez2022does}. On the supply side, physician characteristics and financial incentives directly shape treatment decisions, generating systematic variation in medical expenditures, including overspending and inefficiencies \citep[e.g.,][]{iizuka2012physician,das2016quality,finkelstein2016sources,cutler2019physician,badinski2023geographic,currie2024first}.\footnote{Moreover, \cite{gottschalk2020health} document that patients with higher socioeconomic status are significantly less likely to receive unnecessary treatment recommendations, indicating that physicians’ propensity to overtreat depends on patient characteristics.}

Overutilization, particularly in the form of overprescription, is well documented in China \citep[e.g.,][]{li2012overprescribing,lu2014insurance,fang2021physician}. The healthcare system in China is predominantly state-run, with public hospitals delivering the vast majority of services. Despite ongoing payment reforms, hospitals continue to operate largely under a fee-for-service payment model and receive limited direct government subsidies. As a result, hospital revenue remains closely tied to service volume. Pharmaceutical sales occur primarily within hospitals rather than through independent retail pharmacies. Between 2014 and 2019, drug expenditures represented more than 60 percent of inpatient spending and roughly 70 percent of outpatient spending in public hospitals.\footnote{China Health Statistical Yearbook, various years.} These shares are substantially higher than those observed in many high-income health systems, where prescription drugs account for a considerably smaller fraction of total health expenditures.\footnote{Across OECD countries, pharmaceutical spending typically accounts for roughly 15–20\% of total health expenditures. \url{https://www.oecd.org/en/data/indicators/pharmaceutical-spending.html}.}

Within this broad pattern of overprescription, three drug categories draw particular policy scrutiny: Traditional Chinese Medicine (TCM), antibiotics, and opioids. TCM is widely prescribed in Chinese public hospitals but is characterized by less standardized evidence on effectiveness and side effects,\footnote{Under new regulations effective July 1, 2026, TCMs that do not clearly disclose contraindications, adverse reactions, or precautions will not be approved upon re-registration. Reports indicate that over 70 percent of currently approved products may be affected. \url{https://www.news.cn/politics/20260129/853dc80d18fb43d2b1d61676c46b5675/c.html}.} and often higher or more variable pricing. Antibiotic overprescription has been well documented in China \citep{currie2013social,currie2014addressing} and in the United States \citep{fleming2016prevalence}, raising concerns about antimicrobial resistance. Opioids, while less prevalent in China than in the United States \citep{maclean2020economic}, have drawn substantial international attention due to concerns about misuse and addiction. We focus on these three categories because they are highly policy relevant and have been central to regulatory efforts and public health debates.

Motivated by this institutional setting, we examine how patient access to generative AI affects clinical decisions in China. With access to AI, patients can arrive better informed, less likely to request unnecessary treatments, and better positioned to constrain physicians' overprescription. The information they obtain may not be neutral, however. Shaped by developer guardrails, it may push clinical decisions away from medications and toward diagnostic testing rather than simply reducing unnecessary care uniformly.

Patient access to AI is likely to be especially consequential in the Chinese outpatient setting, where patients have few institutional sources of guidance. In contrast to systems organized around general practitioners, patients in China schedule directly with specialists, without referral or triage by a family doctor who helps interpret their condition or provides continuity of guidance across visits.\footnote{In credence-good markets, repeated interaction is one of the few mechanisms that can discipline expert behavior even absent formal oversight \citep{dulleck2006doctors,fong2022trust}. In the Chinese outpatient system, specialist visits offer no substitute for this missing continuity, as the same patient rarely sees the same specialist repeatedly.} Outpatient consultations are brief, averaging roughly nine minutes per visit in our partnering hospital, which constrains explanation, counseling, and patient engagement. With neither an intermediating generalist nor sufficient consultation time, patients have little means of obtaining or interpreting medical information on their own. In this setting, the potential demand for AI advice is high, and so is the scope for its influence on the consultation.

\section{Experimental Design and Data}
\label{sec:design}
We conducted the experiment in partnership with a large public hospital in Sichuan Province. The hospital employs over 200 outpatient physicians who handle approximately 900 visits per day on average. As a city-level tertiary hospital, it is representative of the mid-tier public institutions that serve roughly 40\% of China's population and account for 35\% of nationwide outpatient visits, supporting the external validity of our findings. The patient population is predominantly rural and has limited exposure to digital technologies, providing a low baseline of AI adoption that minimizes the risk of background usage diluting the treatment effect. The hospital's electronic medical record system captures detailed visit-level data on diagnoses, prescriptions, diagnostic test orders, and other clinical details. Its online appointment platform allows us to randomize chatbot access at the time patients book their appointments.

\subsection{Intervention}
We partnered with a startup that specializes in AI solutions for the healthcare industry. Our intervention is a generative AI chatbot powered by a large language model built on a widely used general-purpose architecture, adapted for healthcare-related interactions, and integrated into the hospital's online appointment system. Immediately after a patient completes an online appointment booking, which occurs on average one day before the visit, those randomly assigned to the treatment group receive a personalized link to the chatbot. Simultaneously, the hospital sends a separate message encouraging patients to use the tool prior to their visit. Panels (a) and (b) of Figure \ref{fig:experiment} present the chatbot's initial landing screen and consultation interface.

The chatbot supports pre-visit patient preparation by eliciting symptom descriptions, interpreting and analyzing patient health concerns, and providing general medical information. Throughout all interactions, the interface prominently reminds users that the chatbot does not replace professional medical advice and that clinical decisions should follow physicians' recommendations. The system automatically logs all user interactions, including timestamps and message content. These records provide an objective measure of engagement with the intervention, which we link to administrative and clinical outcomes using unique patient identifiers. This structure enables estimation of both intent-to-treat and treatment-on-the-treated effects.

The intervention may affect both communication and clinical decision-making through several channels. First, by prompting patients to organize and describe symptoms and to review relevant health information before the visit, the chatbot can improve pre-visit preparation and facilitate clearer communication during the consultation. This is particularly important in our setting, where the average outpatient visit lasts approximately nine minutes, making efficient and precise communication essential. Beyond communication, access to structured medical information can improve patients' understanding of their health conditions, typical diagnostic pathways, and the role of medications and tests. Better-informed patients may therefore be less likely to request unnecessary treatments or prescriptions, reducing demand-driven overuse. In addition, the advice patients receive may not be neutral. Because guardrails encoded in AI training caution against medication use, patients may arrive carrying advice that tilts away from medications and toward diagnostic testing. Finally, by narrowing informational gaps, the chatbot may shift the balance of authority within the consultation. Patients who arrive with greater knowledge and more structured questions are better positioned to question recommendations and constrain physicians' discretion to overprescribe or overtreat. Through these channels, the intervention can reshape the clinical interaction between patients and physicians.

\subsection{Two-Layer Randomization}

We implement a two-layer randomized experimental design to identify the direct effects of providing patients with access to the chatbot while creating variation to detect within-physician spillovers. We first randomize physicians into an \textit{Exposed} group (two-thirds) or an \textit{Unexposed} group (one-third), stratified by seniority, education, and specialty. Among patients scheduling appointments with \textit{Exposed} physicians, half are randomly assigned to the \textit{Treatment} group, with chatbot access, and half to the \textit{Control} group, without access to the chatbot. Patients scheduling appointments with \textit{Unexposed} physicians do not receive chatbot access, so \textit{Unexposed} physicians never interact with AI-assisted patients.

\begin{figure}[h!]
    \centering

    \begin{subfigure}[b]{0.48\textwidth}
        \centering
        \subcaption{Initial Interface}
        \includegraphics[width=\textwidth]{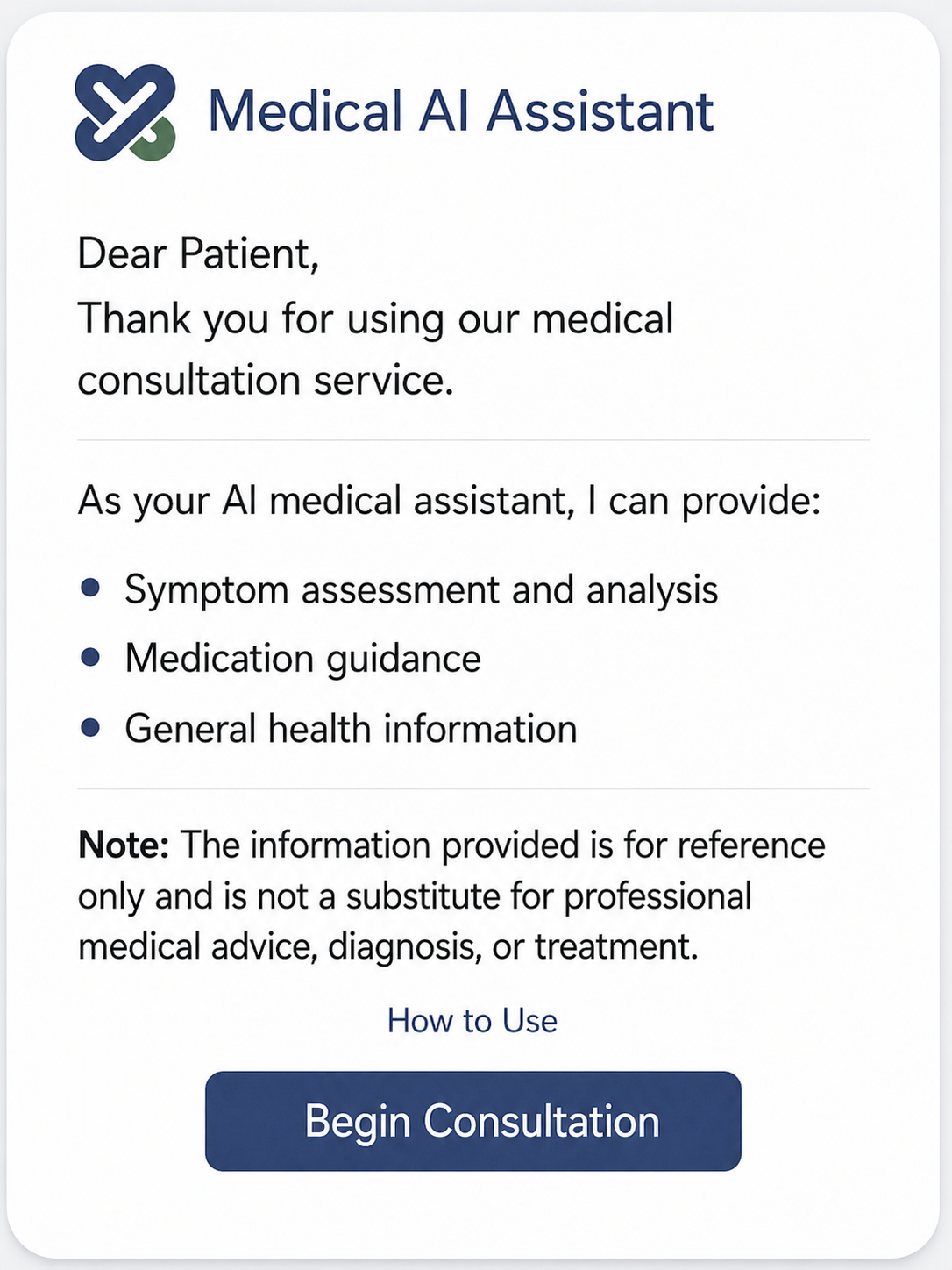}
        \label{fig:interface1}
    \end{subfigure}
    \hfill
    \begin{subfigure}[b]{0.48\textwidth}
        \centering
        \subcaption{Consultation Interface}
        \includegraphics[width=\textwidth]{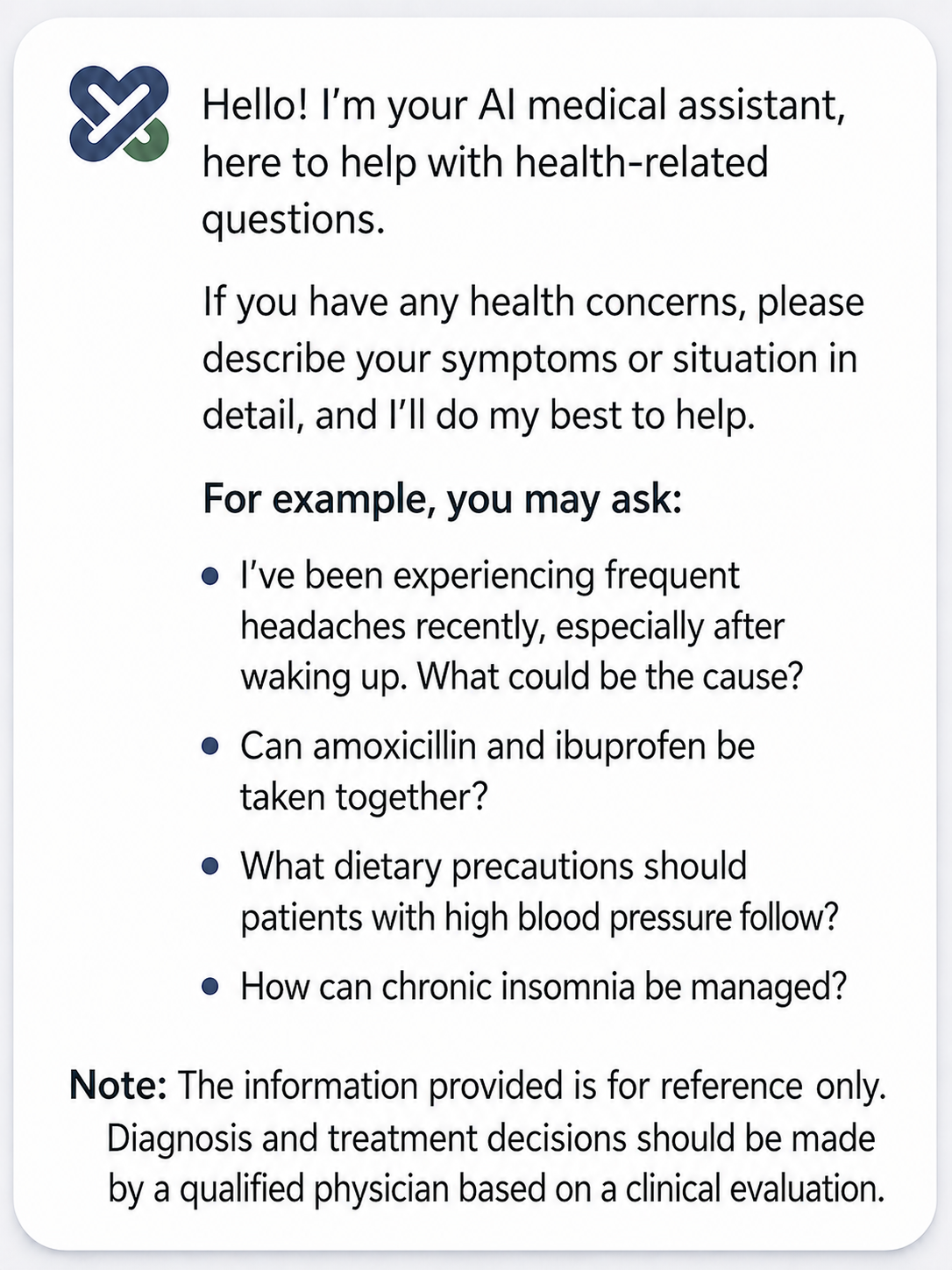}
        \label{fig:interface2}
    \end{subfigure}
    \medskip

      \begin{subfigure}[b]{1\textwidth}
        \centering
        \subcaption{Experimental Design}
        \includegraphics[width=\textwidth]{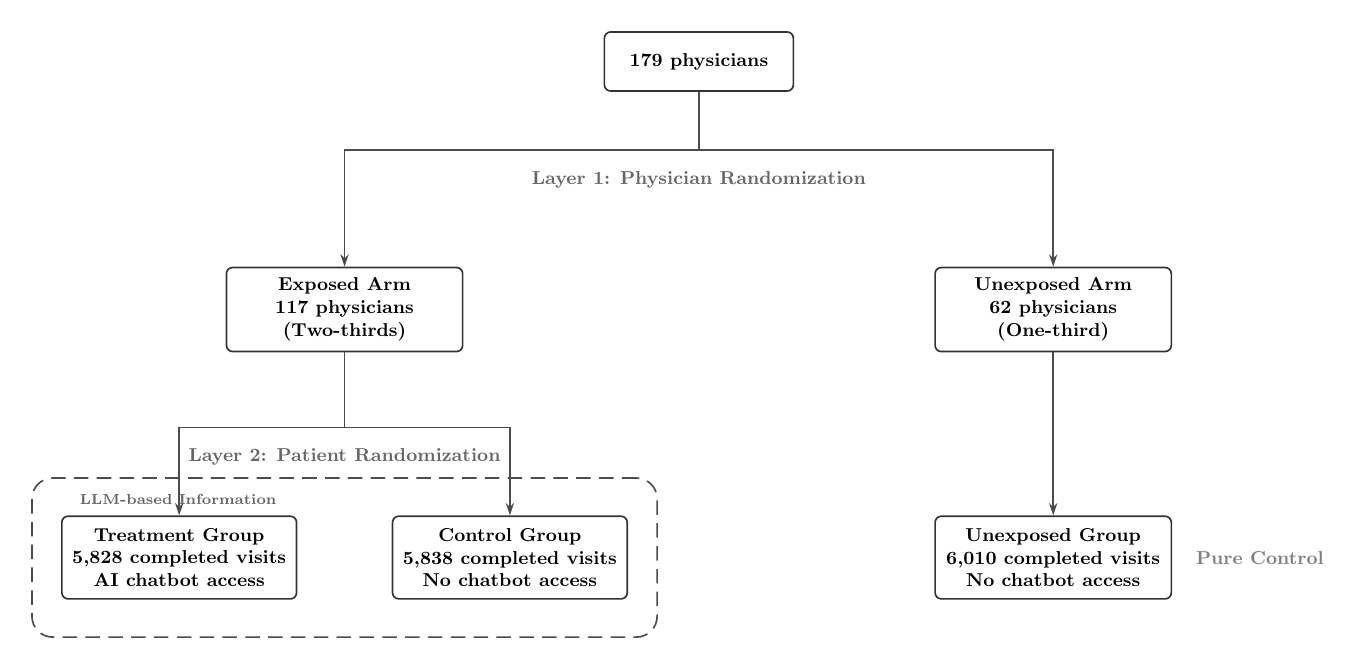}
        \label{fig:flowchart}
    \end{subfigure}

    \caption{Medical AI Chatbot User Interfaces and Experimental Design Flowchart}
    \label{fig:experiment}
\end{figure}

Panel (c) of Figure \ref{fig:experiment} summarizes the two-layer randomization. This design generates three analytically distinct groups. Comparisons between the \textit{Treatment} and \textit{Control} groups identify the direct effect of patient access to the chatbot holding the physician fixed. Comparisons between the \textit{Control} and \textit{Unexposed} groups detect within-physician spillovers arising from physician learning or behavioral adjustment following repeated interactions with AI-prepared patients, since neither group has chatbot access and the two differ only in their physicians' exposure.

The three analytical groups are formally labeled as:
\begin{enumerate}
\singlespacing
\item \textbf{Treatment:} Patients with chatbot access, seeing \textit{Exposed} physicians.
\item \textbf{Control:} Patients without chatbot access, seeing \textit{Exposed} physicians.
\item \textbf{Unexposed:} Patients without chatbot access, seeing \textit{Unexposed} physicians.
\end{enumerate}

\subsection{Data and Sample}

The experiment was conducted between June 2 and July 3, 2025, at our partnering hospital. Because the intervention is delivered through the online appointment system, we restrict the sample to patients who scheduled outpatient visits online during this period. Patients who walked in without an online appointment, approximately one third of total outpatient visits, are excluded from both the experiment and the analysis sample.\footnote{We additionally exclude pediatric patients from the experiment. Approximately 0.6\% of patients who booked an appointment but did not complete payment are also excluded from the analysis sample.}

During the experimental period, 179 physicians provided outpatient care. Of these, 117 (two thirds) were randomized to the \textit{Exposed} group and 62 (one third) to the \textit{Unexposed} group. Panel A of Table \ref{tab:balance_test} reports balance tests at the physician level and shows that the two groups are well balanced across observed characteristics, including age, gender, seniority, education, and specialty, with no statistically significant differences.

During the experimental period, 12,498 online appointments with \textit{Exposed} physicians were randomized with equal probability into the \textit{Treatment} or \textit{Control} group at the time of booking, and 6,435 appointments were scheduled with \textit{Unexposed} physicians. Patients did not show up for 6.6 percent of these appointments. An unattended appointment generates no clinical record and therefore cannot enter the analysis. The analysis sample accordingly comprises 11,666 completed visits with \textit{Exposed} physicians and 6,010 with \textit{Unexposed} physicians. The unit of observation is the outpatient visit rather than the patient, so patients who visited multiple times during the experimental period contribute multiple observations. We show in Section \ref{sec:relationship} that attrition is balanced across groups, indicating that chatbot access did not affect the decision to attend, so the restriction to completed visits does not introduce differential selection.

Panel B of Table \ref{tab:balance_test} reports balance tests at the visit level, comparing the \textit{Treatment} and \textit{Control} groups as well as the \textit{Treatment} and \textit{Unexposed} groups. Across these comparisons, we find no statistically significant differences in age, gender, occupation, first-time visit status, or advance booking time.\footnote{First-visit status equals one if the visit is the patient's first visit to the hospital during the experimental period.} The average patient is approximately 45 years old, and a large share report working as farmers (44\%) or being unemployed or retired (16\%). This demographic profile suggests that baseline familiarity with generative AI tools is likely to be limited in this population. Appointments are booked, on average, approximately 25 hours in advance, giving treated patients adequate opportunity to use the chatbot if they choose to do so.

\paragraph{Medical Records.} We observe administrative outpatient medical records for all visits during the study period. These records include patient and physician identifiers, visit timestamps, diagnoses, detailed drug prescriptions, and orders for diagnostic examinations. For each prescribed drug and diagnostic test, we observe the listed price, which captures medical expenditures prior to insurance reimbursement. These data allow us to construct visit-level measures of clinical decision-making, treatment intensity, and spending outcomes.

\paragraph{Patient Survey.}
We invited patients to complete a brief post-consultation survey on the day of the visit or the following day, offering a monetary incentive of 2 to 10 RMB. The survey elicits patients' overall satisfaction with the visit, perceived quality of communication, and intended adherence to medical advice. For patients in the \textit{Treatment} group, the survey includes additional questions on chatbot usage and perceived usefulness. In total, we collected 2,247 patient responses, corresponding to a 13 percent response rate. The sample includes 640 respondents from the \textit{Treatment} group, 789 from the \textit{Control} group, and 818 from the \textit{Unexposed} group. Response rates are somewhat lower in the \textit{Treatment} group than in the two comparison groups; we therefore interpret survey-based outcomes with caution.

\paragraph{Physician Survey.}
Following the conclusion of the one-month experiment, we surveyed all physicians who conducted outpatient consultations during the study period. Of 179 physicians, 171 completed the questionnaire, yielding a response rate of 95.5 percent. The survey elicits physicians' assessments of their patients along four dimensions: clarity of symptom descriptions, understanding of their own health conditions, ease of communication during the visit, and potential adherence to medical advice. Both \textit{Exposed} and \textit{Unexposed} physicians were also asked to estimate how frequently their patients used AI tools prior to visits.

Taken together, the physician and patient surveys complement the administrative data by validating treatment exposure and providing insight into how AI-provided information affects consultation dynamics and the patient experience.

\subsection{Chatbot Take-up}
Although all patients in the \textit{Treatment} group received free access to the chatbot prior to their consultation, actual usage was voluntary. Linking the usage logs to the hospital administrative system, we find that the chatbot was used before 998 of the 5,828 visits in the \textit{Treatment} group (17.1 percent). Patients in the \textit{Control} group were not given access for their visit and therefore could not use the tool.\footnote{Because randomization occurs at the appointment level, a patient with multiple visits can be assigned to different groups across visits; chatbot links, however, are appointment-specific and expire after the visit.} Appendix Table \ref{tab:IV_first_stage} reports first-stage estimates of the relationship between treatment assignment and chatbot usage. Assignment to the \textit{Treatment} group increases the probability of chatbot use by 17 percentage points.

Take-up among treated patients is self-selected and varies systematically with observable characteristics. Appendix Table \ref{tab:adopter_char} shows that, conditional on being assigned access, younger patients, men, employed patients, and first-time visitors are significantly more likely to use the chatbot. This selective adoption pattern has two implications for the analyses that follow. First, because only a subset of treated patients actively use the chatbot, intent-to-treat estimates reflect the effect of offering access at this take-up rate. We also use random assignment as an instrument for actual usage to estimate treatment-on-the-treated effects in Section \ref{sec:ATT}. Second, the demographic profile of adopters suggests that the benefits of generative AI tools may not be evenly distributed across patient populations, a distributional concern we revisit in Section \ref{sec:spillover}.

\section{Main Results}
\subsection{The Directional Content of Patient--AI Conversations\label{sec:conv}}

We begin by examining how adopters in the \textit{Treatment} group used the chatbot and what their conversations reveal about the advice they received.

Our analytic sample consists of 1{,}192 conversation turns from 956 visits, where each turn comprises one patient message and the chatbot's reply, and a single conversation may span multiple turns.\footnote{The usage logs record chatbot use before 998 visits. Conversations for 42 of these visits could not be matched to hospital records and are excluded, leaving 1{,}192 conversation turns from 956 visits.} Two-thirds of patient messages (66.7\%) describe symptoms and request a diagnosis or underlying cause. The remainder inquire about treatment options (11.7\%), logistical matters (6.6\%), lifestyle and self-care (5.5\%), whether to undergo specific tests (4.9\%), the interpretation of test results or prior diagnoses (3.1\%), and a small residual of greetings and uninformative messages (1.4\%). Appendix~\ref{sec:appendix-intent} reports the full distribution of intent labels along with example messages and details the classification procedure.

We next examine how the AI engages with four clinically relevant topics that map directly onto our outcomes of interest: diagnostic testing, general medication (excluding TCM and antibiotics), TCM, and antibiotics. We identify which topics a response covers through sentence-level keyword matching, and a single response typically spans several of them. For each topic a response covers, we use GPT-4o to classify the AI's stance along two independent dimensions. The first is whether the response \emph{recommends} the test or treatment, that is, whether it endorses or suggests its use. The second is whether the response raises a \emph{caution}, such as a warning about risks or side effects, a contraindication, or advice against use. We keep the two dimensions separate because they often co-occur. Beyond outright advising against a treatment or recommending a treatment or test with no reservation, the chatbot frequently recommends a treatment while warning about its side effects and cautioning against self-use. Separating recommendation from caution lets us distinguish a clean recommendation from a hedged one. Every mention therefore falls into one of four categories: recommend without caution, recommend with caution, caution without recommending, or neither. Appendices~\ref{sec:appendix-topics} and~\ref{sec:appendix-stance} provide the prompts and the full classification procedure, and Table~\ref{tab:conversation} reports the results.

Two patterns stand out. First, it is the AI, not the patient, that typically introduces these topics into the conversation. Patients mention general medication in only 11.4\% of turns and diagnostic testing in 8.5\% of turns, yet the AI raises medication in 58.9\% of replies and testing in 58.2\% of replies. The chatbot is not merely responding to patient queries; it actively shapes the informational content that patients carry into the clinical encounter.

Second, the AI's stance differs sharply between medications and diagnostic testing. When the AI discusses medication, it usually raises cautions, warning about side effects, contraindications, dosing concerns, or the risks of self-medication. Such warnings appear in 69.7\% of general medication mentions, 90.8\% of TCM mentions, and 87.6\% of antibiotic mentions. Clean recommendations, those without any accompanying caution, are rare for medications, particularly for TCM (3.8\%) and antibiotics (7.8\%). Diagnostic testing is treated very differently: when the AI discusses testing, it issues clean recommendations in 94.5\% of cases, with cautions appearing in only 3.3\% of mentions.

These patterns are consistent with the incentives facing LLM developers in high-stakes domains. Safety guardrails in health-related interactions serve both to protect users from the risks of acting on AI medical advice without professional oversight and to limit developers' legal exposure. Both rationales weigh against recommending specific treatments or medications, where adverse outcomes carry clinical and legal consequences. Neither weighs against recommending diagnostic testing, whose costs fall on patients and the health system rather than on the developer. As a result, AI responses are highly cautious about medications and tend to encourage diagnostic testing, a stance that need not align with the objectives of the health system.

The directional pattern documented above could in principle be specific to the model our partner deployed rather than a general feature of LLM-based medical advice. To probe this, we submit the same first-turn patient messages to six widely used large language models: GPT-4o, GPT-5.5, DeepSeek-V3, Claude Sonnet 4.6, Qwen3.5-plus, and Gemini 3.5 Flash. To approximate our setting, we use two system prompts, one framing the model as a pre-visit assistant and one as a direct medical consultant, and classify the replies with the same topic-detection and stance-classification pipeline used for our chatbot.\footnote{Appendix~\ref{sec:appendix-crossmodel} details the procedure and prompts. Because we cannot reconstruct the follow-up turns of each conversation, the comparison uses first-turn messages only, and restricts our chatbot to its first-turn responses to match. The resulting baseline shares therefore differ slightly from Table~\ref{tab:conversation}, which uses the full analytic sample.} Figures~\ref{fig:crossmodel_previst} and~\ref{fig:crossmodel_direct} report the share of clean recommendations by topic.

The comparison shows the same directional asymmetry across models. Every model, under both prompts, recommends diagnostic testing far more cleanly than medications, with testing recommended without caution in 92\% to 99\% of mentions, close to the rate for our chatbot. The degree of caution toward medications nonetheless varies substantially: clean-recommendation rates for medications range from 10\% to 56\% across models, differing across developers and even across models from the same developer.\footnote{GPT-4o, for example, recommends medications without caution more than half the time. This is partly driven by its frequent use of deferral phrasing such as ``the doctor may prescribe\ldots,'' which we classify as a clean recommendation even though it defers the decision to the physician.} Among the comparison models, our chatbot's stance is closest to that of GPT-5.5, a leading frontier model by OpenAI, across all four topics.

The variation across models is itself informative. If the stance toward medications were a neutral reflection of clinical evidence, it would not differ so widely across models built by different developers. That it does indicates that the strength of the caution reflects design choices rather than the underlying medical evidence. The information patients receive from the chatbot is, in this sense, not a neutral distillation of medical knowledge but a view shaped by the priorities of those who build and govern these systems. Treated patients carry the chatbot's advice into their visits. Whether the encoded caution toward medications and the encouragement of diagnostic testing propagate into clinical decisions is the question we turn to next, using the medical records.

\subsection{Clinical Outcomes Across Experimental Groups}
\begin{figure}[t]
    \centering

    \begin{subfigure}[b]{0.48\textwidth}
        \centering
        \subcaption{Medication Prescription}
        \includegraphics[width=\textwidth]{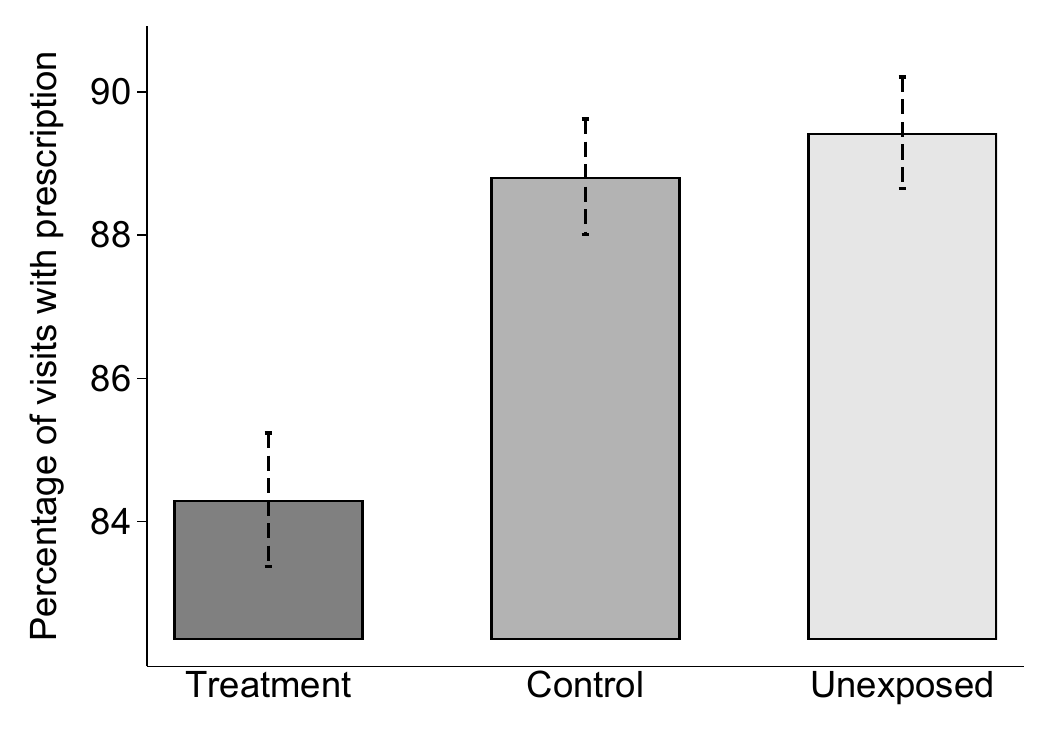}
    \end{subfigure}
    \hfill
    \begin{subfigure}[b]{0.48\textwidth}
        \centering
        \subcaption{TCM}
        \includegraphics[width=\textwidth]{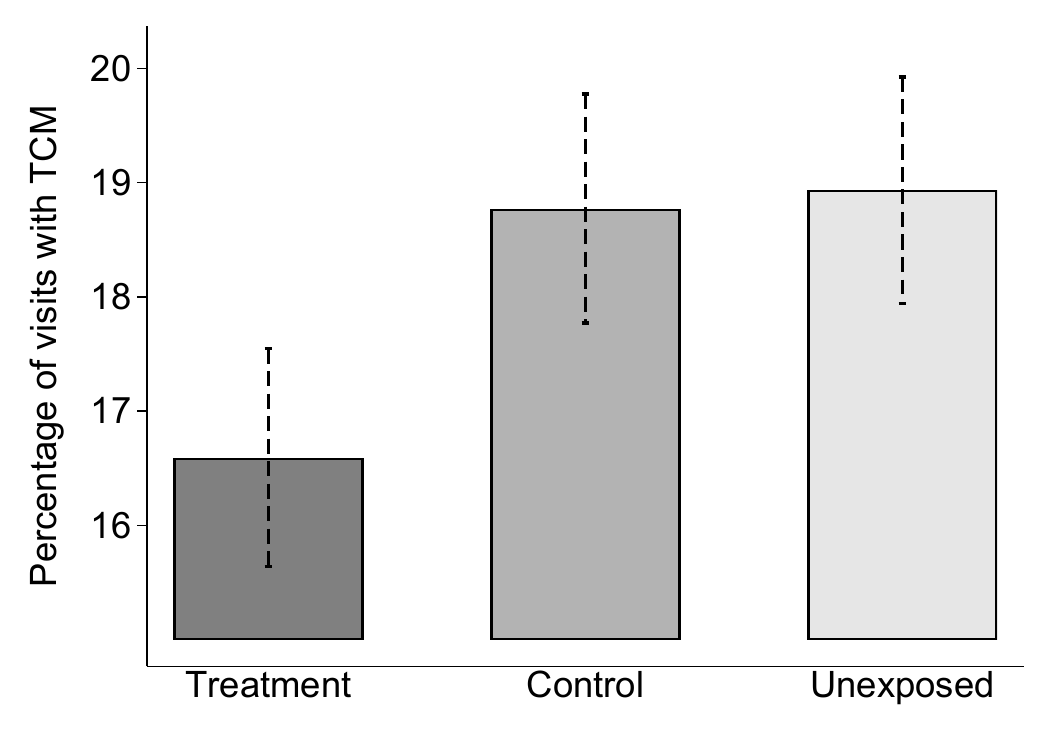}
    \end{subfigure}
    \medskip

    \begin{subfigure}[b]{0.48\textwidth}
        \centering
        \subcaption{Antibiotics}
        \includegraphics[width=\textwidth]{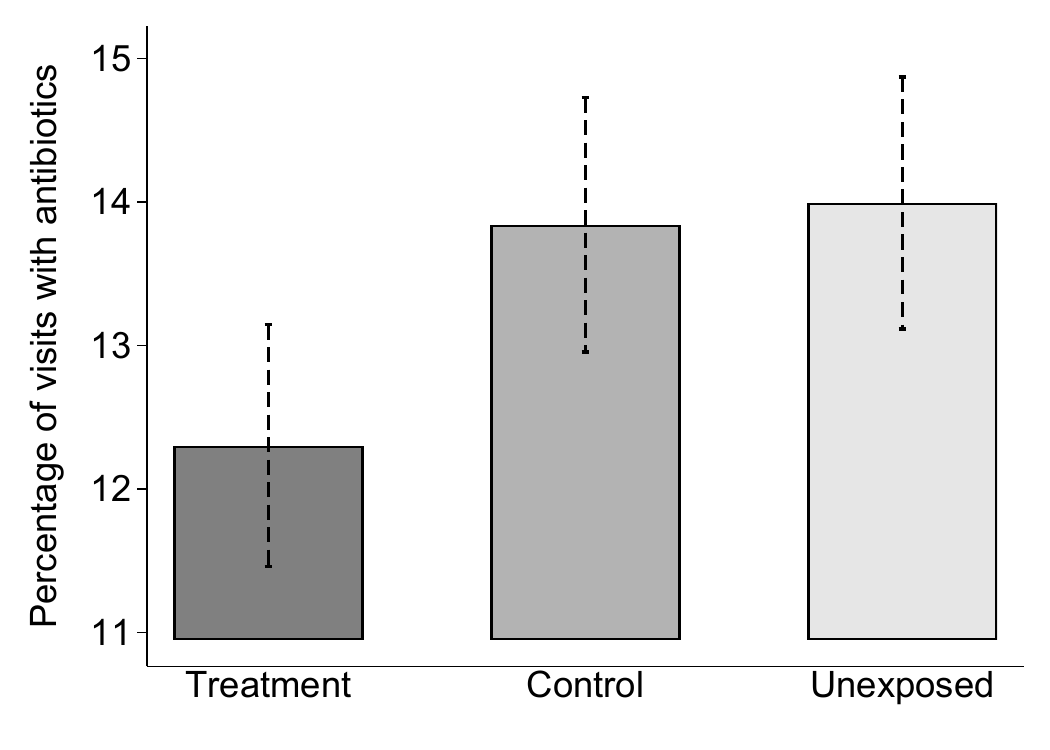}
    \end{subfigure}
    \hfill
    \begin{subfigure}[b]{0.48\textwidth}
        \centering
        \subcaption{Opioids}
        \includegraphics[width=\textwidth]{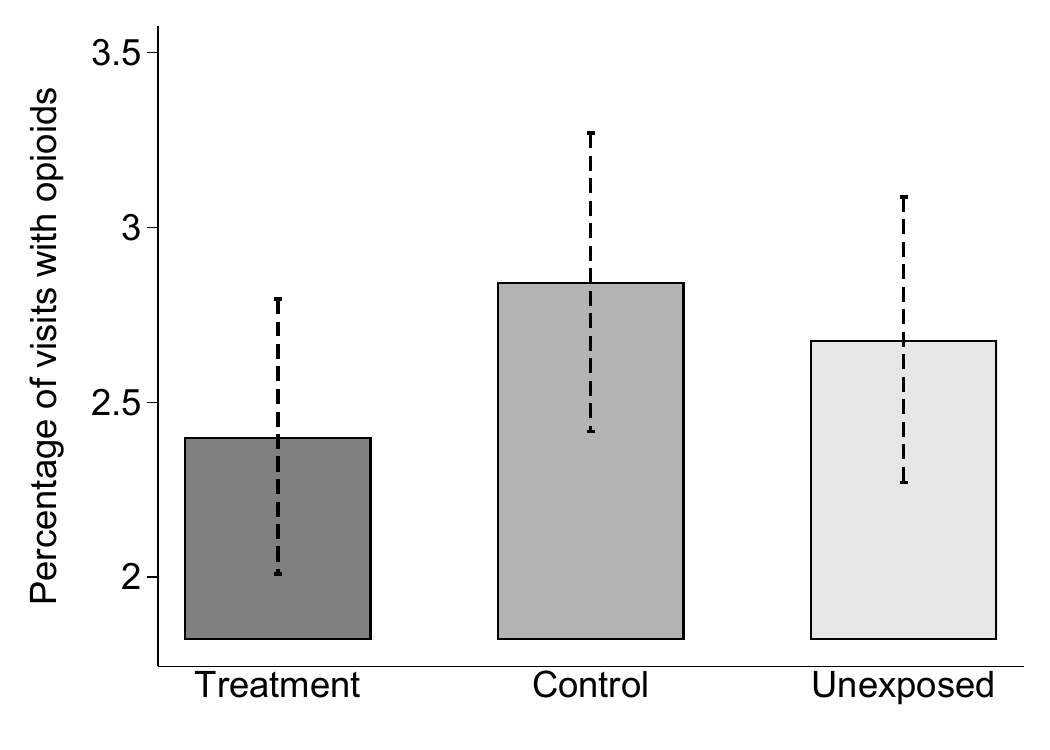}
    \end{subfigure}
    \medskip

    \begin{subfigure}[b]{0.48\textwidth}
        \centering
        \subcaption{Diagnostic Tests}
        \includegraphics[width=\textwidth]{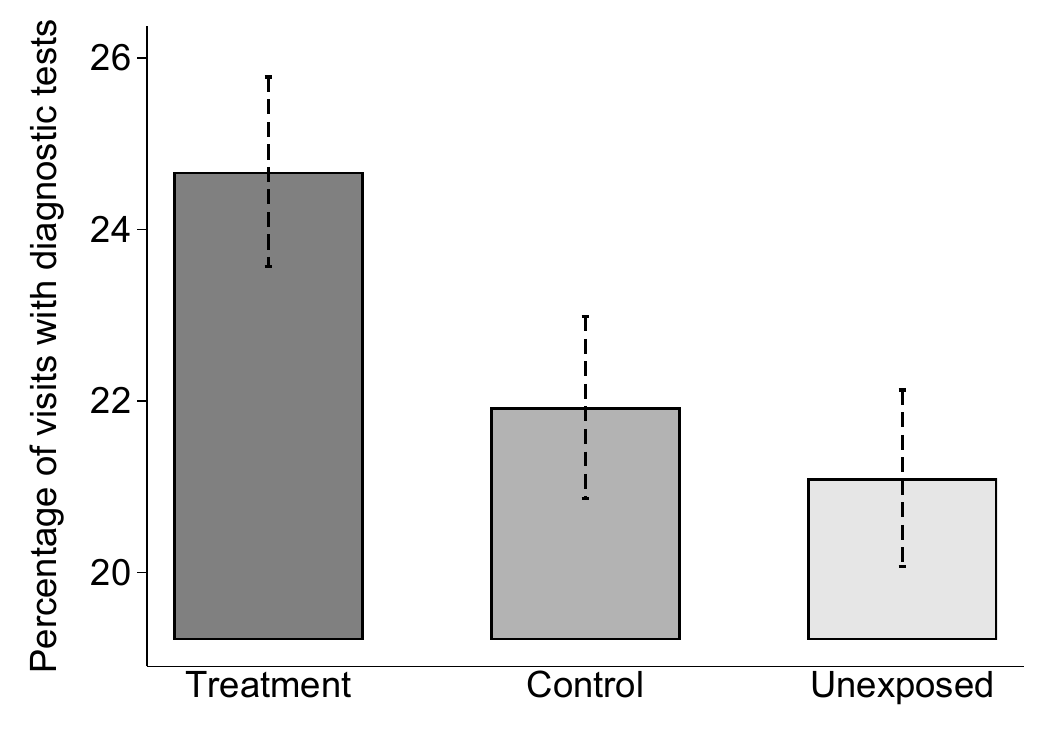}
    \end{subfigure}
    \hfill
    \begin{subfigure}[b]{0.48\textwidth}
        \centering
        \subcaption{Revisit within Two Weeks}
        \includegraphics[width=\textwidth]{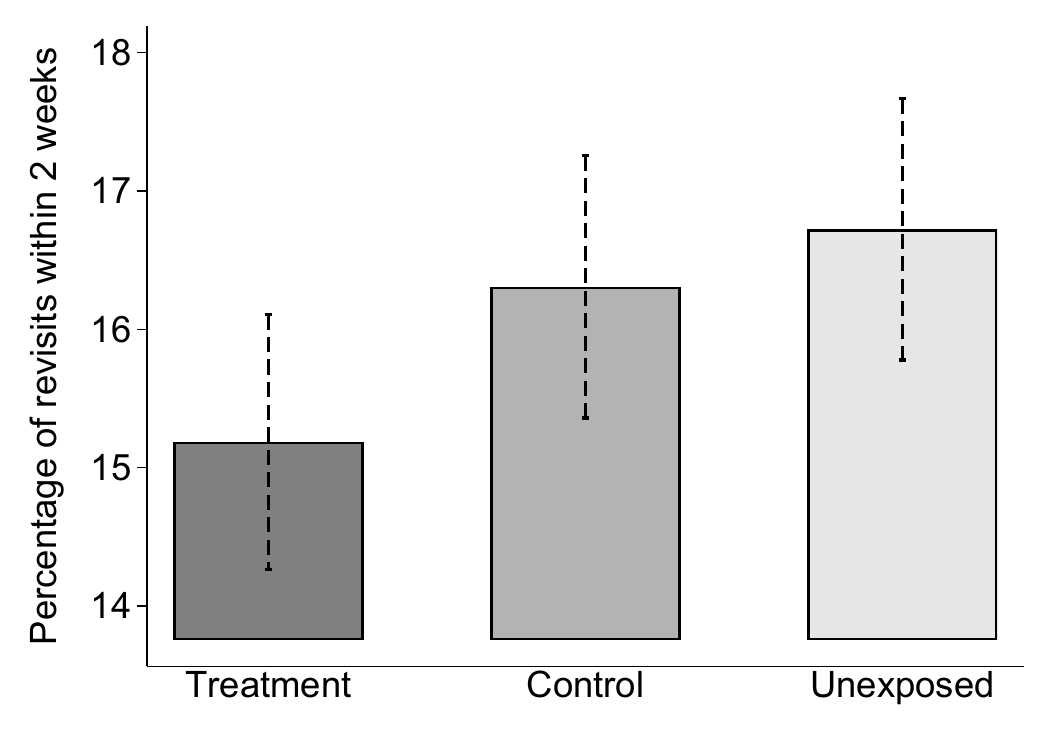}
    \end{subfigure}

    \caption{Prescriptions and Clinical Practices by Treatment Status}
    \label{fig:outcomes_practice}
    \medskip
    \footnotesize
\begin{minipage}{\textwidth}
\textit{Notes:} This figure reports mean clinical outcomes for patients in the \textit{Treatment}, \textit{Control}, and \textit{Unexposed} groups, with 95\% confidence intervals (dashed lines).
\end{minipage}
\end{figure}

We now turn to the clinical records to test whether this directional advice is reflected in what happens at the visit. We graphically present clinical outcomes for the \textit{Treatment}, \textit{Control}, and \textit{Unexposed} groups over the experimental period, beginning with prescribing behavior. Panel (a) of Figure \ref{fig:outcomes_practice} shows that access to the chatbot reduces the likelihood that a visit results in any prescription. The share of visits with at least one prescribed medication is 4.5 percentage points lower in the \textit{Treatment} group than in the \textit{Control} group (significant at the 5\% level), where the baseline prescribing rate is 88.8\%. The prescribing rate in the \textit{Treatment} group is also significantly lower than in the \textit{Unexposed} group, while rates in the \textit{Control} and \textit{Unexposed} groups are statistically indistinguishable.

We next examine prescribing patterns across major medication categories, including Traditional Chinese Medicine (TCM), antibiotics, and opioids. Panels (b) and (c) of Figure \ref{fig:outcomes_practice} show sizable reductions in TCM and antibiotic prescribing in the \textit{Treatment} group, with statistically significant differences relative to both comparison groups. Opioid prescribing is also lower among treated patients, but the differences relative to the \textit{Control} and \textit{Unexposed} groups are not statistically significant. This is expected given the low baseline rate of opioid prescribing in China, below 3\% in both the \textit{Control} and \textit{Unexposed} groups, as shown in Panel (d).

In contrast, physicians seeing treated patients are more likely to order diagnostic tests. Panel (e) shows that 21--22\% of visits in the \textit{Control} and \textit{Unexposed} groups involve at least one diagnostic test, compared with 24.7\% in the \textit{Treatment} group. The increase in testing is statistically significant in both comparisons. This pattern of reduced prescribing and increased diagnostic testing is consistent with the directionality in the AI advice we observe from the conversation logs.

We next assess whether these changes in clinical management are associated with differences in patient outcomes. One outcome observed in our administrative data is whether a patient revisits the same hospital within two weeks of the initial consultation. Panel (f) of Figure \ref{fig:outcomes_practice} shows that patients in the \textit{Treatment} group are 1.1 percentage points less likely to revisit within two weeks than patients in the \textit{Control} group, a 7\% reduction that is statistically significant. The same conclusion holds when comparing treated patients to the \textit{Unexposed} group, whose two-week revisit rate is similar to that of the \textit{Control} group. The reduction in short-term revisits is consistent with improvements in the effectiveness of initial consultations, but it may also reflect other mechanisms. Better-prepared patients may feel less need to seek follow-up care even absent changes in clinical quality. We therefore do not interpret this reduction as direct evidence of improved clinical outcomes. In this rural setting, where patients often face substantial travel costs, reductions in revisits nonetheless have welfare implications, even if the underlying channels remain unclear.

\subsection{Within-Physician Intent-to-Treat Effects}
Because outcomes are similar in the \textit{Control} and \textit{Unexposed} groups, our main regression analyses restrict the sample to visits with \textit{Exposed} physicians.\footnote{We return to the \textit{Unexposed} group in later analyses to formally test for spillovers.}
Within this sample, we estimate the causal effects of offering access to the LLM-based chatbot on clinical outcomes using the following specification:
\vspace{-0.3em}
\begin{equation}
Y_{i} = \alpha + \beta\,\text{Treated}_{i} + \gamma X_{i} + \delta_{p(i)} + \varepsilon_{i}
\end{equation}
\noindent where $i$ indexes an outpatient visit and $p(i)$ denotes the physician seen at that visit. A patient who books more than one appointment during the experiment period contributes more than one visit, and can fall into different treatment groups.\footnote{Some patients visit the hospital more than once during the experiment period. To address the concern that the same patient could be assigned to different treatment conditions across visits, Appendix Table \ref{tab:itt_practice_first} restricts the sample to each patient's first visit during the experimental period, which retains 70\% of visits. The main results are robust: chatbot access continues to significantly reduce overall prescribing and TCM prescribing and to increase diagnostic testing. The point estimates shift in line with the lower baseline prescribing rate and higher baseline testing rate of first visits, and the smaller estimates for antibiotics and two-week revisits are not statistically significant in this reduced sample.} The outcome variable $Y_{i}$ represents outcomes from the consultation: prescriptions, diagnostic tests, and healthcare expenditures. The key explanatory variable, $\text{Treated}_{i}$, is a binary indicator equal to one if visit $i$ was randomly assigned to chatbot access prior to the consultation. We include a vector of patient characteristics $X_{i}$ (age, gender, occupation, and first-time visit status) and physician fixed effects $\delta_{p(i)}$. The error term $\varepsilon_{i}$ captures unobserved factors affecting individual visits. All regressions are estimated using robust standard errors.\footnote{Standard errors are not clustered because randomization occurs at the individual appointment level \citep{abadie2023should}. Appendix Tables \ref{tab:itt_practice_cluster}--\ref{tab:iv_costs_cluster} report estimates with standard errors clustered at the physician level. The core results on prescribing and diagnostic testing remain statistically significant. The two-week revisit effect is no longer marginally significant under clustering, and the reduction in Western Medicine expenditure is no longer significant once patient characteristics and physician fixed effects are added.}

The coefficient of interest, $\beta$, identifies the \textit{Intent-to-Treat} (ITT) effect of being assigned chatbot access, regardless of actual usage. This approach preserves the integrity of the randomization and provides policy-relevant estimates of the effect of \textit{offering} AI tools in a real-world clinical setting. It is important to note that if within-physician spillovers occur, for example, if physicians adjust their general practice patterns for all patients after interacting with AI-informed ones, the control group may also be indirectly affected. In that case, our estimate of $\beta$ understates the true effect of chatbot exposure.

Table \ref{tab:itt_practice} reports within-physician intent-to-treat estimates of the effects of offering chatbot access on clinical practice patterns among visits with \textit{Exposed} physicians. Consistent with the graphical evidence, the regression results show that access to the chatbot significantly reduces the likelihood that a visit results in any prescription by 4.6 percentage points relative to a sample mean of 87\%. This decline is driven partly by reductions in prescribing of Traditional Chinese Medicine and antibiotics, which fall by 2.2 and 1.5 percentage points, respectively, while opioid prescribing declines but the effect is not statistically significant. At the same time, chatbot access increases the probability that a visit involves any diagnostic test by 2.7 percentage points relative to a sample mean of 23\%. Finally, treated patients are about 1.2 percentage points less likely to revisit the same hospital within two weeks; the effect is marginally statistically significant. Estimates are nearly identical with and without controls for patient characteristics and physician fixed effects, confirming that the randomization successfully balanced patient and physician characteristics.

Table \ref{tab:itt_costs} reports intent-to-treat effects on healthcare expenditures, measured as the sum of listed prices for prescribed drugs and diagnostic tests before insurance reimbursement. Consistent with the prescribing changes documented in Table \ref{tab:itt_practice}, chatbot access reduces medication expenditures, with reductions in both TCM and Western Medicine expenditures significant at the 10\% level. In the specification that controls for patient characteristics and physician fixed effects, TCM expenditures decline by approximately 1.76 RMB, a 13\% reduction relative to the sample mean, while Western Medicine expenditures decline by about 5.0 RMB relative to a sample mean of 64.8 RMB. Diagnostic expenditures increase by roughly 6.3 RMB, though this estimate is not statistically significant. Because these changes offset, we find no statistically significant effect of chatbot access on total healthcare expenditures.

\subsection{Treatment-on-the-Treated Effects}\label{sec:ATT}

The results above report intent-to-treat effects based on randomized access to the chatbot, regardless of whether patients actively used the tool. We now estimate treatment-on-the-treated effects, which recover the average effect of chatbot usage among patients who complied with the assignment. We instrument actual chatbot usage with randomized assignment to the \textit{Treatment} group, using the same within-physician specification as in the intent-to-treat analysis.

Table \ref{tab:iv_practice} reports the effects of chatbot usage on clinical practice patterns. Consistent with the intent-to-treat results, chatbot usage is associated with a 27 percentage point reduction in the probability that a visit results in any prescription, partly driven by declines of about 13 percentage points in Traditional Chinese Medicine prescribing and 9 percentage points in antibiotic prescribing. At the same time, chatbot usage increases the likelihood of diagnostic testing by 16.1 percentage points and reduces the probability of a two-week revisit by 7 percentage points. These magnitudes are approximately five times larger than the corresponding intent-to-treat estimates, reflecting partial take-up of the chatbot (17.1\% in the \textit{Treatment} group). The instrumental variable estimates therefore represent local average treatment effects for patients who use the chatbot when offered access. Overall, the scale of these effects suggests that generative AI tools can meaningfully alter clinical decision-making in ways that, if adopted broadly, could translate into substantial changes in healthcare utilization at the system level.

Table \ref{tab:iv_costs} reports the local average treatment effects on healthcare expenditures. Chatbot usage is associated with sizable reductions in medication expenditures, with TCM expenditures declining by 10 RMB and Western Medicine expenditures declining by roughly 30 RMB, both statistically significant at the 10\% level. These reductions are offset by an increase in diagnostic expenditures of approximately 37 RMB, although this estimate is imprecisely estimated. For both Tables \ref{tab:iv_practice} and \ref{tab:iv_costs}, first-stage F-statistics exceed 1,100 across specifications, indicating a strong first stage.

\subsection{Heterogeneous Treatment Effects}
The effects documented above could vary across physicians. If the chatbot operates by changing what patients know and ask for, its influence on clinical decisions could depend on how receptive each physician is to patient input and on whether their baseline practice involves overprescribing that treated patients can help constrain. We examine heterogeneity along both dimensions, using measures from the physician survey and from pre-experiment prescribing records.

We first split physicians by their stance toward patient input, using two measures from the physician survey. We asked physicians whether they view medical knowledge acquired by patients from external sources as helpful and whether patients should strictly follow physicians' recommendations. These measures capture variation in physician openness and authority style. Panel A of Table \ref{tab:hte_physician_attitudes} shows that the reduction in medication prescribing and increase in diagnostic testing are substantially larger among physicians who report being highly open to patient-acquired health knowledge. In contrast, Panel B shows that the treatment effects are significantly smaller among physicians who place strong emphasis on physician authority. These patterns suggest that the chatbot moves clinical decisions mainly when physicians are receptive to what patients bring to the visit, and far less when physicians emphasize their own authority.\footnote{We note one alternative channel for the increase in diagnostic testing. Rather than responding to patient requests for testing, physicians facing better-informed and less deferential patients may order additional tests as a defensive response to perceived legal and dispute risk \citep[e.g.,][]{kessler1996doctors,studdert2005defensive,carroll2021physicians,liu2023short,fang2025fear}. Our design cannot separate the two channels directly. That the testing increase is concentrated among physicians receptive to patient input and absent among those emphasizing authority is, however, more consistent with physicians' accommodation of patient requests than with a defensive response.}

We next ask whether the effects are concentrated where baseline practice is most intensive.\footnote{One potential explanation for heterogeneity in prescribing intensity is financially motivated physician-induced demand \citep[e.g.,][]{gruber1994physician,liu2009financial,iizuka2012physician,lu2014insurance,fang2021physician}. However, such incentives are limited in our institutional setting. Physicians' salaries and bonuses are not tied to drug sales, and hospitals operate under a zero-markup policy for medications \citep{yi2015intended,yip201910,fang2021physician}.} Table \ref{tab:hetero_medication_attitudes} examines whether the treatment effects are larger among physicians who express more favorable views toward prescribing specific medications. In the physician survey, we elicit attitudes toward the use of Traditional Chinese Medicine, antibiotics, and opioids, and classify physicians according to whether they report more favorable or more conservative views toward each category. In column (1) of Table \ref{tab:hetero_medication_attitudes}, the interaction between treatment assignment and a more pro-TCM attitude is negative and statistically significant, indicating that the decline in TCM prescribing is substantially larger among physicians who are more inclined to prescribe TCM. A similar, though imprecisely estimated, pattern appears for antibiotics in column (2). We find no differential effects for opioids, consistent with their low baseline prescribing rates in China.

We then turn to physicians' revealed prescribing behavior prior to the experiment. Specifically, we classify physicians according to whether their prescribing rates for TCM, antibiotics, and opioids between January and May 2025, before the start of the experiment, exceed the median within their department. This measure captures baseline treatment intensity based on actual prescribing patterns rather than self-reported attitudes. Table \ref{tab:hetero_dept_high_prescriber} reports the results. The reduction in medication use is significantly larger among physicians with higher baseline prescribing intensity. In particular, the decline in TCM prescribing is concentrated among physicians whose pre-experiment TCM prescribing rates were above the departmental median. A similar pattern appears for antibiotics, while we find no differential effects for opioids, again consistent with their low baseline prescribing rates in China.

Taken together, the heterogeneity indicates that the treatment effects are concentrated where the intervention has the greatest potential to alter decisions. They are larger among physicians who are receptive to patient input, for whom both the reduction in prescribing and the increase in diagnostic testing are more pronounced, and among physicians with the most intensive baseline prescribing, where there is more scope to reduce medication use.

\section{Dynamic, Spillover, and Persistence Effects \label{sec:spillover}}
The effects documented so far arise from patients' direct use of the chatbot during the experiment. We now ask whether they reach further: whether they persist in physician practice after the experiment, spill over to patients without access, or endure for treated patients at later visits.

\subsection{Dynamic Effects}
To study the temporal dynamics of the intervention, we implement an event-study framework. This approach serves three purposes. First, it provides a diagnostic check on the physician-level randomization by testing for differential pre-trends between \textit{Exposed} and \textit{Unexposed} physicians. Second, it characterizes how outcomes for \textit{Exposed} physicians evolve over the course of the intervention. Third, it allows us to assess whether any effects persist after the formal intervention period ends.

We aggregate the data to the physician-week level and estimate the following specification:
\begin{equation}\label{eq:dynamic}
Y_{pt} = \alpha + \sum_{k \neq -1} \beta_k \cdot \bigl(\text{Exposed}_{p} \times \mathds{1}\{\text{Week}_t = k\}\bigr) + \delta_p + \tau_t + \varepsilon_{pt},
\end{equation}
where $Y_{pt}$ denotes the outcome for physician $p$ in week $t$, such as the rate of Traditional Chinese Medicine prescribing or diagnostic testing. The variable $\text{Exposed}_{p}$ equals one for physicians randomized to the \textit{Exposed} group. The indicator $\mathds{1}\{\text{Week}_t = k\}$ indexes $k$ weeks relative to the intervention start date of June 2, 2025. The coefficient for $k=-1$, corresponding to the week from May 26 to June 1, 2025, is normalized to zero.

The coefficients of interest are $\{\beta_k\}_{k=-12}^{12}$. For $k < -1$, the estimated coefficients capture differences in outcomes between \textit{Exposed} and \textit{Unexposed} physicians prior to the intervention and provide a test for pre-existing differential trends. For $k \geq 0$, the coefficients trace the dynamic response of outcomes for \textit{Exposed} physicians relative to \textit{Unexposed} physicians. The intervention period extends through day 5 of week $k=4$; coefficients for subsequent weeks therefore capture post-intervention dynamics. Weeks earlier than $k=-12$ and later than $k=+12$ are binned at the respective endpoints. All specifications include physician fixed effects $\delta_p$, which absorb time-invariant differences across physicians, and week fixed effects $\tau_t$, which control for common temporal shocks. Standard errors are clustered at the physician level to account for serial correlation within physicians over time. Physician-week observations are weighted by the number of patient visits.

Figure \ref{fig:dynamic_weighted} presents the results for medication prescribing rates (overall, Traditional Chinese Medicine, antibiotics, and opioids), diagnostic testing, and revisit rates. For all outcomes, we observe no evidence of pre-trends prior to the experiment, suggesting balance between \textit{Exposed} and \textit{Unexposed} physicians. Weeks 0 through 3 correspond to the four full weeks of the intervention period, so the coefficients $\beta_0$--$\beta_3$ capture the treatment effect and its evolution during the experiment. Because the analysis compares outcomes between \textit{Exposed} and \textit{Unexposed} physicians, the estimated effects are mechanically attenuated. Only half of the patients seen by \textit{Exposed} physicians are assigned access to the chatbot, and take-up among eligible patients is incomplete (17.1\%). As a result, the magnitudes of the estimates are expected to be small and can be interpreted as roughly one-half of the corresponding intent-to-treat effects and approximately one-tenth of the average treatment-on-the-treated effects.

As shown in Figure \ref{fig:dynamic_weighted}, we observe clear reductions in overall medication prescribing and Traditional Chinese Medicine prescribing beginning in week 1, while the estimated effects on antibiotic and opioid prescribing are substantially noisier. Starting in week 1, we also observe a clear increase in diagnostic testing rates, along with a noisy and statistically insignificant decrease in revisit rates. While point estimates vary across weeks, including a null effect in week 0, we find no statistically significant evidence that the treatment effects change between week 1 and week 3. The experiment concludes in the middle of week 4. Beginning in week 4, we find no differences in any of these outcomes between \textit{Exposed} and \textit{Unexposed} physicians. This pattern indicates no persistence in physicians' clinical practices following the intervention. Once the experiment ends, the gap between \textit{Exposed} and \textit{Unexposed} physicians fully closes.

\subsection{Within-Physician Spillovers}
Having shown that any effect on physician practice disappears once the experiment ends, we next test for within-physician spillovers during the intervention period itself. Specifically, we examine whether interactions with patients who have access to the chatbot affect outcomes for patients without access who see the same physician. To do so, we adopt the event study framework in equation (\ref{eq:dynamic}) with one modification. We exclude all visits assigned access to the chatbot. This restriction allows us to compare outcomes for patients without chatbot access who see \textit{Exposed} physicians (\textit{Control} group) with outcomes for patients without chatbot access who see \textit{Unexposed} physicians (\textit{Unexposed} group). If within-physician spillovers or rebound effects are present, we would expect to observe differences in outcomes between these two groups of patients, even though neither group has access to the chatbot.

Appendix Figure \ref{fig:placebo_weighted} reports the results for the same set of clinical outcomes. Across all outcomes, we find no statistically significant differences between patients randomly assigned to the \textit{Control} group and those assigned to the \textit{Unexposed} group. This finding is consistent with the patterns observed in Figure \ref{fig:outcomes_practice}. Combined with the absence of persistent effects documented in Figure \ref{fig:dynamic_weighted}, we are able to conclude that interactions with AI-prepared patients do not generate spillovers to other patients. In particular, physicians do not appear to update their clinical practices in ways that affect patients who do not have access to the chatbot, either contemporaneously or after the intervention ends. This pattern suggests that the effects of the chatbot operate primarily through direct patient use rather than through lasting changes in physician behavior.

One potential concern is that exposure to AI-prepared patients may generate spillovers across physicians, for example through informal communication or peer learning. Such spillovers would bias our estimates of persistence and within-physician spillovers toward zero by contaminating the comparison group if \textit{Unexposed} physicians indirectly update their practice over time. Several features of our setting mitigate this concern. First, physicians conduct consultations independently in separate rooms, limiting opportunities to observe or discuss specific patient interactions. Second, prescribing patterns among \textit{Unexposed} physicians who never interact with treated patients remain stable over time. Third, for across-physician spillovers to drive the within-physician estimates to zero, the magnitude of across-physician learning would need to be comparable to the magnitude of within-physician learning generated by direct exposure to treated patients. We consider across-physician learning of that scale unlikely in our setting. Taken together, these patterns indicate that across-physician spillovers are unlikely to change the qualitative interpretation of our results.

From a policy perspective, these results raise concerns about unequal access and uptake. Because chatbot adopters are disproportionately younger, male, and employed, the benefits of generative AI tools may flow primarily to already advantaged groups, potentially widening existing health disparities. Policies aimed at expanding access, improving usability, or targeting underserved populations may therefore be necessary to ensure that generative AI tools benefit a broader patient population.

\subsection{Patient-Level Persistence}

The intervention produces no lasting change in physician practice and does not spill over to patients without access. We now ask whether it persists for treated patients at their later visits. We track the same patients over October to December 2025, three to six months after the intervention ended.\footnote{The chatbot link expires immediately after the patient's visit, so treated patients have no further access during the later period.} We first ask whether access to the chatbot affects patients' subsequent outpatient utilization, measured by whether they have any visit and by their number of visits. We then ask whether, for patients with at least one visit, their prescriptions and diagnostic tests differ.\footnote{A patient may visit more than once during the experiment, with different visits falling into different treatment groups, so we classify each patient by the assignment of their first experimental-period visit.}

\begin{figure}[ht]
    \centering

    \begin{subfigure}[b]{0.48\textwidth}
        \centering
        \subcaption{At Least One Visit}
        \includegraphics[width=\textwidth]{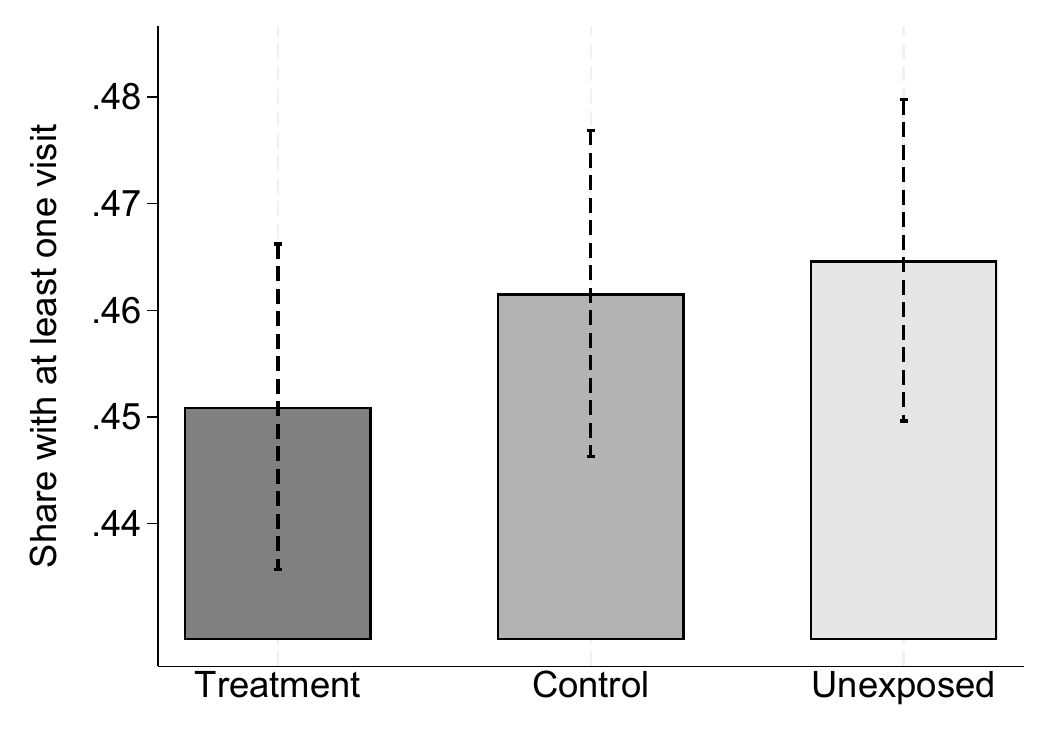}
        \label{fig:patient_visitrate}
    \end{subfigure}
    \hfill
    \begin{subfigure}[b]{0.48\textwidth}
        \centering
        \subcaption{Number of Visits}
        \includegraphics[width=\textwidth]{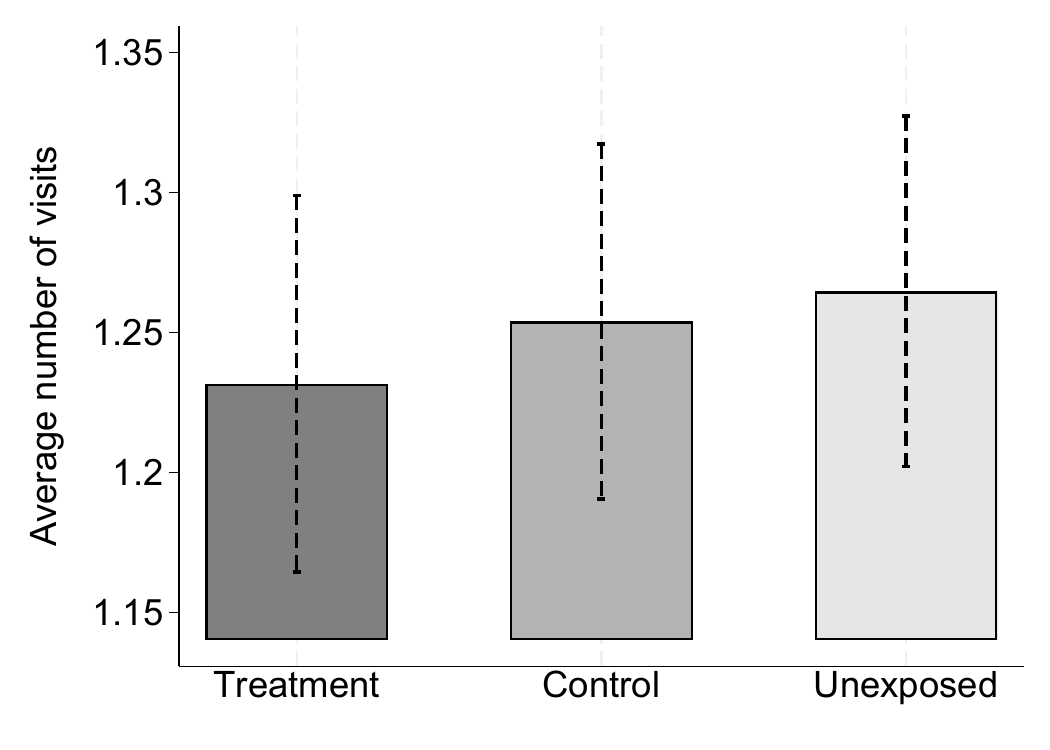}
        \label{fig:patient_visitcount}
    \end{subfigure}
    \medskip

    \begin{subfigure}[b]{0.48\textwidth}
        \centering
        \subcaption{Prescribed Medication}
        \includegraphics[width=\textwidth]{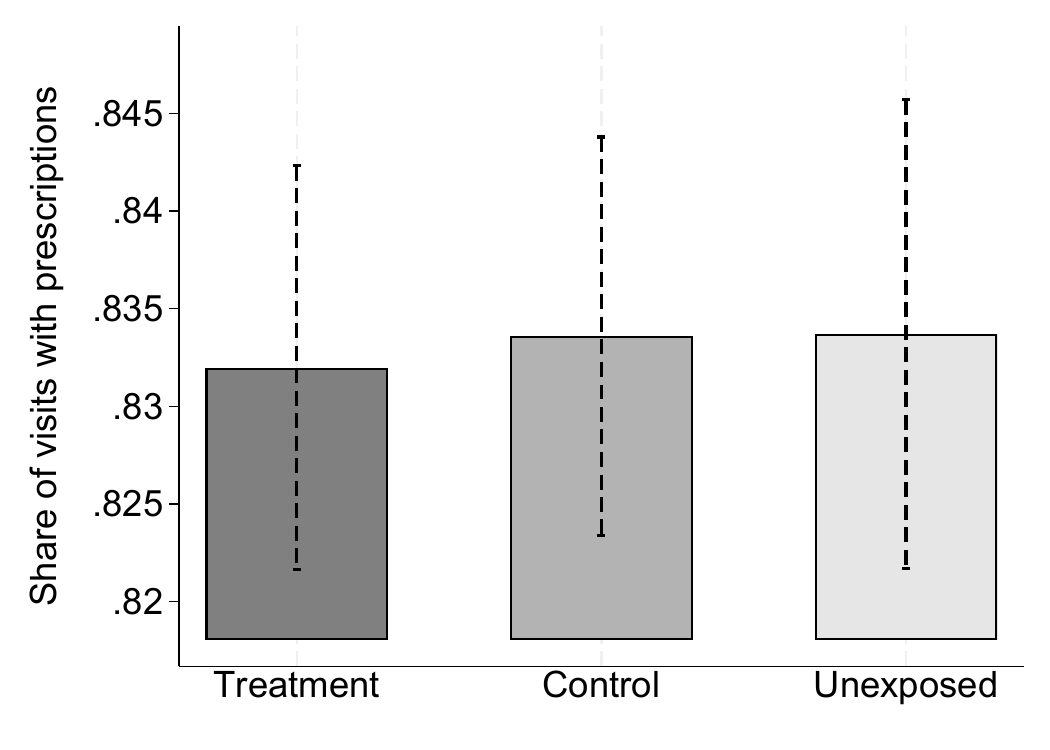}
        \label{fig:visit_rx}
    \end{subfigure}
    \hfill
    \begin{subfigure}[b]{0.48\textwidth}
        \centering
        \subcaption{Diagnostic Tests}
        \includegraphics[width=\textwidth]{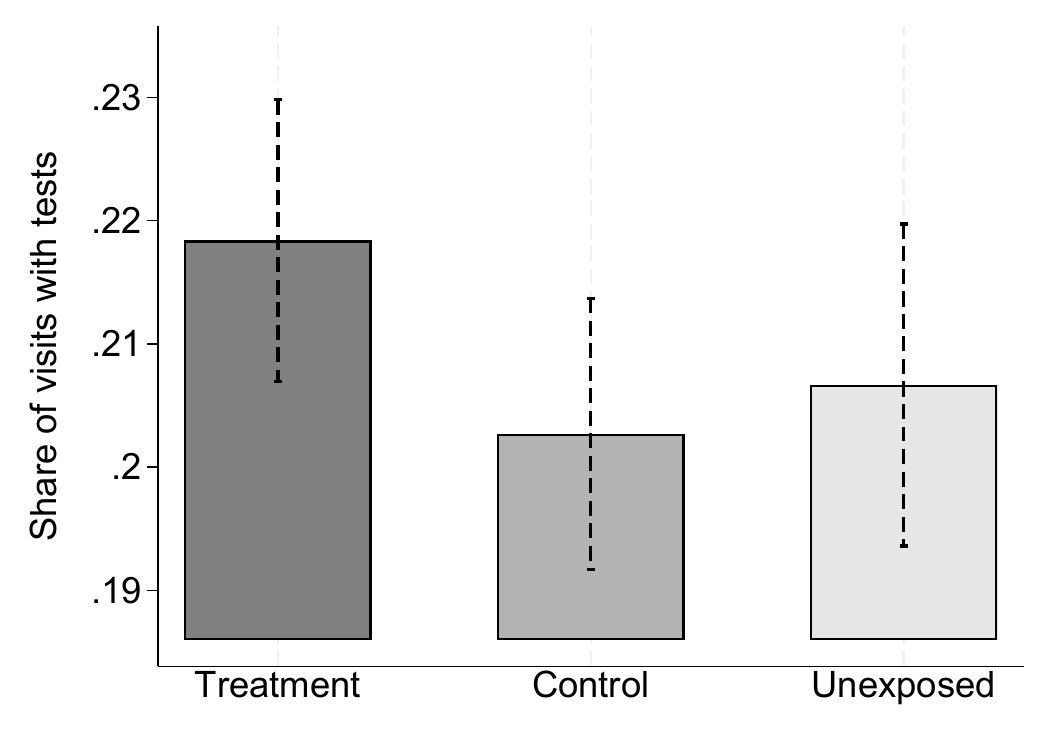}
        \label{fig:visit_test}
    \end{subfigure}
    \caption{Persistent Effects of Treatment Assignment on Outpatient Utilization and Clinical Practice in the Post-Experiment Period (October–December 2025)}
    \label{fig:patient_outcomes_oct_dec}
    \medskip
    \begin{minipage}{\textwidth}
    \footnotesize
    \textit{Notes:} This figure examines whether assignment to chatbot access during the experiment continues to affect outpatient utilization and clinical practice after the experiment ended. The experiment ran from June 2 to July 3, 2025. We track the same patients from October to December 2025 by matching their subsequent outpatient visits through unique patient identifiers. Because a patient may appear in more than one experimental-period visit with differing assignments, we group each patient by their first such visit. Among patients whose first visit was with an \textit{Exposed} physician, those whose first visit received chatbot access form the \textit{Treatment} group and the remainder form the \textit{Control} group, while patients whose first visit was with an \textit{Unexposed} physician form the \textit{Unexposed} group. Panels (a) and (b) report patient-level outcomes computed for every patient in each group. Panel (a) is the share with at least one outpatient visit between October and December 2025, and Panel (b) is the average number of visits. Panels (c) and (d) report visit-level outcomes among patients with at least one outpatient visit between October and December 2025. Panel (c) is the share of visits with a prescribed medication, and Panel (d) is the share with a diagnostic test. Bars denote group means, and dashed lines denote 95\% confidence intervals.
    \end{minipage}
\end{figure}

We begin with subsequent use of outpatient care. Panels (a) and (b) of Figure \ref{fig:patient_outcomes_oct_dec} show that access to the chatbot does not change patients' later use of care. Between October and December 2025, treated patients are slightly less likely to have any visit and make slightly fewer visits than control patients, but neither difference is statistically significant, and Appendix Table \ref{tab:persistent_patient_level} confirms these null results. Although chatbot access marginally reduces two-week revisits, it has no detectable effect on utilization months later, when visits are more likely to be for unrelated conditions. The absence of a difference in utilization also matters for the clinical-practice results that follow. Because treated and control patients have later visits at similar rates, any difference in their prescriptions and tests is less likely to be driven by selection.

We then examine clinical practice at the later visits, reported in Panels (c) and (d) of Figure \ref{fig:patient_outcomes_oct_dec}. The reduction in prescribing has faded, and treated and control patients are now equally likely to receive a medication. The increase in diagnostic testing, however, remains. Appendix Table \ref{tab:persistent_visit_level} shows that visits by treated patients are still 1.6 percentage points more likely to include a diagnostic test. Because the effect is marginally significant and persists for testing but not for prescribing, we read the evidence as suggestive rather than conclusive. Even so, that the testing effect endures suggests that the chatbot's influence may be long-lasting, whether because patients continue to consult other AI tools on their own or because a single exposure durably shifts the preferences and requests they bring to later consultations.

\section{Patient Behavior and Physician--Patient Relationship \label{sec:relationship}}

\begin{figure}[ht]
    \centering

    \begin{subfigure}[b]{0.48\textwidth}
        \centering
        \subcaption{Patient Attrition}
        \includegraphics[width=\textwidth]
        {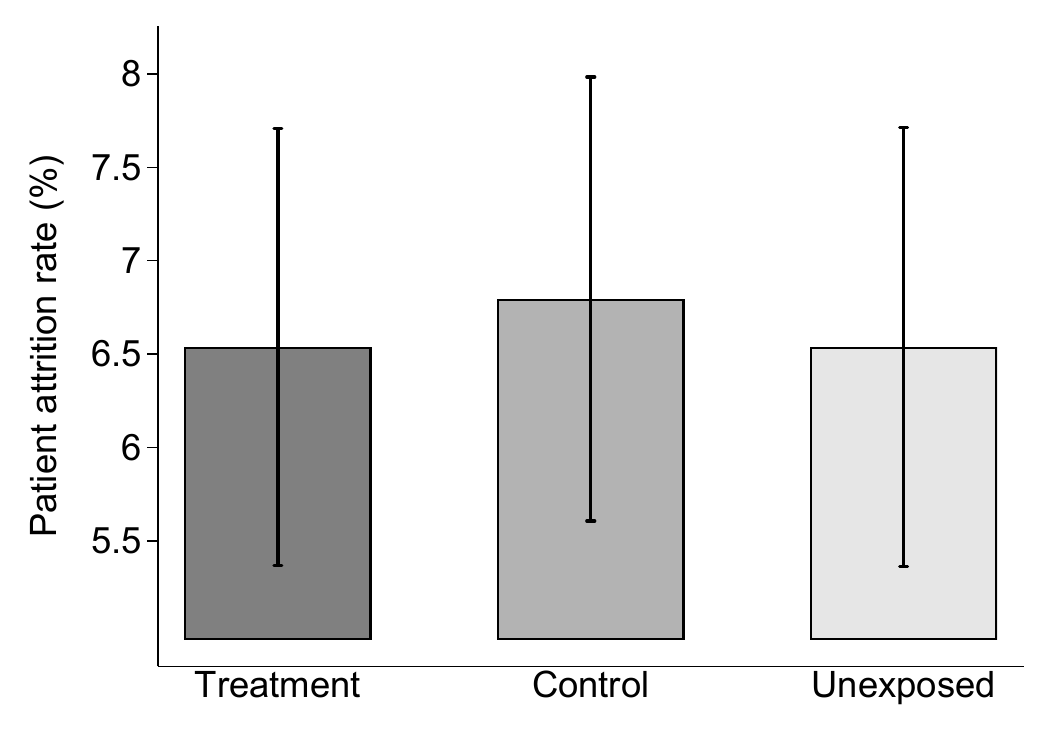}
    \end{subfigure}
        \hfill
    \begin{subfigure}[b]{0.48\textwidth}
        \centering
        \subcaption{Medication Non-Purchase}
        \includegraphics[width=\textwidth]
        {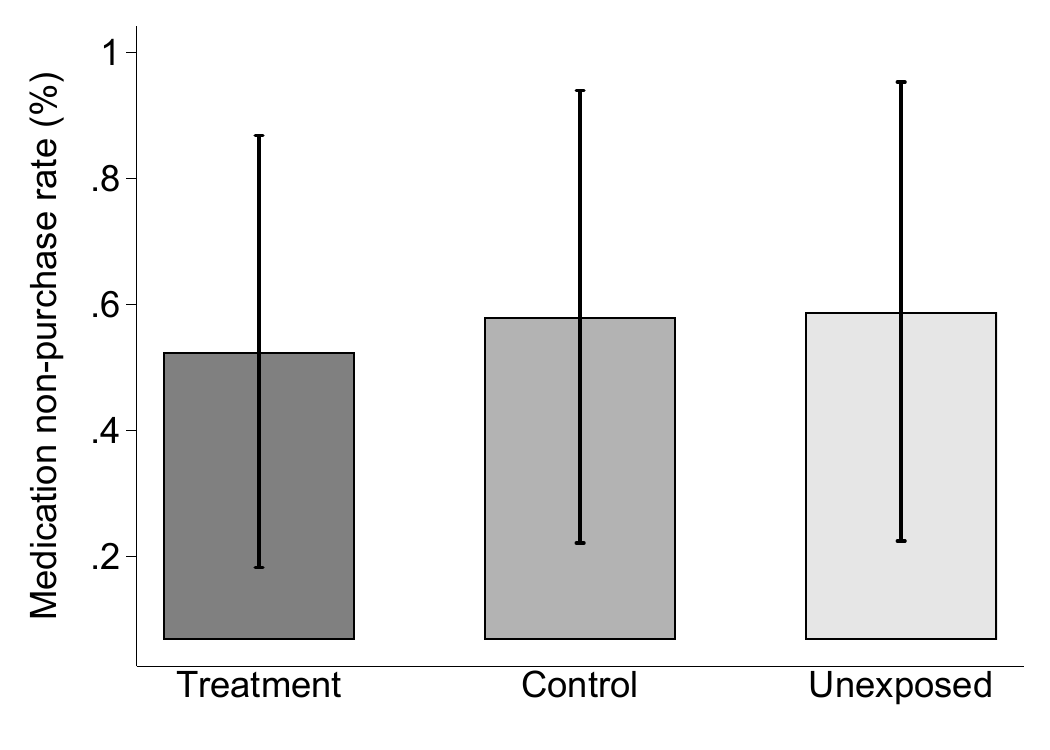}
    \end{subfigure}
    \medskip
    \begin{subfigure}[b]{0.48\textwidth}
        \centering
        \subcaption{Diagnostic Test Non-Completion}
        \includegraphics[width=\textwidth]
        {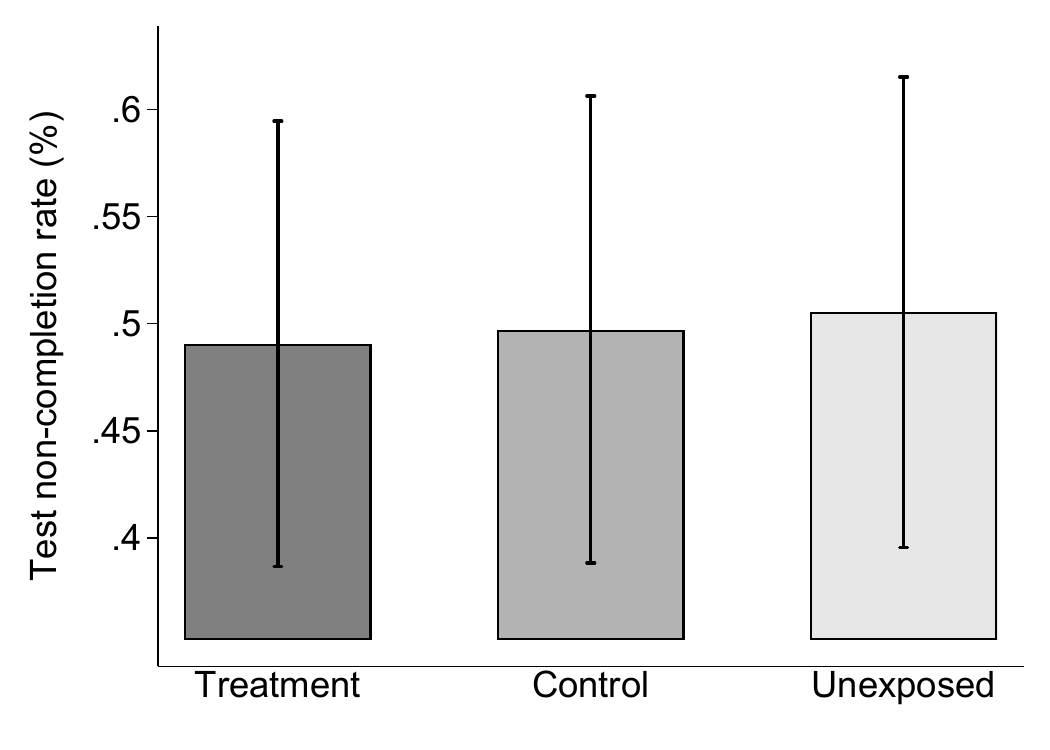}
    \end{subfigure}
    \hfill
    \begin{subfigure}[b]{0.48\textwidth}
        \centering
        \subcaption{Overall Satisfaction}
        \includegraphics[width=\textwidth]{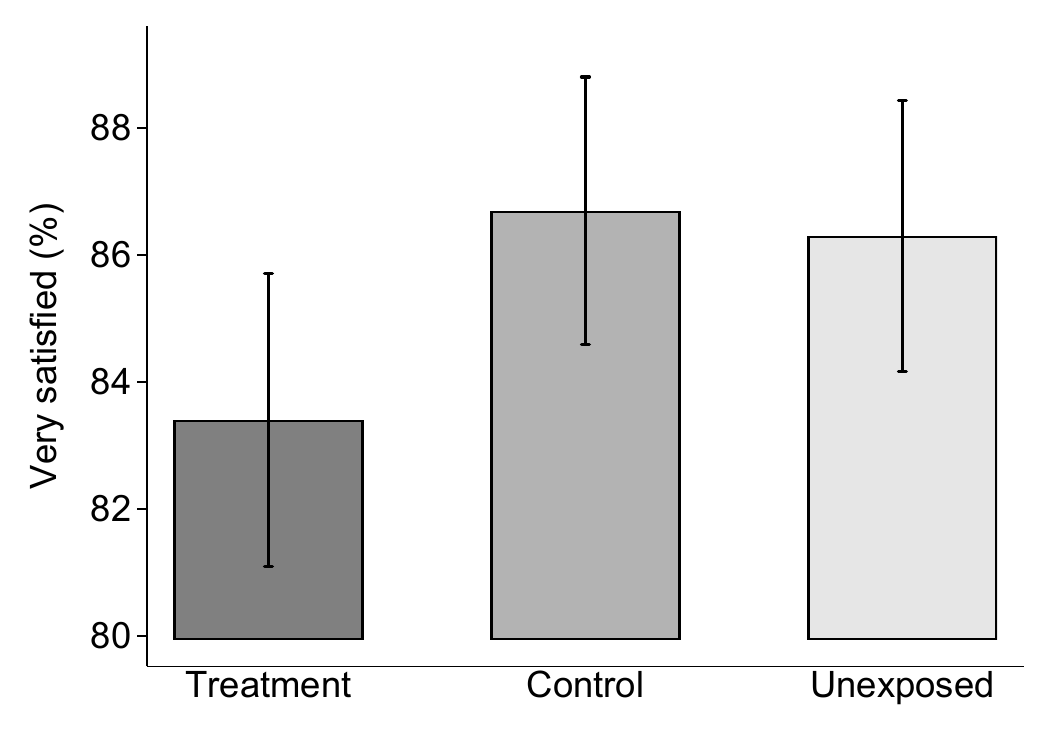}
    \end{subfigure}
    \medskip

    \begin{subfigure}[b]{0.48\textwidth}
        \centering
        \subcaption{Physician--Patient Communication}
        \includegraphics[width=\textwidth]{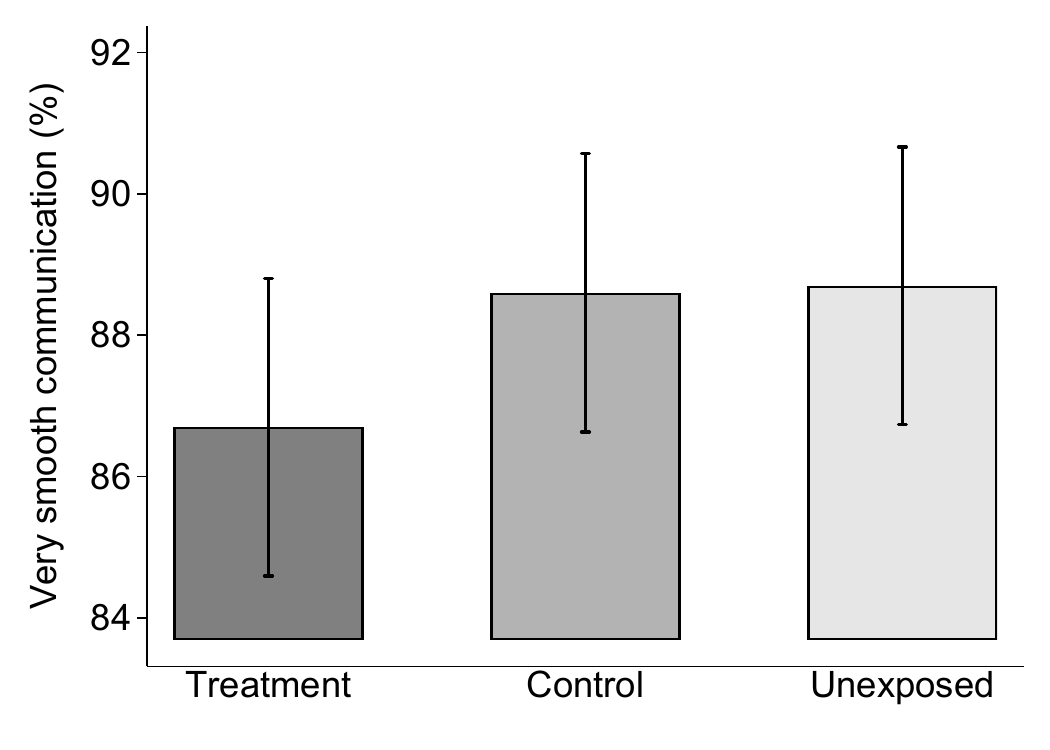}
    \end{subfigure}
    \hfill
    \begin{subfigure}[b]{0.48\textwidth}
        \centering
        \subcaption{Intended Compliance}
        \includegraphics[width=\textwidth]{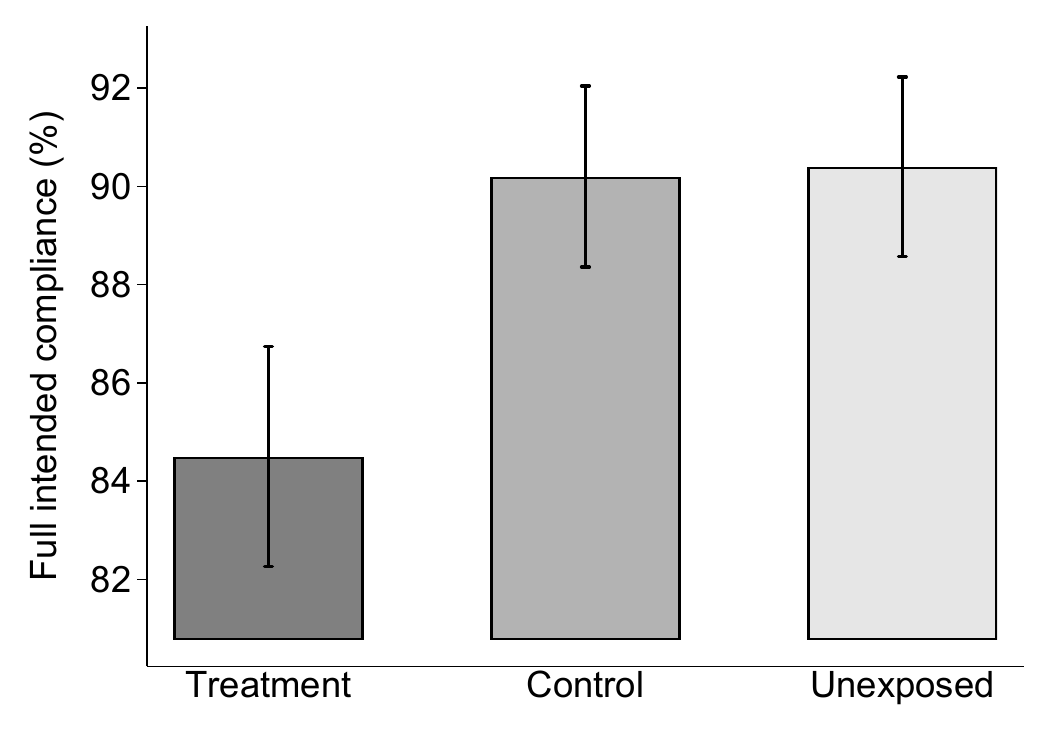}
    \end{subfigure}
    \caption{Patient Behavior and Perceptions by Treatment Status}
    \label{fig:patient_perception}
    \medskip
    \begin{minipage}{\textwidth}
    \footnotesize
    \textit{Notes:} Panels (a)–(c) report outcomes constructed from administrative medical records and capture patient attrition and realized post-consultation behavior. Panels (d)–(f) report outcomes from a post-consultation patient survey, comparing respondents in the \textit{Treatment}, \textit{Control}, and \textit{Unexposed} groups. Percentages denote the share of patients with the indicated outcome or selecting the indicated response category, with 95\% confidence intervals.
    \end{minipage}
\end{figure}

Having documented changes in clinical decisions, we now examine whether chatbot access also affects patient behavior and perceptions of the consultation. Figure \ref{fig:patient_perception} summarizes these outcomes by treatment status, drawing on both administrative medical records and the patient survey. Panels (a) through (c) use administrative medical records. Panel (a) reports patient attrition, measured as whether patients who scheduled an appointment failed to attend their consultation. We find no evidence that access to the chatbot discourages patients from seeing physicians or substitutes for in-person care, as attendance rates are similar across the \textit{Treatment}, \textit{Control}, and \textit{Unexposed} groups. Panels (b) and (c) capture realized post-consultation behavior, measuring whether patients purchased prescribed medications and whether they completed prescribed diagnostic tests following the visit. For both outcomes, we find no statistically meaningful differences across the \textit{Treatment}, \textit{Control}, and \textit{Unexposed} groups. This suggests that access to the chatbot does not materially affect patients' decisions to follow through on physicians' recommendations along these observable dimensions.

Panels (d) through (f) are based on patient survey responses and therefore reflect a more selected sample. While these survey-based measures should be interpreted with caution, they provide insight into how access to the chatbot may have affected the consultation experience. Treated patients report lower overall satisfaction, with 83.3\% of respondents reporting being very satisfied, compared with approximately 86\% in both the \textit{Control} and \textit{Unexposed} groups. Treated patients also perceive communication with physicians as less smooth, as reflected in a 2 percentage point reduction in the likelihood of rating the interaction as very smooth. Finally, intended compliance with physicians' recommendations declines substantially among treated patients, with the share reporting an intention to fully comply falling from 90\% in the \textit{Control} and \textit{Unexposed} groups to 84.5\% in the \textit{Treatment} group. Appendix Table \ref{tab:patient_survey} confirms these patterns in regression estimates. Patients with chatbot access report lower satisfaction, less smooth communication with physicians, and lower intended compliance, with the communication and compliance effects statistically significant despite the relatively small sample size.

Taken together, these patterns suggest that while the chatbot does not meaningfully alter observable patient behavior, it may negatively affect patients' perceptions of the clinical encounter and their interactions with physicians. One possibility is that when patients act more knowledgeable, they receive less polite responses from physicians, a pattern documented by \cite{currie2011patient}. Another possibility is that access to additional health information prior to the visit changes patients' expectations about the consultation, including which questions should be addressed or how much explanation they anticipate from the physician. When these expectations are not fully met, patients may report lower satisfaction or perceive communication as less smooth, even if clinical decisions and follow-through remain largely unchanged. This interpretation is consistent with evidence from \cite{ayers2023comparing}, which shows that in response to patient questions posted on social media, AI chatbot replies are rated more favorably than physician responses overall and on measures of empathy and clarity.\footnote{The survey additionally asks chatbot users to rate the helpfulness of the AI assistant on a five-point scale, from ``not helpful at all'' to ``very helpful.'' Among respondents, 84 percent rated the chatbot as very helpful and an additional 11 percent as somewhat helpful, with fewer than 1 percent reporting it was not helpful.}

While the patient survey provides insight into the consultation experience from the patient’s perspective, it reflects a selected sample of respondents. We therefore complement this evidence with a physician survey that captures the consultation from the physicians’ perspective. The physician survey achieves a high completion rate and thus does not suffer from the same selection concerns, although inference is limited by the smaller sample size of physicians rather than visits. Accordingly, we interpret the physician survey results as exploratory and suggestive.

\begin{figure}[ht]
    \centering

    \begin{subfigure}[b]{0.48\textwidth}
        \centering
            \subcaption{Patients' Symptom Descriptions}
        \includegraphics[width=\textwidth]{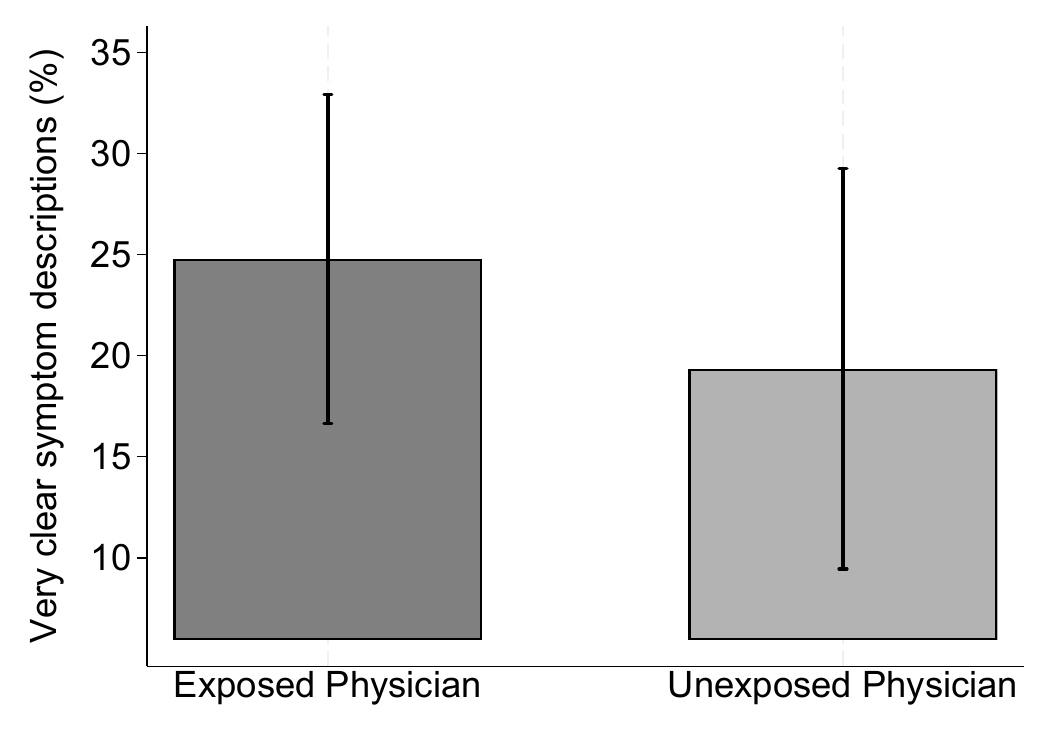}
        \label{fig:physician_very_clear}
    \end{subfigure}
    \hfill
    \begin{subfigure}[b]{0.48\textwidth}
        \centering
        \subcaption{Patients' Understanding of Health Conditions}
        \includegraphics[width=\textwidth]{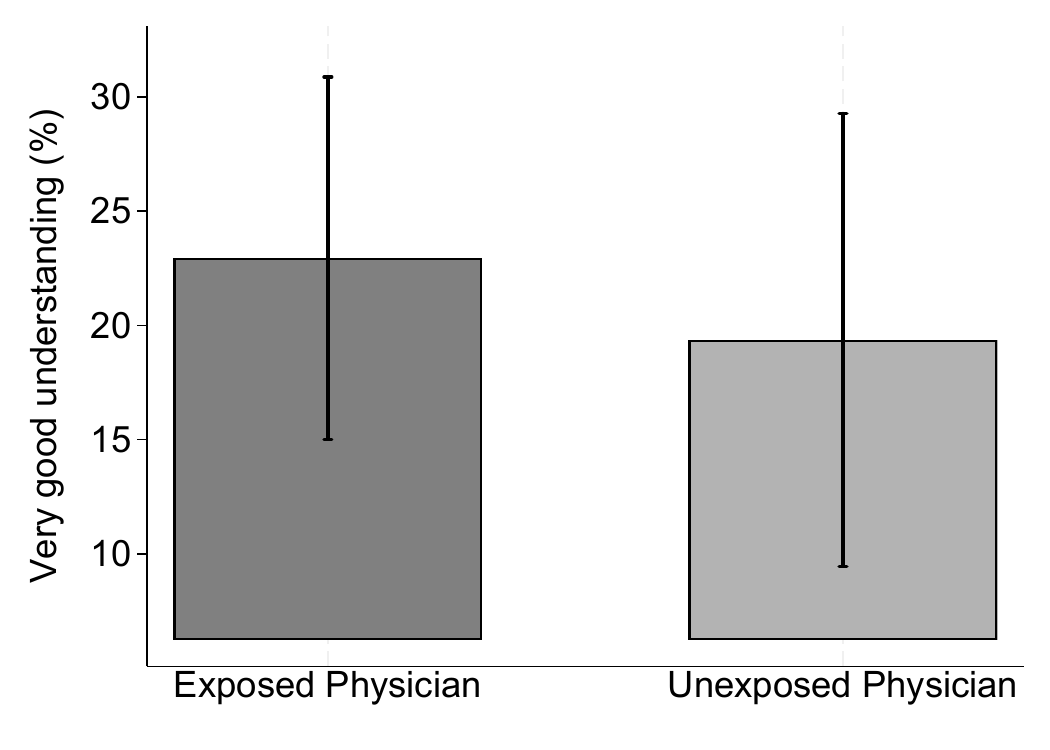}
        \label{fig:physician_very_good}
    \end{subfigure}
    \medskip

    \begin{subfigure}[b]{0.48\textwidth}
        \centering
        \subcaption{Patients' Communication with Physicians}
        \includegraphics[width=\textwidth]{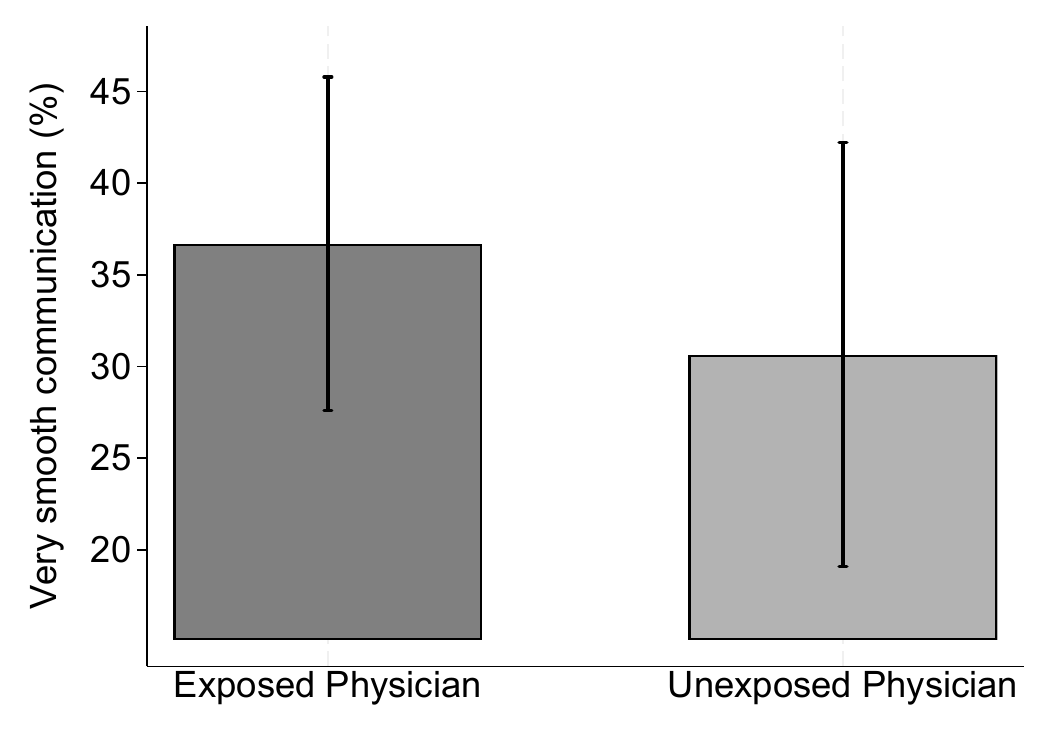}
        \label{fig:physician_very_smooth}
    \end{subfigure}
    \hfill
    \begin{subfigure}[b]{0.48\textwidth}
        \centering
        \subcaption{Patients' Adherence to Medical Advice}
        \includegraphics[width=\textwidth]{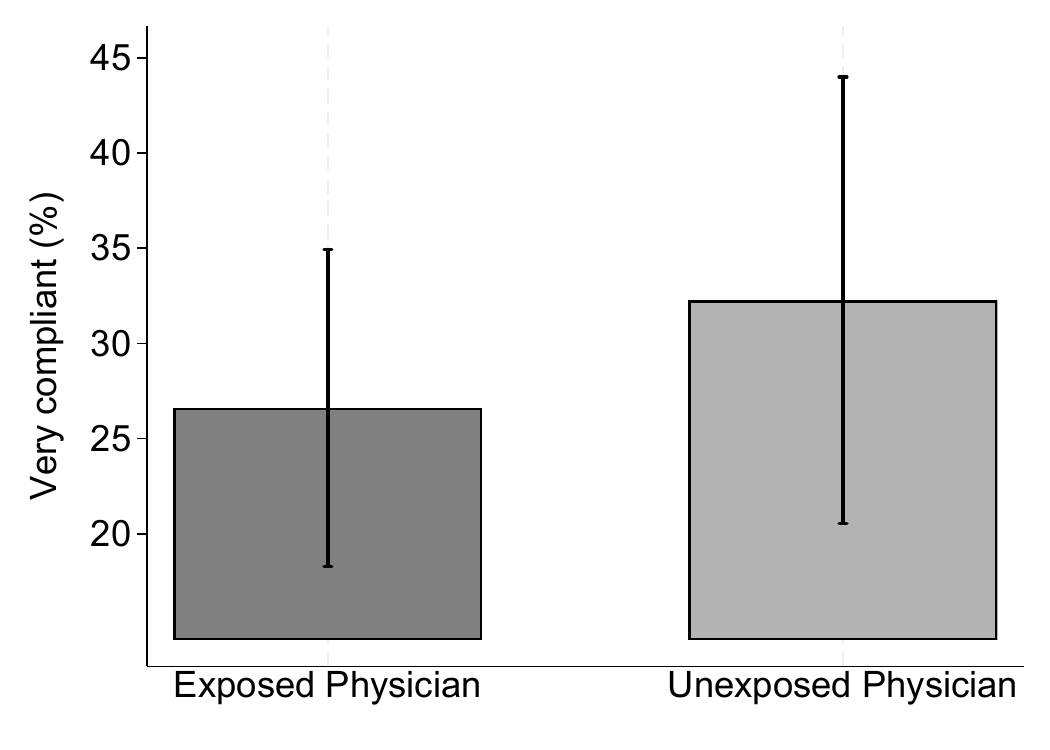}
        \label{fig:physician_very_compliant}
    \end{subfigure}
    \caption{Physicians' Perceptions of Patient Behaviors during Consultations}
    \label{fig:physician_perceptions}
    \medskip
    \begin{minipage}{\textwidth}
    \footnotesize
    \textit{Notes:} This figure reports physicians' survey responses collected after the conclusion of the experiment, comparing physicians randomized to the \textit{Exposed} and \textit{Unexposed} groups. Outcomes reflect physicians' perceptions of patient behavior during consultations. Percentages denote the share of physicians selecting the indicated response category, with 95\% confidence intervals.
    \end{minipage}
\end{figure}

Figure \ref{fig:physician_perceptions} summarizes physicians' perceptions of patient behavior along four dimensions, comparing responses from physicians in the \textit{Exposed} and \textit{Unexposed} groups. In Panel (a), physicians in the \textit{Exposed} group are more likely to report that patients describe their symptoms clearly, with 24.8\% of physicians reporting that patients very clearly describe their symptoms, compared with 19.4\% among \textit{Unexposed} physicians. Panel (b) shows a similar pattern for perceived patient understanding of their condition, with 22.9\% of \textit{Exposed} physicians reporting great understanding, compared with 19.4\% in the \textit{Unexposed} group. Panel (c) indicates smoother perceived physician–patient communication in the \textit{Exposed} group, with 36.7\% reporting very smooth communication, relative to 30.6\% among \textit{Unexposed} physicians. Taken together, the patient and physician surveys reveal an important divergence in perspectives. From the patient's perspective, access to the chatbot is associated with worse perceived communication and lower satisfaction with the visit, whereas from the physician’s perspective, patients appear better prepared and communicate more clearly.

Despite these opposing assessments of communication quality, both patients and physicians report lower adherence or intended compliance with medical advice among treated patients. Panel (d) of Figure \ref{fig:physician_perceptions} shows a 5.7 percentage point reduction in physician-reported patient adherence in the \textit{Exposed} group. This pattern mirrors the patient survey findings and highlights a potential tension between improved patient preparedness and reduced adherence to physicians' recommendations. This is consistent with prior evidence showing that patients with access to expertise exhibit lower adherence \citep{finkelstein2022taste}. From a policy perspective, this pattern underscores a potential trade-off: while AI tools may enhance patient understanding and improve aspects of the consultation process, they may also alter authority dynamics in ways that affect adherence, with potential implications for both care delivery and health outcomes.

Appendix Table \ref{tab:physician_survey_share} reports regression estimates. The independent variable, \textit{Share of AI Adopters}, is the fraction of a physician's patients who used the chatbot, which equals zero by construction for \textit{Unexposed} physicians and varies across \textit{Exposed} physicians based on patient take-up. The point estimates are consistent with Figure \ref{fig:physician_perceptions}. Physicians with higher patient AI adoption report greater symptom clarity, better patient understanding, smoother communication, and lower patient compliance. However, the estimates are imprecise, with only the clarity and compliance coefficients reaching marginal statistical significance in Panel B. The small sample size of 171 physicians limits statistical power.

As a final sanity check, we elicit physicians' beliefs about the share of their patients who used generative AI tools prior to consultations during the experimental month. Importantly, physicians were not informed of the experiment or its nature throughout the study period, so these perceptions are formed only through interacting with their patients.\footnote{In our post-visit patient survey, 80.6\% of \textit{Treatment}-group respondents who reported using the chatbot also reported telling their physician about it during the consultation. While survey respondents are a selected subset of treated patients, this disclosure rate offers a plausible channel through which physicians inferred patients' chatbot use without being informed of the experiment.} Appendix Figure \ref{fig:physician_perceived_ai} reports the results. \textit{Exposed} physicians report an average perceived patient AI adoption rate of 25.2 percent, compared with 16.2 percent among \textit{Unexposed} physicians, a difference of 9 percentage points consistent with the direction of our experimental design. The perceived rates are considerably higher than the actual take-up recorded in our chatbot usage logs, which is unsurprising given that physician recall is subject to recollection biases and their estimates may also capture patients' use of other AI tools beyond our chatbot. Nonetheless, the gap between \textit{Exposed} and \textit{Unexposed} physicians provides additional confirmation that the treatment assignment meaningfully altered the information environment patients brought into the consultation.

\section{Conclusion}
\label{sec:conclusion}
This paper provides the first large-scale randomized evidence on how patient access to generative AI affects healthcare delivery in a real-world clinical setting. We show that providing patients with access to a generative AI chatbot prior to outpatient visits meaningfully alters clinical practice. Medication prescribing declines by 4.6 percentage points, with particularly large reductions for Traditional Chinese Medicine and antibiotics, while diagnostic testing increases by 2.7 percentage points, without raising total medical expenditures. We find no evidence of spillovers in physicians' behavior toward patients without chatbot access, indicating that the effects operate primarily through direct patient use rather than persistent changes in physician practice. Because chatbot adopters are disproportionately younger, male, and employed, the absence of spillovers also implies that unequal access to generative AI tools may translate directly into unequal benefits, which could widen existing health disparities.

The specific pattern of clinical change we document, reduced prescribing and increased diagnostic testing, mirrors the directional advice we observe in the conversation logs. Consistent with the defensive, liability-driven guardrails encoded in AI training, the chatbot cautions heavily against medications while recommending diagnostic testing freely. Treated patients carry this directionality into their consultations, and part of the effect appears to persist in their later visits months after the experiment ended. More broadly, when AI systems are deployed as consumer-facing tools at scale, the norms encoded in their training propagate into real clinical decisions, with the potential to shift aggregate outcomes such as population-level healthcare utilization.

At the same time, our survey results reveal a tension between the use of generative AI tools and the physician--patient relationship. While physicians report clearer communication and better patient preparation when interacting with AI-assisted patients, patients themselves report lower satisfaction and reduced intended adherence to medical advice. These patterns suggest that generative AI may reshape expectations about clinical encounters and shift the balance of authority within the physician--patient interaction in ways that affect compliance.

Whether these shifts improve welfare is not clear. On one hand, the reductions in prescribing fall disproportionately on Traditional Chinese Medicine and antibiotics, categories long viewed as overprescribed in China. The added testing may also surface conditions that would otherwise be missed. On the other hand, less medication may withhold beneficial treatment, and the added testing imposes costs on patients and the health system. Lower adherence is similarly ambiguous, reflecting either informed skepticism or the disregard of sound advice. Importantly, the direction of these changes is set by the objectives of those who build and govern the AI, which need not coincide with the objectives that optimize healthcare delivery.

Taken together, our findings speak to a broader transformation in expert--client markets. Healthcare is a canonical credence good, but it is far from the only one. Legal services, financial advice, and skilled trades share the same underlying information asymmetry that generative AI is now positioned to reduce. Our results suggest that generative AI does more than narrow that asymmetry. The priorities and cautions embedded in AI training travel with the information into real decisions, introducing new frictions in expert--client relationships along the way. While we document this in healthcare, the same mechanism may operate wherever clients consult generative AI before turning to an expert. As people increasingly rely on these tools, from health questions to financial decisions, the design choices of a small number of developers can propagate into decision-making across many domains of life.

\newpage
\begin{spacing}{1}
\renewcommand{\bibfont}{\small}
\bibliographystyle{aer}
\bibliography{reference}
\end{spacing}
\pagebreak

\newpage
\section*{Figures and Tables}

\begin{figure}[t]
    \centering
    \begin{subfigure}{0.6\textwidth}
        \centering
        \subcaption{Medication Prescription}
        \includegraphics[width=\textwidth]{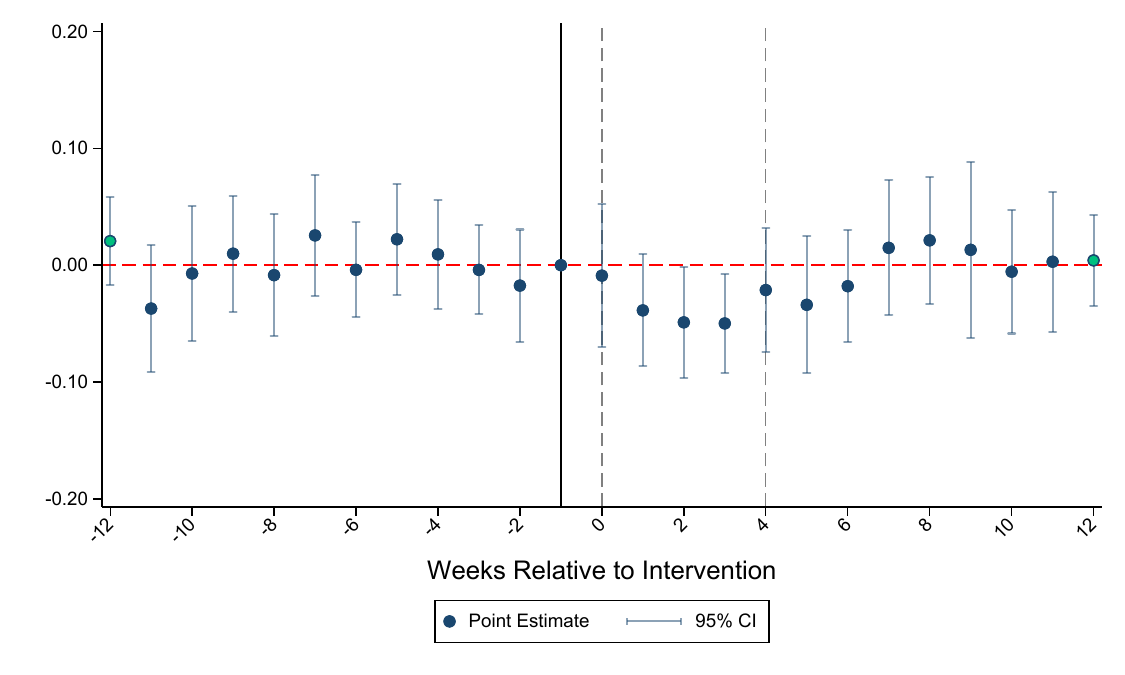}
    \end{subfigure}
    \medskip
    \begin{subfigure}{0.6\textwidth}
        \centering
        \subcaption{TCM}
        \includegraphics[width=\textwidth]{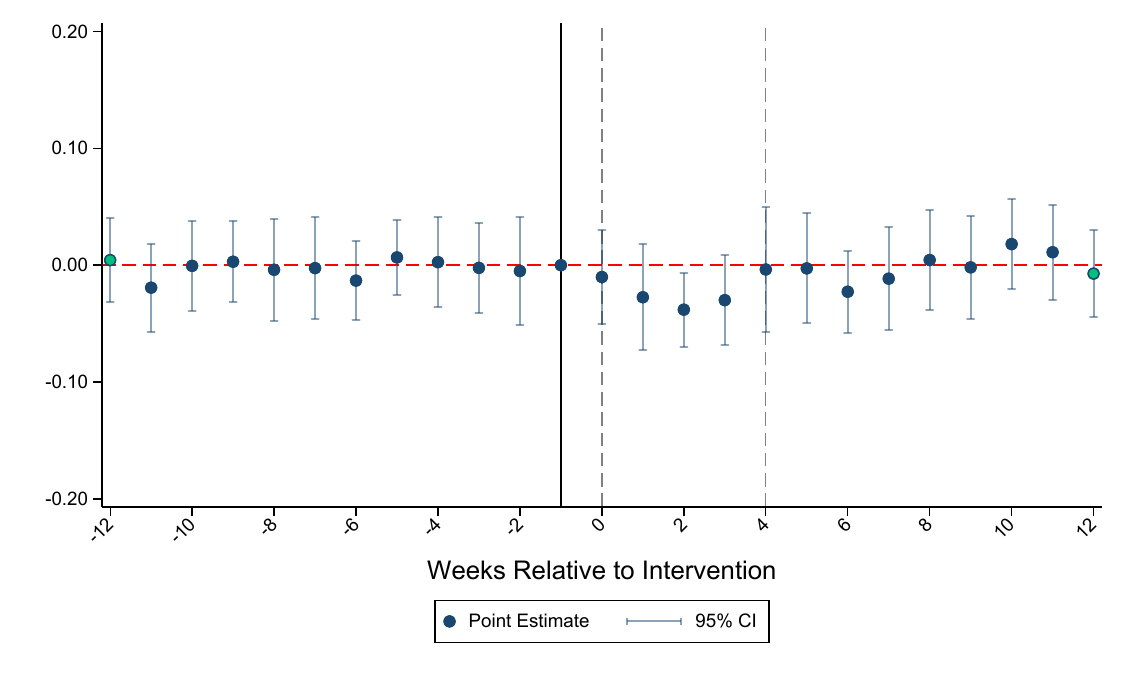}
    \end{subfigure}
    \medskip
    \begin{subfigure}{0.6\textwidth}
        \centering
        \subcaption{Antibiotics}
        \includegraphics[width=\textwidth]{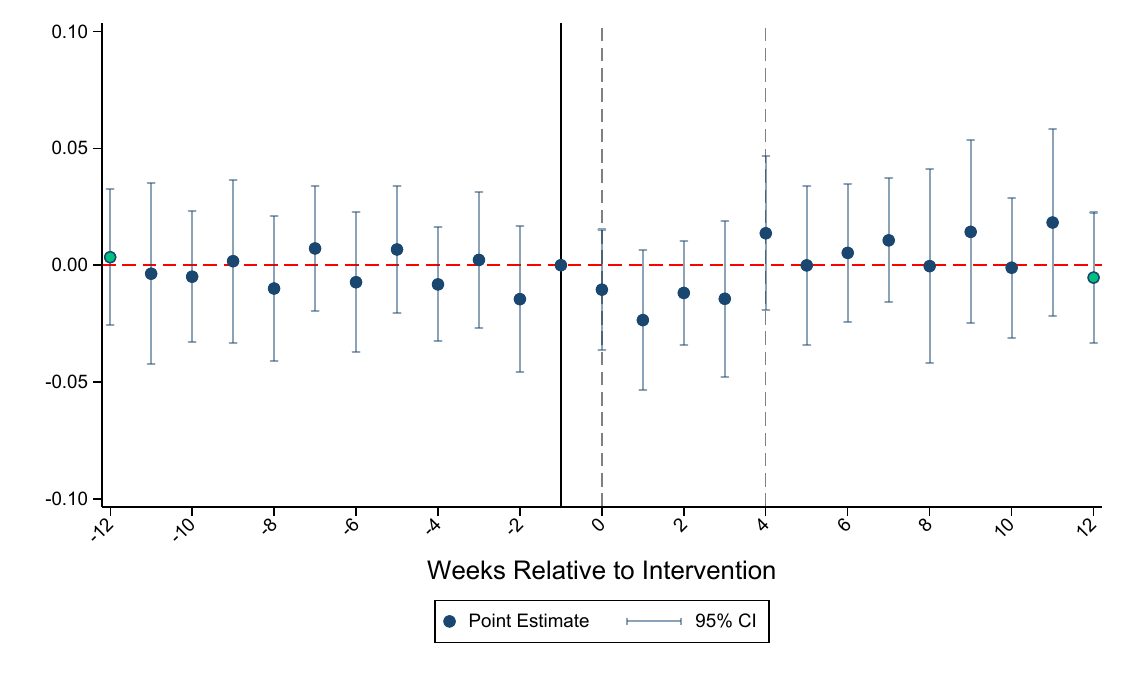}
    \end{subfigure}
    \caption{Dynamic Effects of Exposure to Treated Patients on Prescriptions and Clinical Practices}
    \label{fig:dynamic_weighted}
\end{figure}

\begin{figure}[t]\ContinuedFloat
    \centering
    \begin{subfigure}{0.6\textwidth}
        \centering
        \subcaption{Opioids}
        \includegraphics[width=\textwidth]{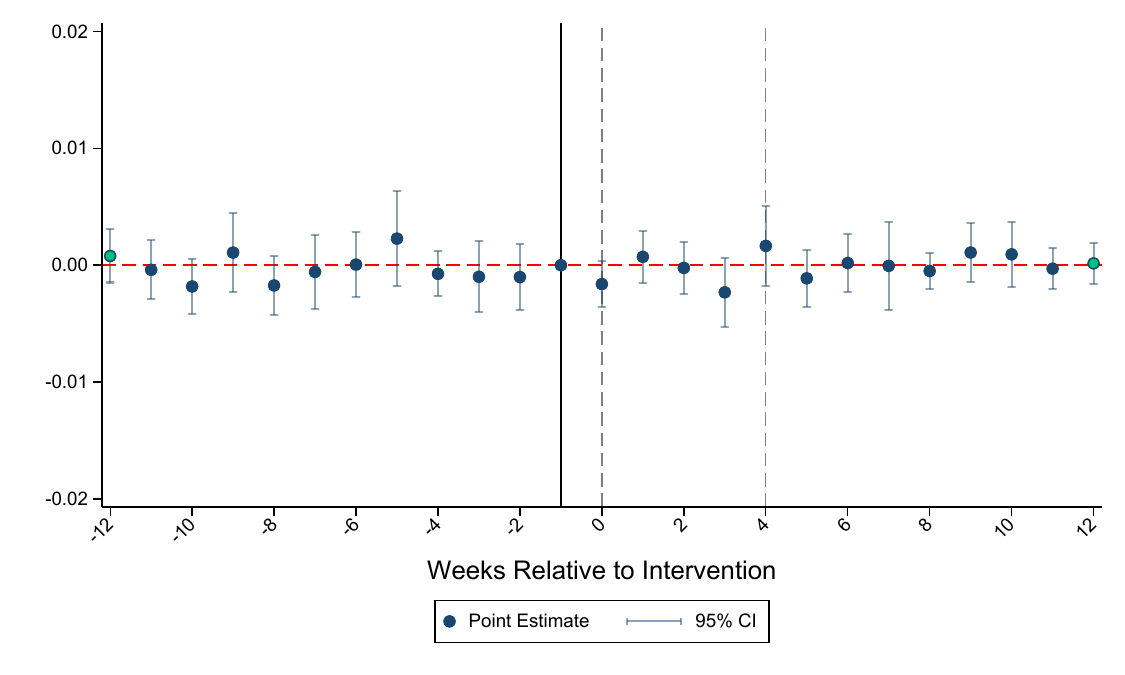}
    \end{subfigure}
    \medskip
    \begin{subfigure}{0.6\textwidth}
        \centering
        \subcaption{Diagnostic Tests}
        \includegraphics[width=\textwidth]{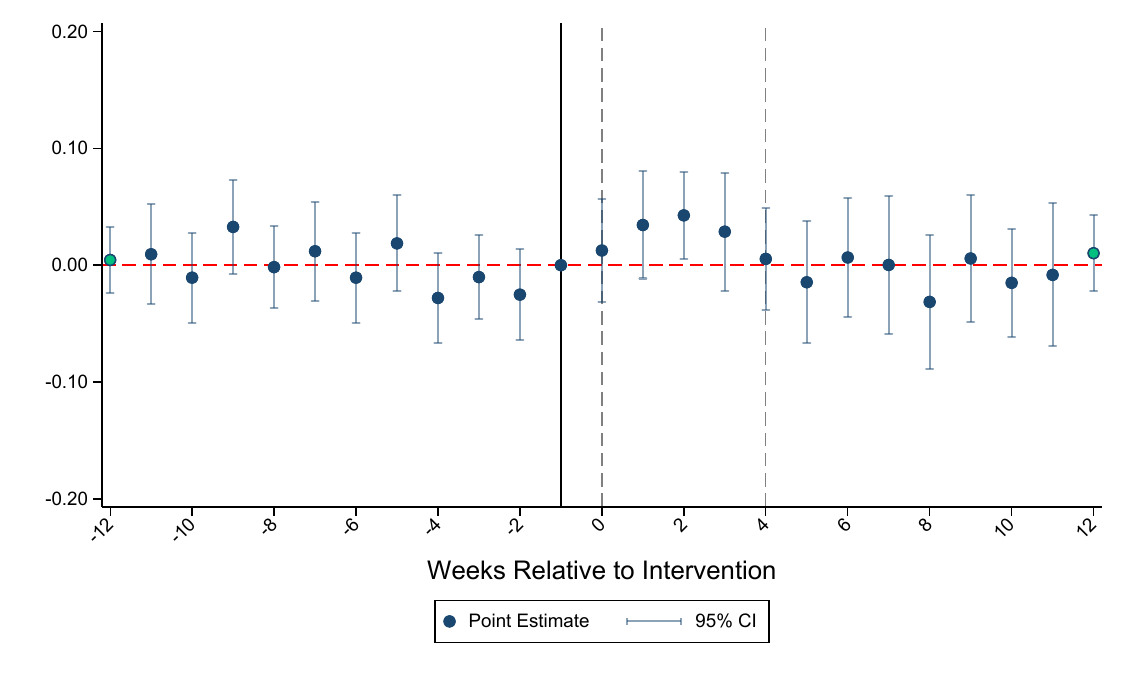}
    \end{subfigure}
    \medskip
    \begin{subfigure}{0.6\textwidth}
        \centering
        \subcaption{Revisit within Two Weeks}
        \includegraphics[width=\textwidth]{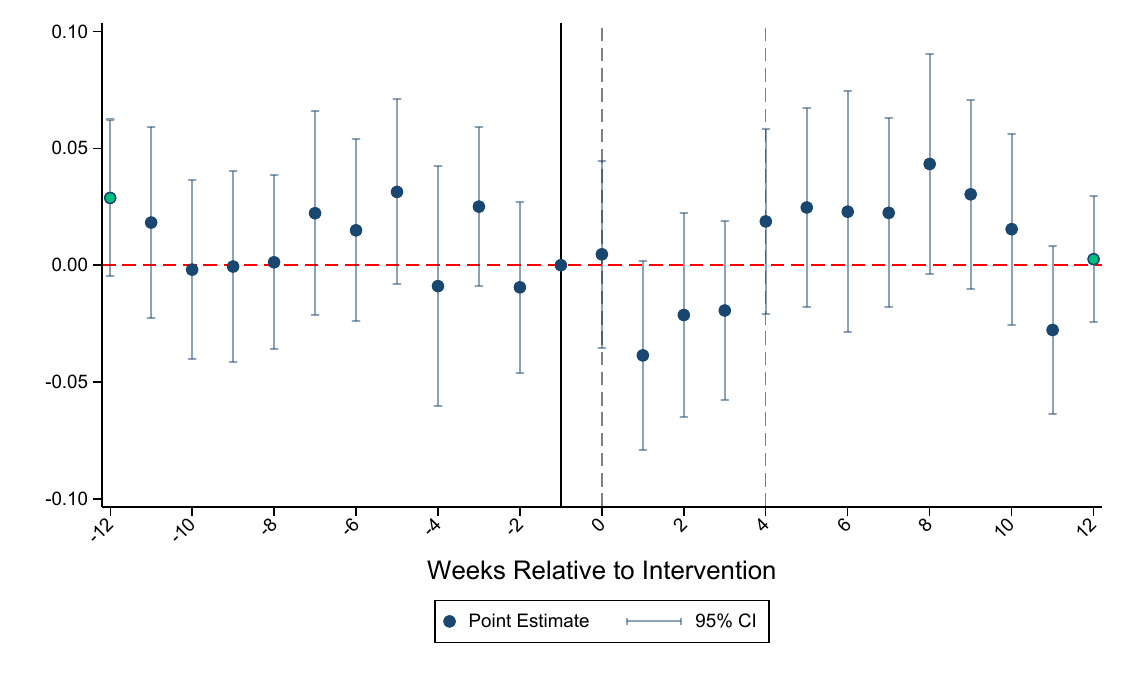}
    \end{subfigure}

    \caption{Dynamic Effects of Exposure to Treated Patients on Prescriptions and Clinical Practices (Continued)}
     \begin{minipage}{\textwidth}
    \footnotesize
    \textit{Notes:} This figure presents the point estimates with 95\% confidence intervals from equation (\ref{eq:dynamic}). Data are aggregated to the physician-week level using all patient visits from 2025, weighted by the number of visits per physician per week. Relative weeks beyond $-12$ and $+12$ are grouped at the endpoints. The baseline period is Week $-1$ (May 26--June 1, 2025). Standard errors are clustered at the physician level.
    \end{minipage}
\end{figure}

\clearpage

\begin{table}[t]
\centering
\caption{Balance Test: Physician and Patient Characteristics}
\label{tab:balance_test}

\scriptsize
\begin{tabular*}{\textwidth}{@{\extracolsep{\fill}}lcccccc@{}}
\toprule
\multicolumn{7}{l}{\textbf{Panel A: Physician Characteristics}} \\
\midrule
Variables & \multicolumn{2}{c}{Exposed Arm} & \multicolumn{2}{c}{Unexposed Arm} & Diff & \textit{p}-value \\
\cmidrule(lr){2-3} \cmidrule(lr){4-5}
& Mean & S.D. & Mean & S.D. & & \\
\midrule
Age & 37.94 & 8.52 & 39.48 & 9.61 & $-1.54$ & 0.29 \\
Male & 0.61 & 0.49 & 0.69 & 0.46 & $-0.09$ & 0.25 \\
Senior Physician & 0.30 & 0.46 & 0.29 & 0.46 & 0.01 & 0.90 \\
College Degree or Above & 0.91 & 0.29 & 0.92 & 0.27 & $-0.01$ & 0.76 \\
Specialty: Internal Medicine & 0.34 & 0.48 & 0.34 & 0.48 & $-0.00$ & 0.97 \\
Specialty: Surgery & 0.26 & 0.44 & 0.23 & 0.42 & 0.04 & 0.56 \\
Specialty: Psychiatry & 0.24 & 0.43 & 0.27 & 0.45 & $-0.04$ & 0.62 \\
Specialty: Emergency & 0.05 & 0.22 & 0.08 & 0.27 & $-0.03$ & 0.47 \\
Specialty: Other Dept. & 0.12 & 0.33 & 0.11 & 0.32 & 0.01 & 0.89 \\
\midrule
Num of Physicians & \multicolumn{2}{c}{117} & \multicolumn{2}{c}{62} & & \\
\end{tabular*}

\scriptsize
\begin{tabular*}{\textwidth}{@{\extracolsep{\fill}}lcccccccccc@{}}
\toprule
\multicolumn{11}{l}{\textbf{Panel B: Patient Characteristics}} \\
\midrule
& \multicolumn{2}{c}{Treatment} & \multicolumn{2}{c}{Control} & \multicolumn{2}{c}{Unexposed} & \multicolumn{2}{c}{Treat vs Ctrl} & \multicolumn{2}{c}{Treat vs Unexp} \\
\cmidrule(lr){2-3} \cmidrule(lr){4-5} \cmidrule(lr){6-7} \cmidrule(lr){8-9} \cmidrule(lr){10-11}
& Mean & S.D. & Mean & S.D. & Mean & S.D. & Diff & \textit{p}-value & Diff & \textit{p}-value \\
\midrule
Age & 45.36 & 16.92 & 45.33 & 16.93 & 45.49 & 17.24 & 0.02 & 0.95 & $-0.15$ & 0.63 \\
Male & 0.36 & 0.48 & 0.36 & 0.48 & 0.36 & 0.48 & $-0.00$ & 0.96 & 0.00 & 0.93 \\
Occupation: Farmer & 0.44 & 0.50 & 0.44 & 0.50 & 0.44 & 0.50 & 0.00 & 0.98 & $-0.00$ & 0.97 \\
Occupation: Formally Employed & 0.24 & 0.43 & 0.23 & 0.42 & 0.24 & 0.43 & 0.01 & 0.20 & $-0.01$ & 0.18 \\
Occupation: Student & 0.09 & 0.29 & 0.09 & 0.29 & 0.09 & 0.29 & $-0.00$ & 0.92 & 0.00 & 0.89 \\
Occupation: Other & 0.07 & 0.26 & 0.07 & 0.26 & 0.07 & 0.26 & $-0.00$ & 0.72 & 0.00 & 0.69 \\
Unemployed or Retired & 0.16 & 0.36 & 0.17 & 0.37 & 0.16 & 0.37 & $-0.01$ & 0.23 & $-0.00$ & 0.82 \\
First-time Visit & 0.70 & 0.46 & 0.70 & 0.46 & 0.70 & 0.46 & $-0.00$ & 0.95 & 0.00 & 0.95 \\
Advance Booking Time (hours) & 24.63 & 27.08 & 24.90 & 28.02 & 24.96 & 27.59 & -0.28 & 0.55 & -0.32 & 0.48 \\
\midrule
Num of Visits & \multicolumn{2}{c}{5,828} & \multicolumn{2}{c}{5,838} & \multicolumn{2}{c}{6,010} & \multicolumn{2}{c}{} & \multicolumn{2}{c}{} \\
\bottomrule
\end{tabular*}

\end{table}

\begin{table}[htbp]
\centering
\caption{Patient--AI Conversation Patterns by Topic}
\label{tab:conversation}
\begin{threeparttable}
\resizebox{\textwidth}{!}{\begin{tabular}{lcccccc}
\toprule
& \multicolumn{2}{c}{Mention Rate} & \multicolumn{4}{c}{AI Stance (\% of AI mentions)} \\
\cmidrule(lr){2-3} \cmidrule(lr){4-7}
Topic & $\;\ $Patient $\;\ $ & $\;\ $ AI $\;\ \;\ $ &$\,$ Rec Only $\,$& Rec+Caution & Caution Only & $\;\ $ Neither $\;\ $ \\
\midrule
Diagnostic testing & 8.5\% & 58.2\% & 94.5\% & 3.0\% & 0.3\% & 2.2\% \\
General medication & 11.4\% & 58.9\% & 19.1\% & 44.2\% & 25.5\% & 11.3\% \\
Traditional Chinese Medicine & 0.8\% & 11.0\% & 3.8\% & 3.8\% & 87.0\% & 5.3\% \\
Antibiotics & 0.1\% & 18.2\% & 7.8\% & 13.4\% & 74.2\% & 4.6\% \\
\bottomrule
\end{tabular}}
\begin{minipage}{\textwidth}
\footnotesize
\textit{Notes:} The sample consists of 1,192 conversation turns from 956 visits with chatbot access. Mention rates are shares of all turns in which the topic appears, identified through keyword matching at the sentence level. Patient mentions are based on the patient message; AI mentions are based on the chatbot's reply. The four AI attitude categories are constructed from two independent binary flags assigned by GPT-4o for each turn in which the AI mentions the topic: whether the AI recommends the treatment or test, and whether it raises any caution or advises against its use. Attitude shares are computed as percentages of turns in which the AI mentions the topic and therefore sum to 100\% within each row. A single turn can mention multiple topics, in which case it is classified independently for each topic.
\end{minipage}
\end{threeparttable}
\end{table}

\clearpage
\begin{table}[htbp]
\centering
\caption{Within-Physician Intent-to-Treat Effects on Clinical Practice Patterns}
\label{tab:itt_practice}

\begin{tabular}{l*{6}{c}}
\toprule
& \multicolumn{6}{c}{Dependent Variables} \\
\cmidrule(lr){2-7}
& Prescribed & TCM & Antibiotics & Opioids & Diagnostic & Revisit \\
& medication & & & & tests & \\
& (1) & (2) & (3) & (4) & (5) & (6) \\
\midrule
\multicolumn{7}{l}{\textbf{Panel A: No Controls}} \\
Treated & $-0.045^{***}$ & $-0.022^{***}$ & $-0.015^{**}$ & $-0.004$ & $0.027^{***}$ & $-0.011^{*}$ \\
& $(0.006)$ & $(0.007)$ & $(0.006)$ & $(0.003)$ & $(0.008)$ & $(0.007)$ \\
& & & & & & \\
Mean & 0.87 & 0.18 & 0.13 & 0.03 & 0.23 & 0.16 \\
Observations & 11,666 & 11,666 & 11,666 & 11,666 & 11,666 & 11,666 \\
\midrule
\multicolumn{7}{l}{\textbf{Panel B: Patient Characteristics \& Physician FE}} \\
Treated & $-0.046^{***}$ & $-0.022^{***}$ & $-0.015^{**}$ & $-0.004$ & $0.027^{***}$ & $-0.012^{*}$ \\
& $(0.006)$ & $(0.007)$ & $(0.006)$ & $(0.003)$ & $(0.008)$ & $(0.007)$ \\
& & & & & & \\
Patient Char. & Y & Y & Y & Y & Y & Y \\
Physician FE & Y & Y & Y & Y & Y & Y \\
Mean & 0.87 & 0.18 & 0.13 & 0.03 & 0.23 & 0.16 \\
Observations & 11,666 & 11,666 & 11,666 & 11,666 & 11,666 & 11,666 \\
\bottomrule
\end{tabular}

\begin{minipage}{\textwidth}
    \footnotesize
    \textit{Notes:} This table reports within-physician intent-to-treat estimates of the effect of chatbot access on clinical practice patterns. The sample is restricted to outpatient visits with \textit{Exposed} physicians. Panel A reports estimates without controls; Panel B adds patient characteristics (age, gender, occupation, and an indicator for first visit) and physician fixed effects. Robust standard errors are reported in parentheses. $^{*}$ p $<$ 0.10, $^{**}$ p $<$ 0.05, $^{***}$ p $<$ 0.01.
\end{minipage}
\end{table}

\clearpage
\begin{table}[htbp]
\centering
\caption{Within-Physician Intent-to-Treat Effects on Healthcare Expenditures}
\label{tab:itt_costs}

\begin{tabular}{l*{5}{c}}
\toprule
& \multicolumn{5}{c}{Dependent Variables} \\
\cmidrule(lr){2-6}
& TCM costs & WM costs & Diagnostic & Other costs & Total costs \\
& (RMB) & (RMB) & costs (RMB) & (RMB) & (RMB) \\
& (1) & (2) & (3) & (4) & (5) \\
\midrule
\multicolumn{6}{l}{\textbf{Panel A: No Controls}} \\
Treated & $-1.686^{*}$ & $-5.276^{*}$ & $6.542$ & $0.174$ & $-0.245$ \\
& $(0.944)$ & $(2.939)$ & $(5.662)$ & $(3.293)$ & $(7.264)$ \\
& & & & & \\
Mean & 13.10 & 64.81 & 92.69 & 35.88 & 206.48 \\
Observations & 11,666 & 11,666 & 11,666 & 11,666 & 11,666 \\
\midrule
\multicolumn{6}{l}{\textbf{Panel B: Patient Characteristics \& Physician FE}} \\
Treated & $-1.762^{*}$ & $-5.021^{*}$ & $6.313$ & $-0.002$ & $-0.472$ \\
& $(0.945)$ & $(2.952)$ & $(5.622)$ & $(3.388)$ & $(7.279)$ \\
& & & & & \\
Patient Char. & Y & Y & Y & Y & Y \\
Physician FE & Y & Y & Y & Y & Y \\
Mean & 13.10 & 64.81 & 92.69 & 35.88 & 206.48 \\
Observations & 11,666 & 11,666 & 11,666 & 11,666 & 11,666 \\
\bottomrule
\end{tabular}

\begin{minipage}{\textwidth}
    \footnotesize
    \textit{Notes:} This table reports within-physician intent-to-treat estimates of the effect of chatbot access on expenditures. The sample is restricted to outpatient visits with \textit{Exposed} physicians. Panel A reports estimates without controls; Panel B adds patient characteristics (age, gender, occupation, and an indicator for first visit) and physician fixed effects. Robust standard errors are reported in parentheses. $^{*}$ p $<$ 0.10, $^{**}$ p $<$ 0.05, $^{***}$ p $<$ 0.01.
\end{minipage}
\end{table}

\clearpage
\begin{table}[htbp]
\centering
\caption{Instrumental Variable Estimates: Treatment Effects on Clinical Practice Patterns}
\label{tab:iv_practice}

\begin{tabular}{l*{6}{c}}
\toprule
& \multicolumn{6}{c}{Dependent Variables} \\
\cmidrule(lr){2-7}
& Prescribed & TCM & Antibiotics & Opioids & Diagnostic & Revisit \\
& medication & & & & tests & \\
& (1) & (2) & (3) & (4) & (5) & (6) \\
\midrule
\multicolumn{7}{l}{\textbf{Panel A: No Controls}} \\
Chatbot Usage & $-0.265^{***}$ & $-0.128^{***}$ & $-0.090^{**}$ & $-0.026$ & $0.161^{***}$ & $-0.066^{*}$ \\
& $(0.037)$ & $(0.041)$ & $(0.037)$ & $(0.017)$ & $(0.046)$ & $(0.039)$ \\
& & & & & & \\

Mean & 0.87 & 0.18 & 0.13 & 0.03 & 0.23 & 0.16 \\
Observations & 11,666 & 11,666 & 11,666 & 11,666 & 11,666 & 11,666 \\
F-statistic  & \multicolumn{6}{c}{1198.188} \\
\midrule
\multicolumn{7}{l}{\textbf{Panel B: Patient Characteristics \& Physician FE}} \\
Chatbot Usage & $-0.267^{***}$ & $-0.129^{***}$ & $-0.089^{**}$ & $-0.025$ & $0.161^{***}$ & $-0.070^{*}$ \\
& $(0.037)$ & $(0.042)$ & $(0.037)$ & $(0.017)$ & $(0.046)$ & $(0.040)$ \\
& & & & & & \\

Patient Char. & Y & Y & Y & Y & Y & Y \\
Physician FE & Y & Y & Y & Y & Y & Y \\
Mean & 0.87 & 0.18 & 0.13 & 0.03 & 0.23 & 0.16 \\
Observations & 11,666 & 11,666 & 11,666 & 11,666 & 11,666 & 11,666 \\
F-statistic  & \multicolumn{6}{c}{1181.441} \\
\bottomrule
\end{tabular}

\begin{minipage}{\textwidth}
    \footnotesize
    \textit{Notes:} This table reports within-physician instrumental variable estimates of the effect of chatbot usage on clinical practice patterns, where chatbot usage is instrumented with random assignment to chatbot access. The sample is restricted to outpatient visits with \textit{Exposed} physicians. Panel A reports estimates without controls; Panel B adds patient characteristics (age, gender, occupation, and an indicator for first visit) and physician fixed effects. The reported first-stage F-statistic is the Kleibergen-Paap Wald F-statistic. Robust standard errors are reported in parentheses. $^{*}$ p $<$ 0.10, $^{**}$ p $<$ 0.05, $^{***}$ p $<$ 0.01.
\end{minipage}
\end{table}

\clearpage
\begin{table}[htbp]
\centering
\caption{Instrumental Variable Estimates: Treatment Effects on Healthcare Expenditures}
\label{tab:iv_costs}

\begin{tabular}{l*{5}{c}}
\toprule
& \multicolumn{5}{c}{Dependent Variables} \\
\cmidrule(lr){2-6}
& TCM costs & WM costs & Diagnostic & Other costs & Total costs \\
& (RMB) & (RMB) & costs (RMB) & (RMB) & (RMB) \\
& (1) & (2) & (3) & (4) & (5) \\
\midrule
\multicolumn{6}{l}{\textbf{Panel A: No Controls}} \\
Chatbot Usage & $-9.884^{*}$ & $-30.933^{*}$ & $38.359$ & $1.023$ & $-1.436$ \\
& $(5.542)$ & $(17.227)$ & $(33.189)$ & $(19.306)$ & $(42.585)$ \\
& & & & & \\

Mean & 13.10 & 64.81 & 92.69 & 35.88 & 206.48 \\
Observations & 11,666 & 11,666 & 11,666 & 11,666 & 11,666 \\
F-statistic & \multicolumn{5}{c}{1198.188} \\
\midrule
\multicolumn{6}{l}{\textbf{Panel B: Patient Characteristics \& Physician FE}} \\
Chatbot Usage & $-10.334^{*}$ & $-29.444^{*}$ & $37.018$ & $-0.009$ & $-2.770$ \\
& $(5.552)$ & $(17.310)$ & $(32.967)$ & $(19.868)$ & $(42.679)$ \\
& & & & & \\

Patient Char. & Y & Y & Y & Y & Y \\
Physician FE & Y & Y & Y & Y & Y \\
Mean & 13.10 & 64.81 & 92.69 & 35.88 & 206.48 \\
Observations & 11,666 & 11,666 & 11,666 & 11,666 & 11,666 \\
F-statistic & \multicolumn{5}{c}{1181.441} \\
\bottomrule
\end{tabular}

\begin{minipage}{\textwidth}
    \footnotesize
    \textit{Notes:} This table reports within-physician instrumental variable estimates of the effect of chatbot usage on healthcare expenditures, where chatbot usage is instrumented with random assignment to chatbot access. The sample is restricted to outpatient visits with \textit{Exposed} physicians. Panel A reports estimates without controls; Panel B adds patient characteristics (age, gender, occupation, and an indicator for first visit) and physician fixed effects. The reported first-stage F-statistic is the Kleibergen-Paap Wald F statistic. Robust standard errors are reported in parentheses. $^{*}$ p $<$ 0.10, $^{**}$ p $<$ 0.05, $^{***}$ p $<$ 0.01.
\end{minipage}
\end{table}

\clearpage
\begin{table}[!htbp]\centering
\caption{Heterogeneous Treatment Effects by Physicians' Attitudes}
\label{tab:hte_physician_attitudes}
\begin{threeparttable}
\setlength{\tabcolsep}{7pt}
\resizebox{\textwidth}{!}{

\begin{tabular}{lcccccc}
\toprule
& \multicolumn{6}{c}{Dependent variables} \\
\cmidrule(lr){2-7}
& Prescribed & TCM & Antibiotics & Opioids & Diagnostic & Revisit \\
& medication &  &  &  & tests &  \\
& (1) & (2) & (3) & (4) & (5) & (6) \\
\midrule

\multicolumn{7}{l}{\textbf{Panel A. Physician Openness}} \\
\addlinespace[2pt]
Treated $\times$ Openness
& $-0.069^{***}$ & $-0.037^{**}$ & $-0.017$ & $-0.000$ & $0.087^{***}$ & $-0.031$ \\
& $(0.017)$ & $(0.019)$ & $(0.016)$ & $(0.007)$ & $(0.021)$ & $(0.023)$ \\
Treated
& $-0.035^{***}$ & $-0.015$ & $-0.011$ & $0.000$ & $0.007$ & $0.004$ \\
& $(0.007)$ & $(0.008)$ & $(0.007)$ & $(0.003)$ & $(0.009)$ & $(0.009)$ \\
\addlinespace[6pt]
Patient Char. & Y & Y & Y & Y & Y & Y \\
Physician FE & Y & Y & Y & Y & Y & Y \\
\addlinespace[2pt]
Mean & 0.87 & 0.18 & 0.13 & 0.03 & 0.23 & 0.16 \\
Observations & 11,666 & 11,666 & 11,666 & 11,666 & 11,666 & 11,666 \\
\hline
\addlinespace[6pt]

\multicolumn{7}{l}{\textbf{Panel B. Physician Authority}} \\
\addlinespace[2pt]
Treated $\times$ Authority
& $0.036^{**}$ & $0.021^{*}$ & $0.023^{**}$ & $0.004$ & $-0.031$ & $0.007$ \\
& $(0.016)$ & $(0.012)$ & $(0.011)$ & $(0.007)$ & $(0.020)$ & $(0.017)$ \\
Treated
& $-0.068^{***}$ & $-0.031^{***}$ & $-0.026^{***}$ & $-0.003$ & $0.046^{***}$ & $-0.019$ \\
& $(0.014)$ & $(0.011)$ & $(0.010)$ & $(0.006)$ & $(0.018)$ & $(0.016)$ \\

\addlinespace[6pt]
Patient Char. & Y & Y & Y & Y & Y & Y \\
Physician FE & Y & Y & Y & Y & Y & Y \\
\addlinespace[2pt]
Mean & 0.87 & 0.18 & 0.13 & 0.03 & 0.23 & 0.16 \\
Observations & 11,666 & 11,666 & 11,666 & 11,666 & 11,666 & 11,666 \\
\bottomrule
\end{tabular}

}
\begin{minipage}{\textwidth}
    \footnotesize
    \textit{Notes:} This table reports heterogeneous treatment effects by physicians' attitudes measured in the post-experiment survey.
    In Panel A, \textit{Openness} equals one if the physician reported that medical knowledge acquired by patients from external sources is ``very helpful'' to physicians, and zero otherwise. In Panel B, \textit{Authority} equals one if the physician strongly agreed with the statement ``Patients should closely follow physicians' recommendations,'' and zero otherwise. All regressions control for patient characteristics (age, gender, occupation, and first-visit indicator) and include physician fixed effects.
    Robust standard errors are reported in parentheses.
$^{*} p < 0.10$, $^{**} p < 0.05$, $^{***} p < 0.01$.
\end{minipage}
\end{threeparttable}
\end{table}

\clearpage
\begin{table}[htbp]
\centering
\caption{Heterogeneous Treatment Effects by Physicians' Attitudes Toward Medications}
\label{tab:hetero_medication_attitudes}

\begin{tabular}{l*{3}{c}}
\toprule
& \multicolumn{3}{c}{Dependent Variables} \\
\cmidrule(lr){2-4}
& TCM & Antibiotics & Opioids \\
& (1) & (2) & (3) \\
\midrule
Treated $\times$ Pro-TCM & $-0.029^{**}$ & & \\
& $(0.014)$ & & \\
Treated $\times$ Pro-antibiotic & & $-0.017$ & \\
& & $(0.012)$ & \\
Treated $\times$ Pro-opioid & & & $-0.001$ \\
& & & $(0.007)$ \\
Treated & $-0.013$ & $-0.009$ & $0.001$ \\
& $(0.011)$ & $(0.009)$ & $(0.006)$ \\
& & & \\
Patient Char. & Y & Y & Y \\
Physician FE & Y & Y & Y \\
& & & \\
Mean & 0.18 & 0.13 & 0.03 \\
Observations & 11,666 & 11,666 & 11,666 \\
\bottomrule
\end{tabular}

\begin{minipage}{\textwidth}
    \footnotesize
    \textit{Notes:} This table reports heterogeneous treatment effects by physicians' attitudes toward different medications. Pro-TCM equals one if the physician selected ``Equal emphasis on TCM and Western medicine'' when asked about the relationship between TCM and Western medicine, and zero if the physician selected ``Completely Western medicine'' or ``Western medicine primary, TCM supplementary.'' Pro-antibiotic equals one if the physician selected ``Use is generally appropriate'' when asked about antibiotic use in China, and zero if the physician selected ``Obviously overused'' or ``Slightly overused.'' Pro-opioid equals one if the physician selected ``Use is generally appropriate'' when asked about opioid analgesic use in China, and zero otherwise. All regressions control for patient characteristics (age, gender, occupation, and an indicator for first visit) and include physician fixed effects. Robust standard errors are reported in parentheses.
$^{*}$ $p<0.10$, $^{**}$ $p<0.05$, $^{***}$ $p<0.01$.
\end{minipage}
\end{table}

\clearpage
\begin{table}[htbp]
\centering
\caption{Heterogeneous Treatment Effects by Physicians' Baseline Prescribing Intensity}
\label{tab:hetero_dept_high_prescriber}

\begin{tabular}{l*{3}{c}}
\toprule
& \multicolumn{3}{c}{Dependent Variables} \\
\cmidrule(lr){2-4}
& TCM & Antibiotics & Opioids \\
& (1) & (2) & (3) \\
\midrule
Treated $\times$ High-TCM & $-0.023^{**}$ & & \\
& $(0.012)$ & & \\
Treated $\times$ High-antibiotic & & $-0.019^{**}$ & \\
& & $(0.009)$ & \\
Treated $\times$ High-opioid & & & $-0.004$ \\
& & & $(0.004)$ \\
Treated & $-0.005$ & $-0.003$ & $0.004$ \\
& $(0.009)$ & $(0.008)$ & $(0.005)$ \\
& & & \\
Patient Char. & Y & Y & Y \\
Physician FE & Y & Y & Y \\
& & & \\
Mean & 0.18 & 0.13 & 0.03 \\
Observations & 11,666 & 11,666 & 11,666 \\
\bottomrule
\end{tabular}

\begin{minipage}{\textwidth}
    \footnotesize
    \textit{Notes:} This table reports heterogeneous treatment effects based on whether a physician has above-median treatment intensity within their department prior to the experiment between January and May 2025. High TCM (High Antibiotic, High Opioid) is an indicator equal to one if the physician’s pre-experiment mean prescribing rate for TCM (antibiotics, opioids) exceeds the department median, and zero otherwise. All regressions control for patient characteristics (age, gender, occupation, and an indicator for first visit) and include physician fixed effects. Robust standard errors are reported in parentheses.
$^{*}$ $p<0.10$, $^{**}$ $p<0.05$, $^{***}$ $p<0.01$.
\end{minipage}
\end{table}

\appendix
\setcounter{figure}{0}
\setcounter{table}{0}
\renewcommand\thetable{\thesection\arabic{table}}
\renewcommand\thefigure{\thesection\arabic{figure}}

\pagebreak
\newpage
\clearpage

\section*{ONLINE APPENDIX}
\section{Additional Figures and Tables}

\begin{figure}[h!]
\centering
\caption{Share of Clean Recommendations by Topic and Model, System Prompt Framing the Model as a Pre-Visit Assistant}
\label{fig:crossmodel_previst}
\includegraphics[width=\linewidth]{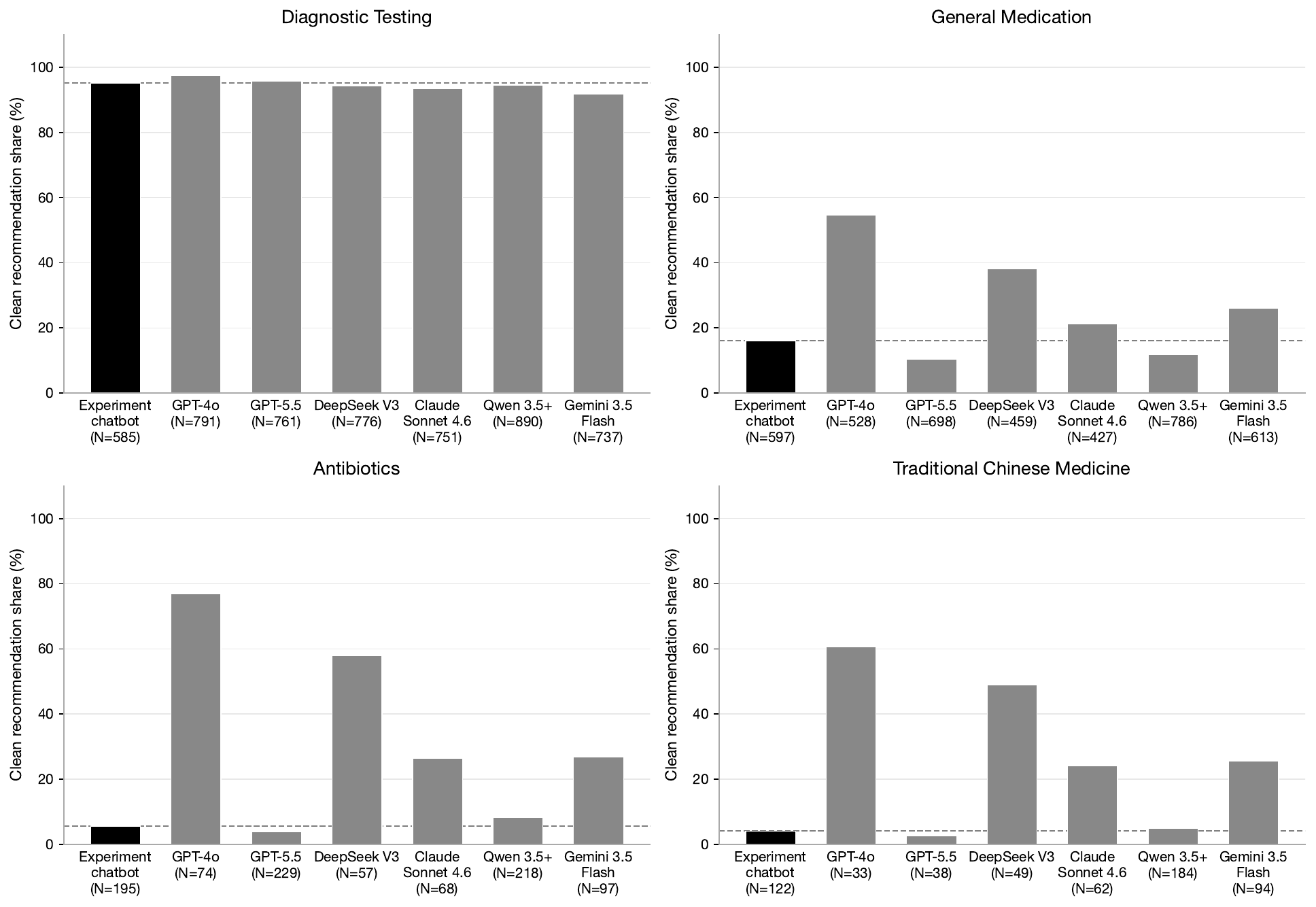}
\begin{minipage}{\textwidth}
\footnotesize
\textit{Notes:} Each panel corresponds to one of four clinical topics. Bars report the
share of topic mentions that the model recommends without any caution (clean
recommendations), classified using the stance pipeline described in
Appendix~\ref{sec:appendix-stance}. The dark bar is our experiment chatbot; the
remaining bars are six comparison models given the same first-turn patient messages
under a system prompt framing the model as a pre-visit assistant. The number of topic
mentions ($N$) appears below each model. Because the follow-up turns of each
conversation cannot be reconstructed for the comparison models, our chatbot is
restricted here to its first-turn responses, so its shares differ slightly from
Table~\ref{tab:conversation}, which uses the full analytic sample. Traditional Chinese
Medicine mention counts are small for some models, and those shares should be
read with caution.
\end{minipage}
\end{figure}

\begin{figure}[h!]
\centering
\caption{Share of Clean Recommendations by Topic and Model, System Prompt Framing the Model as a Direct Medical Consultant}
\label{fig:crossmodel_direct}
\includegraphics[width=\linewidth]{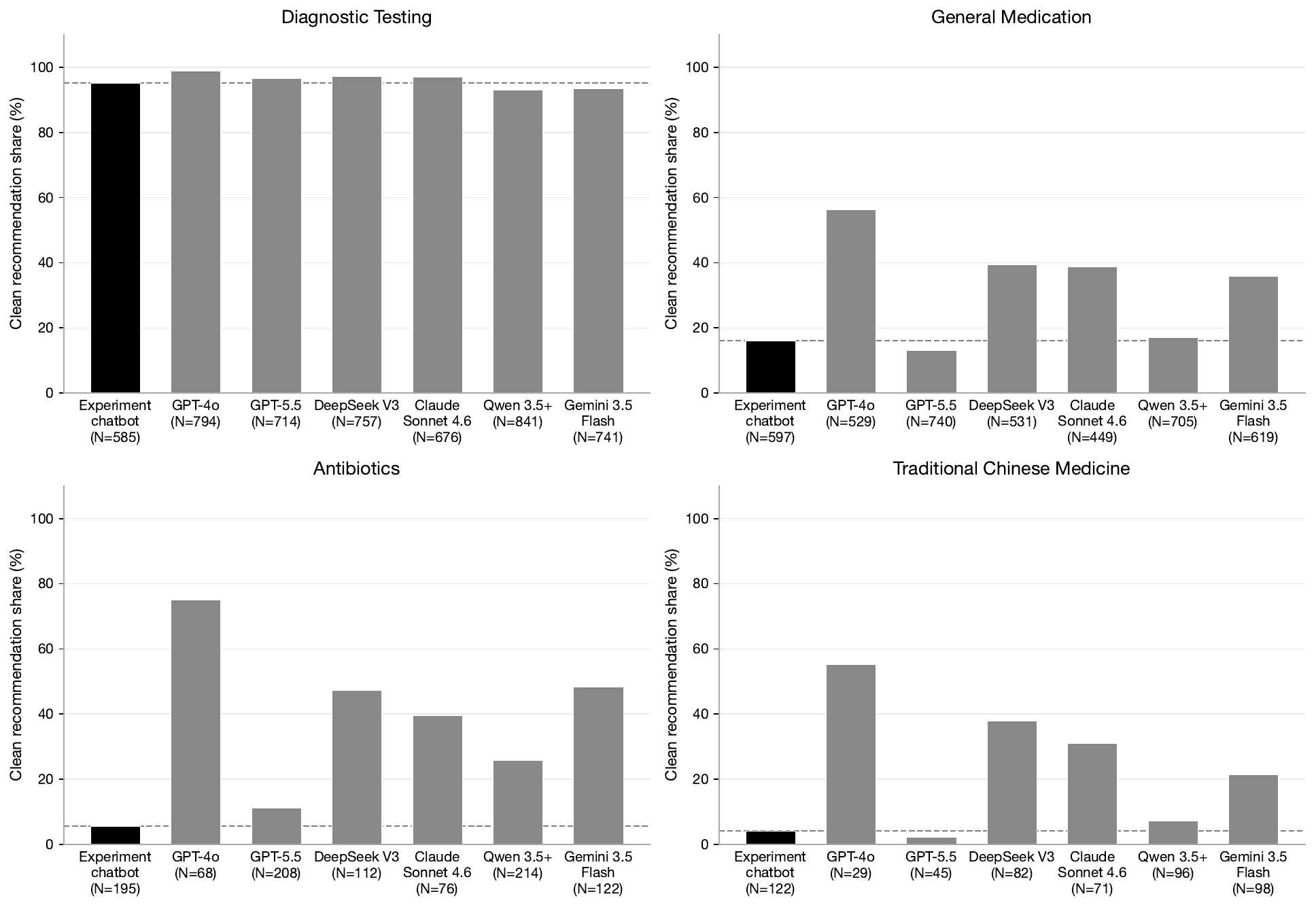}
\begin{minipage}{\textwidth}
\footnotesize
\textit{Notes:} Each panel corresponds to one of four clinical topics. Bars report the
share of topic mentions that the model recommends without any caution (clean
recommendations), classified using the stance pipeline described in
Appendix~\ref{sec:appendix-stance}. The dark bar is our experiment chatbot; the
remaining bars are six comparison models given the same first-turn patient messages
under a system prompt framing the model as a direct medical consultant. The number of topic
mentions ($N$) appears below each model. Because the follow-up turns of each
conversation cannot be reconstructed for the comparison models, our chatbot is
restricted here to its first-turn responses, so its shares differ slightly from
Table~\ref{tab:conversation}, which uses the full analytic sample. Traditional Chinese
Medicine mention counts are small for some models, and those shares should be
read with caution.
\end{minipage}
\end{figure}

\begin{figure}[t]
    \centering
    \begin{subfigure}{0.6\textwidth}
        \centering
        \subcaption{Medication Prescription}
        \includegraphics[width=\textwidth]{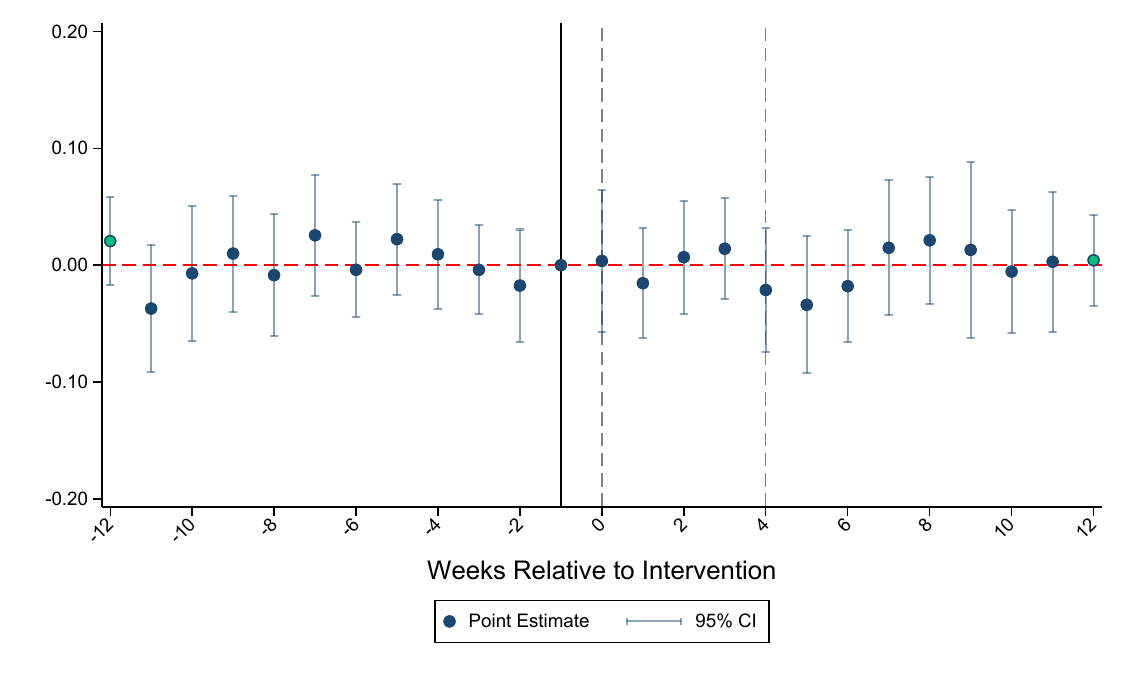}
    \end{subfigure}
    \medskip
    \begin{subfigure}{0.6\textwidth}
        \centering
        \subcaption{TCM}
        \includegraphics[width=\textwidth]{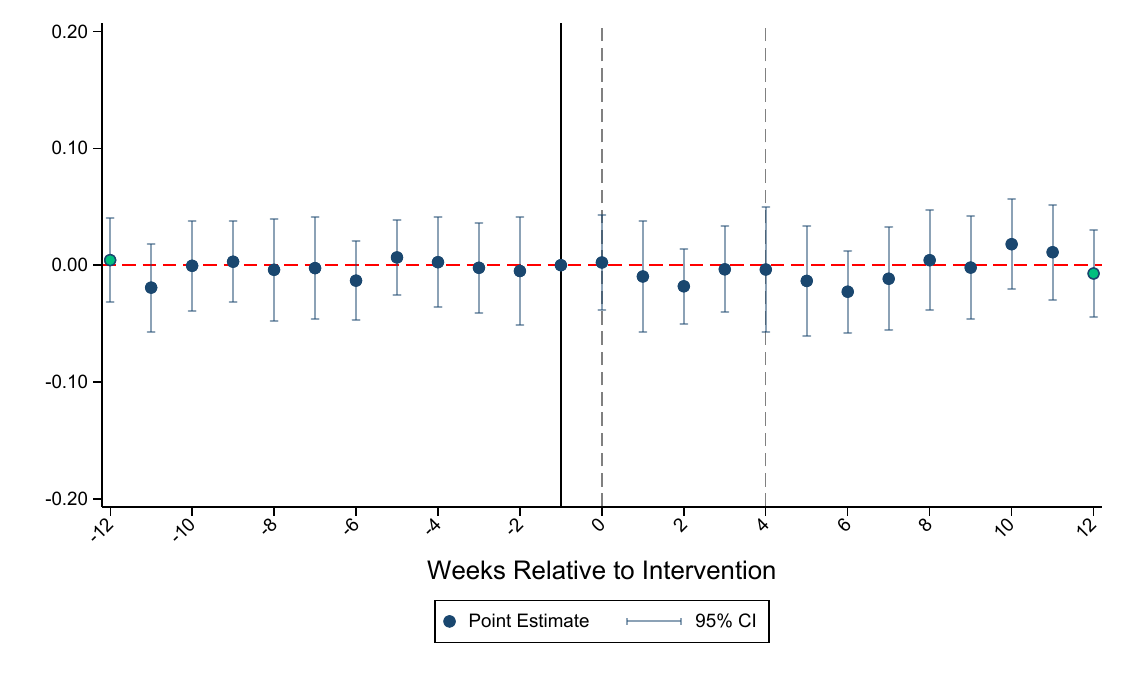}
    \end{subfigure}
    \medskip
    \begin{subfigure}{0.6\textwidth}
        \centering
        \subcaption{Antibiotics}
        \includegraphics[width=\textwidth]{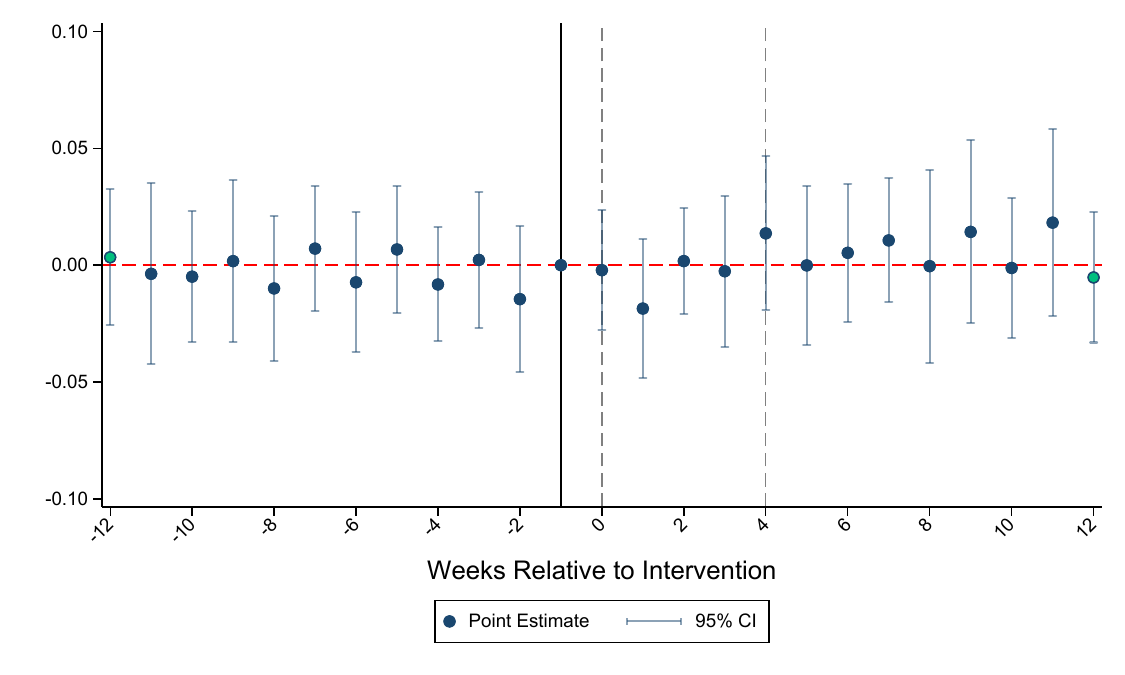}
    \end{subfigure}
    \caption{Dynamic Effects of Exposure to Treated Patients on Prescriptions and Clinical Practices, Excluding Treated Patients}
    \label{fig:placebo_weighted}
\end{figure}

\begin{figure}[t]\ContinuedFloat
    \centering
    \begin{subfigure}{0.6\textwidth}
        \centering
        \subcaption{Opioids}
        \includegraphics[width=\textwidth]{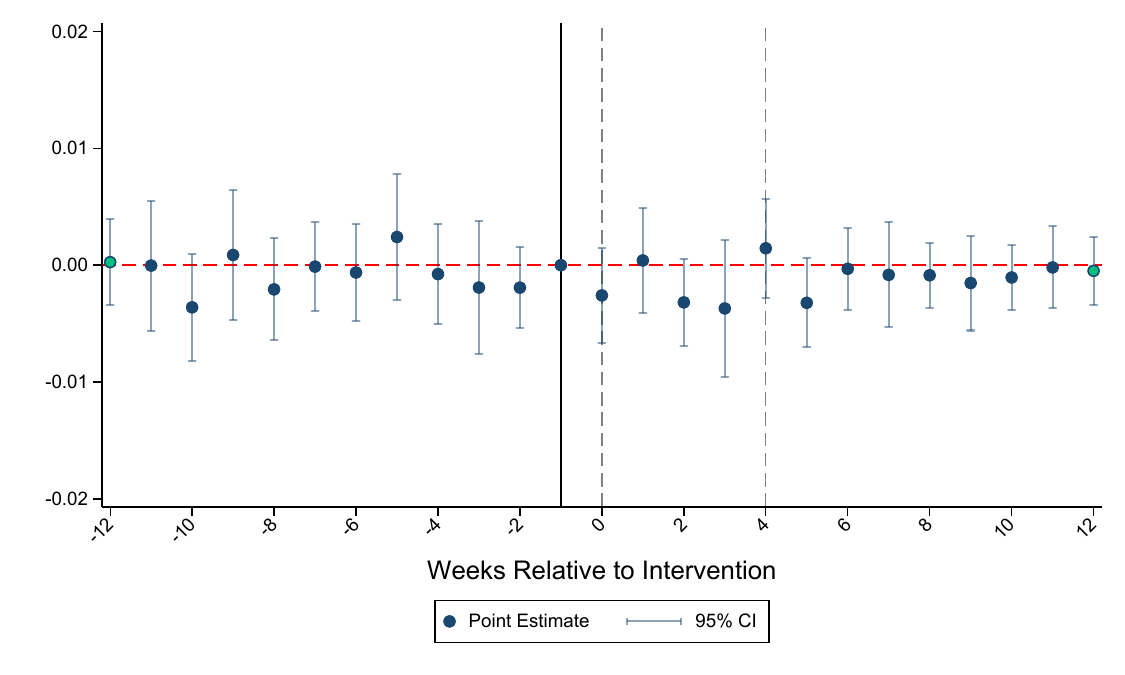}
    \end{subfigure}
    \medskip
    \begin{subfigure}{0.6\textwidth}
        \centering
        \subcaption{Diagnostic Tests}
        \includegraphics[width=\textwidth]{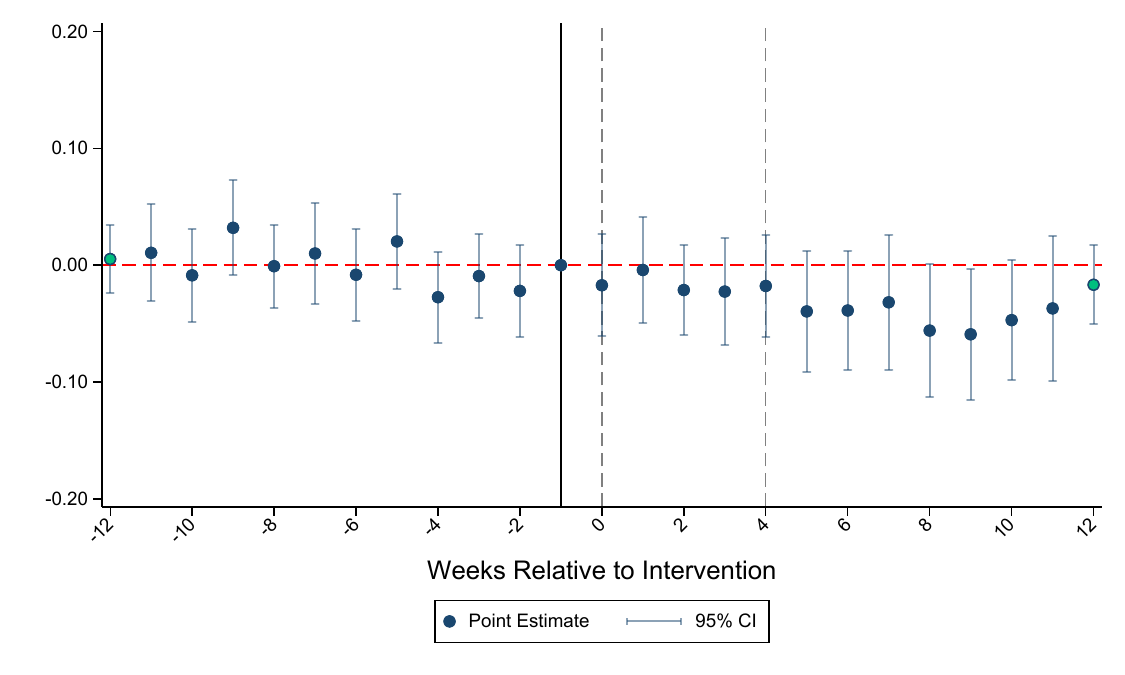}
    \end{subfigure}
    \medskip
    \begin{subfigure}{0.6\textwidth}
        \centering
        \subcaption{Revisit within Two Weeks}
        \includegraphics[width=\textwidth]{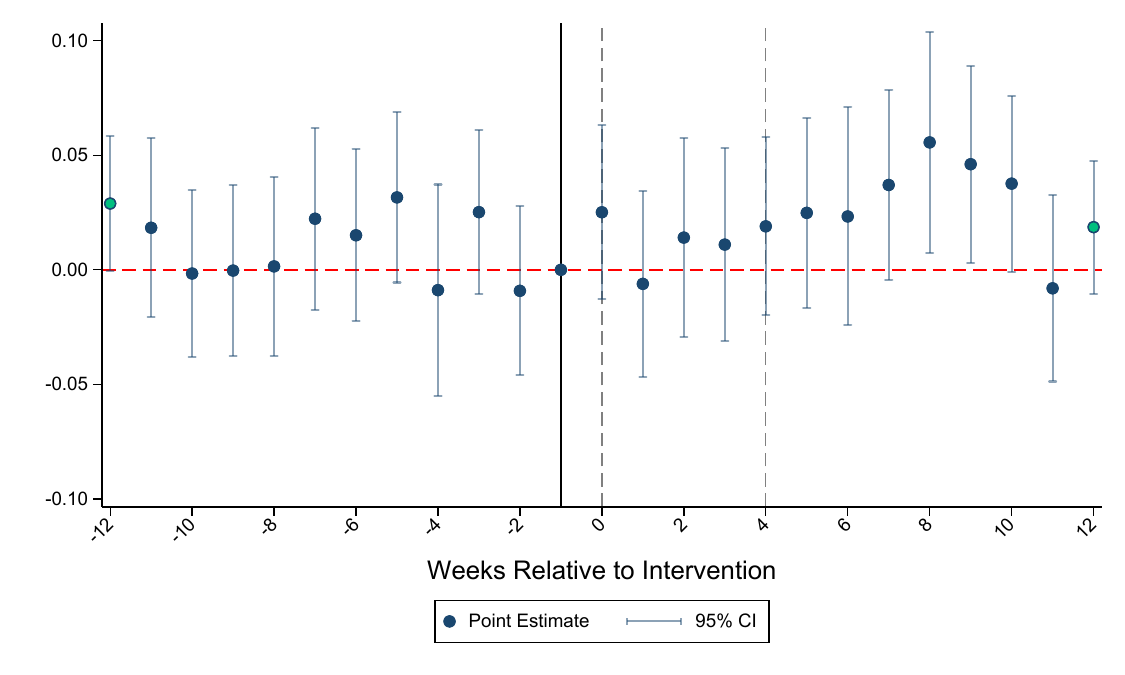}
    \end{subfigure}

    \caption{Dynamic Effects of Exposure to Treated Patients on Prescriptions and Clinical Practices, Excluding Treated Patients (Continued)}
     \begin{minipage}{\textwidth}
    \footnotesize
    \textit{Notes:} This figure presents the point estimates with 95\% confidence intervals from equation (\ref{eq:dynamic}). Data are aggregated to the physician-week level using patient visits from 2025, comparing those from \textit{Control} group and \textit{Unexposed} group, weighted by the number of visits per physician per week. Relative weeks beyond $-12$ and $+12$ are grouped at the endpoints. The baseline period is Week $-1$ (May 26--June 1, 2025). Standard errors are clustered at the physician level.
    \end{minipage}
\end{figure}

\newpage
\begin{figure}[ht]
    \centering
        \caption{Physicians' Perceived Patient AI Adoption Rate}
        \includegraphics[width=0.6\textwidth]{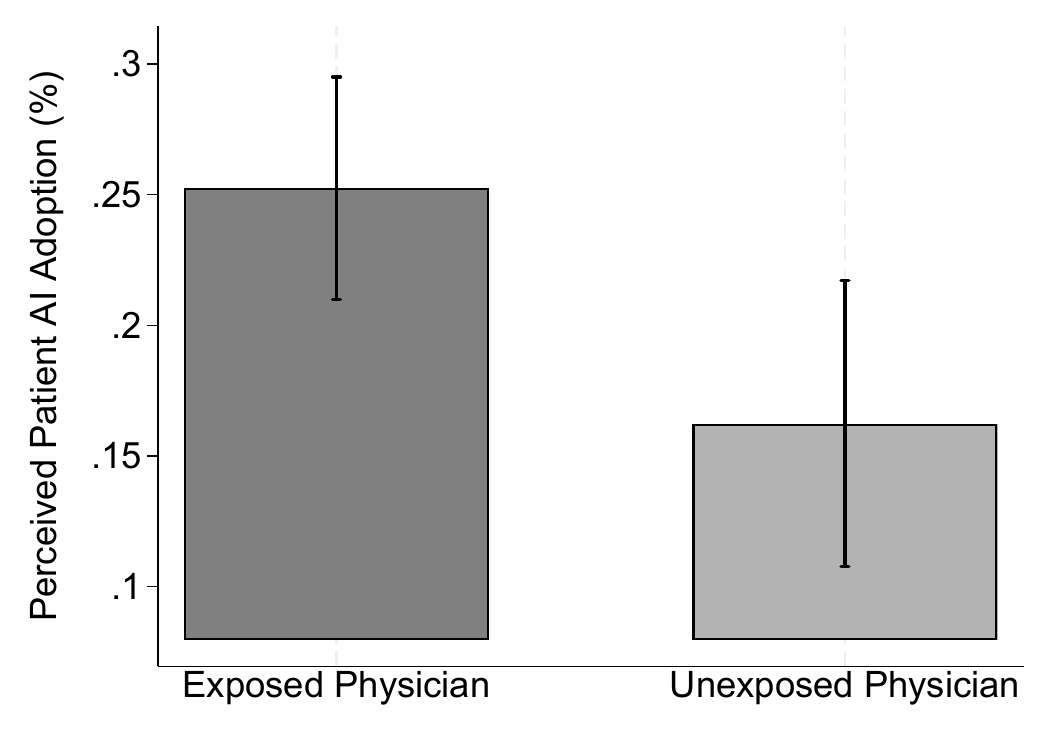}
    \label{fig:physician_perceived_ai}
    \medskip
    \begin{minipage}{\textwidth}
    \footnotesize
    \textit{Notes:} This figure reports physicians' survey responses collected after the conclusion of the experiment, comparing physicians randomized to the \textit{Exposed} and \textit{Unexposed} groups. The outcome variable is physicians' perceived share of patients who used generative AI chatbots prior to consultations during the experimental month, with 95\% confidence intervals (dashed lines).
    \end{minipage}
\end{figure}

\clearpage

\begin{table}[htbp]
\centering
\caption{Effect of Treatment on AI Adoption}
\label{tab:IV_first_stage}

\begin{tabular}{p{5.5cm} @{\hspace{1em}} >{\centering\arraybackslash}p{4cm} >{\centering\arraybackslash}p{4cm}}
\toprule
& \multicolumn{2}{c}{Whether Using AI Consulting Before Visit (1=Yes)} \\
\cmidrule(lr){2-3}
& (1) & (2) \\
\midrule
Treated & $0.171^{***}$ & $0.170^{***}$ \\
& (0.005) & (0.005) \\
\addlinespace
Patient Characteristics & No & Yes \\
Observations & 11,666 & 11,666 \\
R-squared & 0.093 & 0.098 \\
\bottomrule
\end{tabular}

\begin{minipage}{\textwidth}
    \footnotesize
    \textit{Notes:} This table reports linear probability estimates of the effect of treatment assignment on AI adoption, using the \textit{Treatment} and \textit{Control} samples. The dependent variable equals one if the patient used the AI chatbot before the visit, and zero otherwise. Column (1) includes only the treatment indicator, while Column (2) additionally controls for patient characteristics, including age, gender, occupation, and an indicator for first visit. Robust standard errors are reported in parentheses.
$^{*}$ p $<$ 0.10, $^{**}$ p $<$ 0.05, $^{***}$ p $<$ 0.01.
\end{minipage}
\end{table}

\clearpage
\begin{table}[htbp]
\centering
\caption{Patient Characteristics and AI Usage}
\label{tab:adopter_char}

\begin{tabular}{lcccc}
\toprule
& \multicolumn{4}{c}{Whether Using AI Consulting Before Visit (1=Yes)} \\
\cline{2-5}
& (1) & (2) & (3) & (4) \\
\hline
Age & $-0.005^{***}$ & $-0.006^{***}$ & $-0.006^{***}$ & $-0.006^{***}$ \\
& (0.000) & (0.000) & (0.000) & (0.000) \\
Male &  & $0.075^{***}$ & $0.055^{***}$ & $0.055^{***}$ \\
&  & (0.009) & (0.012) & (0.012) \\
Students &  &  & $-0.001$ & $-0.009$ \\
&  &  & (0.019) & (0.020) \\
Farmers &  &  & $-0.010$ & $-0.008$ \\
&  &  & (0.013) & (0.013) \\
Formally Employed &  &  & $0.067^{***}$ & $0.066^{***}$ \\
&  &  & (0.013) & (0.014) \\
Unemployed/Retired &  &  & $0.024$ & $0.022$ \\
&  &  & (0.025) & (0.027) \\
First Visit &  &  &  & $0.024^{**}$ \\
&  &  &  & (0.010) \\
\hline
Mean & 0.171 & 0.171  & 0.171 & 0.171 \\
Observations & 5,828 & 5,828 & 5,828 & 5,828 \\
\bottomrule
\end{tabular}

\begin{minipage}{\textwidth}
    \footnotesize
    \textit{Notes:} This table reports estimates from linear probability models examining the relationship between patient characteristics and AI usage. The dependent variable equals one if the patient used AI consulting before the visit, and zero otherwise. The sample is restricted to patients in the treatment group who have access to the chatbot. Columns (1)–(4) sequentially add patient characteristics, including age, gender, student status, occupation, employment status, and first visit status. Robust standard errors are reported in parentheses. $^{*}$ p $<$ 0.10, $^{**}$ p $<$ 0.05, $^{***}$ p $<$ 0.01.
\end{minipage}
\end{table}

\clearpage
\begin{table}[htbp]
\centering
\caption{Robustness Checks for Intent-to-Treat Effects on Clinical Practice Patterns: Clustered Standard Errors}
\label{tab:itt_practice_cluster}
\begin{tabular}{l*{6}{c}}
		\toprule
		& \multicolumn{6}{c}{Dependent Variables} \\
		\cmidrule(lr){2-7}
		& Prescribed & TCM & Antibiotics & Opioids & Diagnostic & Revisit \\
		& medication & & & & tests & \\
		& (1) & (2) & (3) & (4) & (5) & (6) \\
		\midrule
		\multicolumn{7}{l}{\textbf{Panel A: No Controls}} \\
		Treated & $-0.045^{***}$ & $-0.022^{***}$ & $-0.015^{***}$ & $-0.004$ & $0.027^{***}$ & $-0.011$ \\
		& $(0.007)$ & $(0.005)$ & $(0.006)$ & $(0.003)$ & $(0.008)$ & $(0.008)$ \\
		& & & & & & \\
		Mean & 0.87 & 0.18 & 0.13 & 0.03 & 0.23 & 0.16 \\
		Observations & 11,666 & 11,666 & 11,666 & 11,666 & 11,666 & 11,666 \\
		\midrule
\multicolumn{7}{l}{\textbf{Panel B: Patient Characteristics \& Physician FE}} \\
		Treated & $-0.046^{***}$ & $-0.022^{***}$ & $-0.015^{***}$ & $-0.004$ & $0.027^{***}$ & $-0.012$ \\
		& $(0.007)$ & $(0.005)$ & $(0.006)$ & $(0.003)$ & $(0.008)$ & $(0.008)$ \\
		& & & & & & \\
		Patient Char. & Y & Y & Y & Y & Y & Y \\
        Physician FE & Y & Y & Y & Y & Y & Y \\
		Mean & 0.87 & 0.18 & 0.13 & 0.03 & 0.23 & 0.16 \\
		Observations & 11,666 & 11,666 & 11,666 & 11,666 & 11,666 & 11,666 \\
		\bottomrule
\end{tabular}
\begin{minipage}{\textwidth}
    \footnotesize
    \textit{Notes:} This table reports within-physician intent-to-treat estimates of the effect of chatbot access on clinical practice patterns. The sample is restricted to outpatient visits with \textit{Exposed} physicians. Panel A reports estimates without controls; Panel B adds patient characteristics (age, gender, occupation, and an indicator for first visit) and physician fixed effects. Standard errors are clustered at the physician level.
		$^{*}$ p $<$ 0.10, $^{**}$ p $<$ 0.05, $^{***}$ p $<$ 0.01.
\end{minipage}
\end{table}

\clearpage
\begin{table}[htbp]
\centering
\caption{Robustness Checks for Intent-to-Treat Effects on Healthcare Expenditures: Clustered Standard Errors}
\label{tab:itt_costs_cluster}
\begin{tabular}{l*{5}{c}}
		\toprule
		& \multicolumn{5}{c}{Dependent Variables} \\
		\cmidrule(lr){2-6}
		& TCM costs & WM costs & Diagnostic & Other costs & Total costs \\
		& (RMB) & (RMB) & costs (RMB) & (RMB) & (RMB) \\
		& (1) & (2) & (3) & (4) & (5) \\
		\midrule
		\multicolumn{6}{l}{\textbf{Panel A: No Controls}} \\
		Treated & $-1.686^{*}$ & $-5.276^{*}$ & $6.542$ & $0.174$ & $-0.245$ \\
		& $(1.008)$ & $(3.080)$ & $(7.170)$ & $(3.007)$ & $(8.180)$ \\
		& & & & & \\
		Mean & 13.10 & 64.81 & 92.69 & 35.88 & 206.48 \\
		Observations & 11,666 & 11,666 & 11,666 & 11,666 & 11,666 \\
		\midrule
\multicolumn{6}{l}{\textbf{Panel B: Patient Characteristics \& Physician FE}} \\
		Treated & $-1.762^{*}$ & $-5.021$ & $6.313$ & $-0.002$ & $-0.472$ \\
		& $(0.999)$ & $(3.080)$ & $(7.229)$ & $(3.108)$ & $(8.304)$ \\
		& & & & & \\
		Patient Char. & Y & Y & Y & Y & Y \\
        Physician FE & Y & Y & Y & Y & Y \\
		Mean & 13.10 & 64.81 & 92.69 & 35.88 & 206.48 \\
		Observations & 11,666 & 11,666 & 11,666 & 11,666 & 11,666 \\
		\bottomrule
\end{tabular}
\begin{minipage}{\textwidth}
    \footnotesize
    \textit{Notes:} This table reports within-physician intent-to-treat estimates of the effect of chatbot access on expenditures. The sample is restricted to outpatient visits with \textit{Exposed} physicians. Panel A reports estimates without controls; Panel B adds patient characteristics (age, gender, occupation, and an indicator for first visit) and physician fixed effects. Standard errors are clustered at the physician level.
		$^{*}$ p $<$ 0.10, $^{**}$ p $<$ 0.05, $^{***}$ p $<$ 0.01.
\end{minipage}
\end{table}

\clearpage
\begin{table}[htbp]
\centering
\caption{Robustness Checks for Instrumental Variable Estimates for Clinical Practice Patterns: Clustered Standard Errors}\label{tab:iv_practice_cluster}
\begin{tabular}{l*{6}{c}}
\toprule
    & \multicolumn{6}{c}{Dependent Variables} \\
    \cmidrule(lr){2-7}
    & Prescribed & TCM & Antibiotics & Opioids & Diagnostic & Revisit \\
    & medication & & & & tests & \\
    & (1) & (2) & (3) & (4) & (5) & (6) \\
\midrule

\multicolumn{7}{l}{\textbf{Panel A: No Controls}} \\
    Chatbot Usage
        & $-0.265^{***}$ & $-0.128^{***}$ & $-0.090^{***}$ & $-0.026$ & $0.161^{***}$ & $-0.066$ \\
        & $(0.039)$ & $(0.030)$ & $(0.033)$ & $(0.017)$ & $(0.044)$ & $(0.043)$ \\
    & & & & & & \\
    Mean
        & 0.87 & 0.18 & 0.13 & 0.03 & 0.23 & 0.16 \\
    Observations
        & 11,666 & 11,666 & 11,666 & 11,666 & 11,666 & 11,666 \\
    First Stage F-statistics
        & \multicolumn{6}{c}{964.983} \\

\midrule

\multicolumn{7}{l}{\textbf{Panel B: Patient Characteristics \& Physician FE}} \\
    Chatbot Usage
        & $-0.267^{***}$ & $-0.129^{***}$ & $-0.089^{***}$ & $-0.025$ & $0.161^{***}$ & $-0.070$ \\
        & $(0.040)$ & $(0.031)$ & $(0.033)$ & $(0.017)$ & $(0.045)$ & $(0.044)$ \\
    & & & & & & \\
    Patient Char.
        & Y & Y & Y & Y & Y & Y \\
    Physician FE
        & Y & Y & Y & Y & Y & Y \\
    Mean
        & 0.87 & 0.18 & 0.13 & 0.03 & 0.23 & 0.16 \\
    Observations
        & 11,666 & 11,666 & 11,666 & 11,666 & 11,666 & 11,666 \\
    First Stage F-statistics
        & \multicolumn{6}{c}{947.617} \\

\bottomrule
\end{tabular}
\begin{minipage}{\textwidth}
    \footnotesize
    \textit{Notes:} This table reports within-physician instrumental variable estimates of the effect of chatbot usage on clinical practice patterns, where chatbot usage is instrumented with random assignment to chatbot access. The sample is restricted to outpatient visits with \textit{Exposed} physicians. Panel A reports estimates without controls; Panel B adds patient characteristics (age, gender, occupation, and an indicator for first visit) and physician fixed effects. The reported first-stage F-statistic is the Kleibergen-Paap Wald F statistic. Standard errors are clustered at the physician level.
		$^{*}$ p $<$ 0.10, $^{**}$ p $<$ 0.05, $^{***}$ p $<$ 0.01.
\end{minipage}
\end{table}

\clearpage
\begin{table}[htbp]
\centering
\caption{Robustness Checks for Instrumental Variable Estimates for Healthcare Expenditures: Clustered Standard Errors}
\label{tab:iv_costs_cluster}
\begin{tabular}{l*{5}{c}}
\toprule
    & \multicolumn{5}{c}{Dependent Variables} \\
    \cmidrule(lr){2-6}
    & TCM costs & WM costs & Diagnostic & Other costs & Total costs \\
    & (RMB) & (RMB) & costs (RMB) & (RMB) & (RMB) \\
    & (1) & (2) & (3) & (4) & (5) \\
\midrule

\multicolumn{6}{l}{\textbf{Panel A: No Controls}} \\
    Chatbot Usage
        & $-9.884^{*}$ & $-30.933^{*}$ & $38.359$ & $1.023$ & $-1.436$ \\
        & $(5.911)$ & $(18.072)$ & $(41.813)$ & $(17.560)$ & $(47.759)$ \\
    & & & & & \\
    Mean
        & 13.10 & 64.81 & 92.69 & 35.88 & 206.48 \\
    Observations
        & 11,666 & 11,666 & 11,666 & 11,666 & 11,666 \\
    First Stage F-statistics
        & \multicolumn{5}{c}{964.983} \\

\midrule

\multicolumn{6}{l}{\textbf{Panel B: Patient Characteristics \& Physician FE}} \\
    Chatbot Usage
    & $-10.334^{*}$ & $-29.444$ & $37.018$ & $-0.009$ & $-2.770$ \\
    & $(5.883)$ & $(18.154)$ & $(42.344)$ & $(18.225)$ & $(48.693)$ \\

    & & & & & \\
    Patient Char.
        & Y & Y & Y & Y & Y \\
    Physician FE
        & Y & Y & Y & Y & Y \\
    Mean
        & 13.10 & 64.81 & 92.69 & 35.88 & 206.48 \\
    Observations
        & 11,666 & 11,666 & 11,666 & 11,666 & 11,666 \\
    First Stage F-statistics
        & \multicolumn{5}{c}{947.617} \\

\bottomrule
\end{tabular}
\begin{minipage}{\textwidth}
    \footnotesize
    \textit{Notes:} This table reports within-physician instrumental variable estimates of the effect of chatbot usage on healthcare expenditures, where chatbot usage is instrumented with random assignment to chatbot access. The sample is restricted to outpatient visits with \textit{Exposed} physicians. Panel A reports estimates without controls; Panel B adds patient characteristics (age, gender, occupation, and an indicator for first visit) and physician fixed effects. The reported first-stage F-statistic is the Kleibergen-Paap Wald F statistic. Standard errors are clustered at the physician level.
		$^{*}$ p $<$ 0.10, $^{**}$ p $<$ 0.05, $^{***}$ p $<$ 0.01.
\end{minipage}
\end{table}

\clearpage
\begin{table}[htbp]
	\centering
        \caption{Robustness Checks for Intent-to-Treat Effects on Clinical Practice Patterns: First Visit of the Experimental Period Subsample}
	\label{tab:itt_practice_first}
\begin{tabular}{l*{6}{c}}
\toprule
    & \multicolumn{6}{c}{Dependent Variables} \\
    \cmidrule(lr){2-7}
    & Prescribed & TCM & Antibiotics & Opioids & Diagnostic & Revisit \\
    & medication & & & & tests & \\
    & (1) & (2) & (3) & (4) & (5) & (6) \\
\midrule

\multicolumn{7}{l}{\textbf{Panel A: No Controls}} \\
    Treated
        & $-0.036^{***}$ & $-0.018^{**}$ & $-0.011$ & $-0.006^{*}$ & $0.036^{***}$ & $-0.008$ \\
        & $(0.008)$ & $(0.008)$ & $(0.007)$ & $(0.004)$ & $(0.009)$ & $(0.008)$ \\
    & & & & & & \\
    Mean
        & 0.81 & 0.16 & 0.12 & 0.04 & 0.28 & 0.14 \\
    Observations
        & 8,166 & 8,166 & 8,166 & 8,166 & 8,166 & 8,166 \\

\midrule

\multicolumn{7}{l}{\textbf{Panel B: Patient Characteristics \& Physician FE}} \\
    Treated
        & $-0.037^{***}$ & $-0.018^{**}$ & $-0.011$ & $-0.006$ & $0.036^{***}$ & $-0.009$ \\
        & $(0.008)$ & $(0.008)$ & $(0.008)$ & $(0.004)$ & $(0.009)$ & $(0.008)$ \\
    & & & & & & \\
    Patient Char.
        & Y & Y & Y & Y & Y & Y \\
    Physician FE
        & Y & Y & Y & Y & Y & Y \\
    Mean
        & 0.81 & 0.16 & 0.12 & 0.04 & 0.28 & 0.14 \\
    Observations
        & 8,166 & 8,166 & 8,166 & 8,166 & 8,166 & 8,166 \\

\bottomrule
\end{tabular}
\begin{minipage}{\textwidth}
    \footnotesize
    \textit{Notes:} This table reports within-physician intent-to-treat estimates of the effect of chatbot access on clinical practice patterns. The sample is restricted to patients' first-time outpatient visits with \textit{Exposed} physicians during the experimental period. Panel A reports estimates without controls; Panel B adds patient characteristics (age, gender, and occupation) and physician fixed effects. Robust standard errors are reported in parentheses.
		$^{*}$ p $<$ 0.10, $^{**}$ p $<$ 0.05, $^{***}$ p $<$ 0.01.
\end{minipage}
\end{table}

\clearpage
\begin{table}[htbp]
\centering
\caption{Persistent Effects of Treatment Assignment on Outpatient Utilization in the Post-Experiment Period  (October–December 2025)}
\label{tab:persistent_patient_level}

\begin{tabular}{p{4cm} @{\hspace{1em}} >{\centering\arraybackslash}p{2.6cm} >{\centering\arraybackslash}p{2.6cm} >{\centering\arraybackslash}p{2.6cm} >{\centering\arraybackslash}p{2.6cm}}

\toprule

& \multicolumn{2}{c}{Any visit (dummy)} & \multicolumn{2}{c}{Number of visits} \\

& (1) & (2) & (3) & (4) \\

\midrule

Treated        & $-0.011$ & $-0.009$ & $-0.022$ & $-0.012$ \\

               & (0.011)  & (0.011)  & (0.047)  & (0.046)  \\

               &          &          &          &          \\

Patient Char.  & N        & Y        & N        & Y        \\

Mean           &  0.46        &  0.46         &  1.24        &  1.24        \\

Observations   & 8,166    & 8,166    & 8,166    & 8,166    \\

\bottomrule

\end{tabular}
\begin{minipage}{\textwidth}
    \footnotesize
    \textit{Notes:} This table reports the persistent effects of treatment assignment on patients' use of outpatient care after the experiment ended. The unit of observation is the patient. The sample consists of all patients whose first experimental-period visit was with an \textit{Exposed} physician. We track patients from the experimental period and match their subsequent outpatient visits over October to December 2025 through unique patient identifiers. Because a patient may appear in more than one experimental-period visit with differing assignments, we group each patient by the assignment of their first such visit: \textit{Treatment} if it received chatbot access and \textit{Control} otherwise. The dependent variable in columns (1) and (2) is an indicator equal to one if the patient made at least one outpatient visit during the period; in columns (3) and (4), it is the patient's total visit count over the same period. Columns (2) and (4) include patient characteristics (age, gender, and occupation). Robust standard errors are reported in parentheses. $^{*}$~$p < 0.10$, $^{**}$~$p < 0.05$, $^{***}$~$p < 0.01$.
\end{minipage}
\end{table}

\clearpage
\begin{table}[htbp]
\centering
\caption{Persistent Effects of Treatment Assignment on Clinical Practice in the Post-Experiment Period (October–December 2025)}
\label{tab:persistent_visit_level}
\begin{tabular}{p{4cm} @{\hspace{1em}} >{\centering\arraybackslash}p{2.6cm} >{\centering\arraybackslash}p{2.6cm} >{\centering\arraybackslash}p{2.6cm} >{\centering\arraybackslash}p{2.6cm}}
\toprule
& \multicolumn{2}{c}{Prescribed medication} & \multicolumn{2}{c}{Diagnostic tests} \\
& (1) & (2) & (3) & (4) \\
\midrule
Treated & $-0.002$ & $-0.003$ & $0.016^{*}$ & $0.016^{*}$ \\
& (0.008) & (0.008) & (0.010) & (0.009) \\
& & & & \\
Patient Char. & N & Y & N & Y \\
Mean & 0.83 & 0.83 & 0.21 & 0.21 \\
Observations & 10,149 & 10,149 & 10,149 & 10,149 \\
\bottomrule
\end{tabular}
\begin{minipage}{\textwidth}
    \footnotesize
    \textit{Notes:} This table reports persistent effects of treatment assignment on clinical practice at subsequent visits after the experiment ended. The unit of observation is the visit. The sample consists of all outpatient visits over October to December 2025 by patients whose first experimental-period visit was with an \textit{Exposed} physician. Because a patient may appear in more than one experimental-period visit with differing assignments, we group each patient by the assignment of their first such visit, \textit{Treatment} if it received chatbot access and \textit{Control} otherwise. The dependent variables are indicators equal to one if the visit resulted in a prescribed medication, in columns (1) and (2), and in a diagnostic test, in columns (3) and (4). Columns (2) and (4) include patient characteristics (age, gender, and occupation). Standard errors are clustered at the patient level. $^{*}$~$p < 0.10$, $^{**}$~$p < 0.05$, $^{***}$~$p < 0.01$.
\end{minipage}
\end{table}

\clearpage
\begin{table}[htbp]
\centering
\caption{Effect of Chatbot Access on Patient Perceptions of the Consultation}
\label{tab:patient_survey}
\resizebox{0.6\textwidth}{!}{

\begin{tabular}{l*{3}{c}}
\toprule
 & Satisfaction
 & Communication
 & Compliance \\
  & (1) & (2) & (3) \\[2pt]
\midrule
Treated
    & $-0.032$       & $-0.035^{*}$    & $-0.041^{**}$ \\
    & $(0.020)$      & $(0.019)$      & $(0.019)$     \\
     &  &  &  \\
Patient Char.   & Y & Y & Y \\
Physician FE & Y & Y & Y \\
Mean & 0.86  & 0.85  & 0.87  \\
Observations   & 1,418 & 1,418 & 1,418 \\
\bottomrule
\end{tabular}
}
\begin{minipage}{\textwidth}
    \footnotesize
    \textit{Notes:} This table reports the effect of chatbot access on patient survey responses following the consultation. The dependent variables are binary indicators equal to one if the patient reported (1) being ``very satisfied'' with the consultation, (2) perceiving physician–patient communication as ``very smooth,'' and (3) intending to ``fully comply'' with the physician's medical advice. The sample is restricted to survey respondents in the \textit{Treatment} and \textit{Control} groups. All regressions control for patient characteristics (age, gender, occupation, and an indicator for first visit) and include physician fixed effects. Robust standard errors are reported in parentheses. $^{*}$~$p < 0.10$, $^{**}$~$p < 0.05$, $^{***}$~$p < 0.01$.
\end{minipage}
\end{table}

\clearpage
\begin{table}[htbp]
\centering
\caption{Effect of Patient AI Adoption on Physicians' Perceptions of the Consultation}
\label{tab:physician_survey_share}

\begin{tabular}{
>{\raggedright\arraybackslash}p{3.5cm}
>{\centering\arraybackslash}p{2.4cm}
>{\centering\arraybackslash}p{2.4cm}
>{\centering\arraybackslash}p{2.4cm}
>{\centering\arraybackslash}p{2.4cm}
}
\toprule
 & Clarity
 & Understanding
 & Communication
 & Compliance \\
 & (1) & (2) & (3) & (4) \\
\midrule

\multicolumn{5}{l}{\textbf{Panel A: No Controls}} \\

Share of AI Adopters
    & $0.242$   & $0.228$   & $0.014$  & $-0.215^{*}$ \\
    & $(0.147)$ & $(0.146)$  & $(0.162)$ & $(0.121)$ \\
    &           &           &              &           \\
Mean & 0.23 & 0.22 & 0.35 & 0.29 \\
Observations  & 171  & 171  & 171  & 171  \\

\midrule

\multicolumn{5}{l}{\textbf{Panel B: Physician Characteristics \& Department FE}} \\

Share of AI Adopters
    & $0.267^{*}$ & $0.187$   & $0.063$ & $-0.217^{*}$   \\
    & $(0.146)$   & $(0.142)$  & $(0.163)$  & $(0.125)$ \\
    &             &           &              &           \\
Mean & 0.23 & 0.22 & 0.35 & 0.29 \\
Observations  & 171  & 171  & 171  & 171  \\

\bottomrule
\end{tabular}
\begin{minipage}{\textwidth}
    \footnotesize
    \textit{Notes:} This table reports the effect of patient AI adoption on physicians' perceptions of the consultation. The dependent variables are binary indicators equal to one if the physician reports that (1) the patient described their symptoms ``very clearly''; (2) the patient demonstrated ``very good understanding'' of their health condition; (3) communication with the patient was ``very smooth''; and (4) the patient was ``very compliant'' with physician instructions. The unit of observation is the physician.  \textit{Share of AI Adopters} is the fraction of a physician's patients who used the AI chatbot, equal to zero for \textit{Unexposed} physicians by construction. Panel A includes no controls. Panel B controls for physician characteristics (gender, education, years of experience, and seniority) and includes department fixed effects. Robust standard errors are reported in parentheses. $^{*}$~$p < 0.10$, $^{**}$~$p < 0.05$, $^{***}$~$p < 0.01$.
\end{minipage}
\end{table}

\clearpage

\section{Analyses of Patient--AI Conversation Logs}
\label{sec:appendix-classification}

This appendix provides additional detail on the analyses of the patient--AI conversation logs. The analytic sample contains 1,192 turns from 956 visits. A turn consists of a single patient message and the AI's reply. Most conversations (81.5 percent) consist of a single turn; the remaining 18.5 percent are multi-turn conversations, with two-turn conversations accounting for the majority of these. The analysis proceeds in three steps. We first classify each patient message by its intent, in order to characterize what patients ask the chatbot. We then identify, for each conversation turn, which of four clinical topics the patient and the AI discuss. Finally, for each topic the AI discusses, we classify the AI's stance along two dimensions: whether the AI recommends use of the treatment or test, and whether it raises any caution against use.

\subsection{Patient Intent}
\label{sec:appendix-intent}

We classify the intent of each patient message into one of eight categories. Five categories cover the medical content of the consultation: symptom inquiries, in which the patient describes symptoms and asks for a diagnosis or cause; test result inquiries, in which the patient shares a lab result and asks for an interpretation; treatment inquiries, in which the patient asks how to manage a condition or which medication to take; test inquiries, in which the patient asks whether to undergo a specific test; and lifestyle inquiries, in which the patient asks about diet, exercise, or daily self-care. Two categories cover the practicalities of obtaining care: care-seeking, in which the patient asks whether, when, or where to see a doctor, and administrative inquiries about cost, insurance, or scheduling. The descriptive shares reported in Section \ref{sec:conv} combine these last two into a single ``logistical questions'' share, since both concern how to obtain care rather than its medical content. A residual ``other'' category collects greetings and uninformative messages.

Classification is performed by GPT-4o-mini at temperature zero. For each message, the model is given the eight category definitions and returns a primary label, defined as the label that best captures the message's main intent, and zero or more secondary labels. We instruct the model to assign secondary labels only when the message explicitly raises a second distinct topic, not when a second topic is merely implied by the primary.

Appendix Table~\ref{tab:intent-distribution} reports the distribution of primary intent labels across the 1,192 messages in the analytic sample, together with a representative example of each. Symptom inquiries account for two-thirds of all messages; treatment inquiries are the next largest category, followed by smaller shares of lifestyle, test, care-seeking, result, and administrative inquiries.

\begin{table}[h!]
\centering
\caption{Primary Intent Distribution and Examples}
\label{tab:intent-distribution}
\begin{tabular}{p{0.23\linewidth} r r p{0.50\linewidth}}
\toprule
Intent & N & \% & Example patient message (translated) \\
\midrule
Symptom inquiry & 795 & 66.7 & ``Lately I keep feeling tightness in my chest. It hurts a little when I breathe deeply. It's been about a week.'' \\
\addlinespace
Treatment inquiry & 140 & 11.7 & ``Do I need to take any medicine? Can I take Suxiao Jiuxin Wan?'' \\
\addlinespace
Lifestyle inquiry & 65 & 5.5 & ``One more question: can I exercise during this period, or will it make symptoms worse?'' \\
\addlinespace
Test inquiry & 59 & 4.9 & ``My foot is still swollen. Should I get an X-ray?'' \\
\addlinespace
Care-seeking & 51 & 4.3 & ``How long do I have to wait before I can get a dental implant?'' \\
\addlinespace
Result inquiry & 37 & 3.1 & ``I'm a 35-year-old woman. A recent physical showed an elevated AST/ALT ratio of 1.64.'' \\
\addlinespace
Administrative inquiry & 28 & 2.3 & ``If I get tested at the CDC, do I have to use my real name? Will there be a record?'' \\
\addlinespace
Other & 17 & 1.4 & ``I haven't had that nasal endoscopy.'' \\
\bottomrule
\end{tabular}
\begin{minipage}{\textwidth}
    \footnotesize
\textit{Notes:} Sample of 1,192 patient messages from 956 visits with chatbot access. Each message is assigned exactly one primary intent label. Examples are translated from the original Mandarin.
\end{minipage}
\end{table}

\subsection{Topic Identification}
\label{sec:appendix-topics}

The remaining analysis examines the clinical content of the conversation. We focus on four topics that correspond to the clinical outcomes studied in the main analyses: diagnostic testing, general medication (excluding TCM and antibiotics), Traditional Chinese Medicine, and antibiotics. For each turn, and separately for the patient's message and the AI's reply, we record whether each topic is mentioned.

Topic mentions are identified by keyword matching at the sentence level. We split each message into sentences and check whether any sentence contains a keyword associated with the topic. Sentence-level matching, rather than document-level matching, prevents false positives that arise when distantly related terms co-occur in long structured replies. A turn may match multiple topics, in which case it enters each topic's analysis independently.

Two topic-specific refinements are worth noting. For diagnostic testing, we distinguish always-specific test names, which match on their own (for example, complete blood count, CT, or endoscopy), from condition terms such as liver function, which match only when paired with a testing-action verb in the same sentence; this prevents general references to a clinical condition from being miscoded as references to a test. For general medication, we exclude any sentence whose match is driven by a TCM keyword, an antibiotic keyword, or a marker for non-pharmacological treatment, so that the residual category captures conventional Western pharmaceuticals without overlapping the TCM and antibiotic categories.

Appendix Table~\ref{tab:keyword-scope} summarizes the keyword scope for each topic.

\begin{table}[h!]
\centering
\caption{Keyword Lists for Topic Identification}
\label{tab:keyword-scope}
\begin{tabular}{p{0.22\linewidth} p{0.72\linewidth}}
\toprule
Topic & Keywords \\
\midrule
Diagnostic testing & Always-specific test names (e.g., complete blood count, CT, MRI, ultrasound, endoscopy, biopsy) match directly; condition terms (e.g., liver function, lipid panel, thyroid function) match only when paired with a testing-action verb in the same sentence.  \\
\addlinespace
General medication & Sentence contains the generic term for medication or a specific drug name (e.g., ibuprofen, acetaminophen, aspirin, metformin, omeprazole, statins, progesterone); excludes sentences also matching TCM, antibiotic, or non-pharmacological keywords. \\
\addlinespace
TCM & Chinese herbal medicine, proprietary Chinese medicines, traditional Chinese medicine practice, herbal decoctions, acupuncture, and tuina massage. \\
\addlinespace
Antibiotics & Generic antibiotic terms and named agents (penicillin, amoxicillin, cephalosporins, azithromycin, erythromycin, levofloxacin, ofloxacin, metronidazole, tinidazole, clarithromycin, roxithromycin, doxycycline, tetracycline). The colloquial Chinese term for ``anti-inflammatory medicine,'' which patients use loosely, is treated as general medication rather than antibiotics. \\
\bottomrule
\end{tabular}
\end{table}

\subsection{AI Stance}
\label{sec:appendix-stance}

For every turn--topic pair in which the AI mentions a topic, we classify the AI's stance along two dimensions: whether the AI recommends the treatment or test, and whether it raises a caution against use. We treat these as independent binary flags. Clinical guidance often combines both: a passage may recommend acetaminophen for fever above 38.5\textdegree C while warning against self-administration of cough suppressants, or recommend physician-supervised use of an antibiotic while discouraging patient-initiated use. Coding the two dimensions separately yields four interpretable cells per topic: recommend without caution, recommend with caution, caution without recommending, and neither.

Classification is performed by GPT-4o at temperature zero. For each turn--topic pair, we extract the topic-matching sentences from the AI reply, concatenate them into a single passage, and submit the passage together with the topic name. The model is instructed to evaluate the two flags only with respect to how the AI frames the specified medical topic.

Three features of the prompt warrant emphasis. First, the prompt specifies that providing dosage, timing, or usage instructions for a treatment counts as recommending it, even when no explicit directive appears, since these instructions carry the same operative meaning as an explicit recommendation. By the same logic, listing tests with diagnostic purposes counts as recommending the tests, and urging the patient to seek care urgently counts as a recommendation rather than a caution. Second, the prompt distinguishes deferring to a physician from recommending physician-supervised use: telling the patient to consult a physician about whether to use a treatment is not coded as a recommendation, while instructing the patient to use a treatment under physician guidance is. Third, the prompt treats negative statements as cautions: stating that a treatment is not needed, not indicated, or not effective for the condition is coded as cautioning, since these statements function as discouragements of use even when the AI does not invoke risk or side effects.

Appendix Table~\ref{tab:classification-examples} reports representative AI replies illustrating each of the main attitude cells.

\begin{table}[h!]
\centering
\caption{Illustrative Classification Examples}
\label{tab:classification-examples}
\begin{tabular}{p{0.2\linewidth} p{0.10\linewidth} p{0.62\linewidth}}
\toprule
Topic & Label & Excerpt from AI reply \\
\midrule
Diagnostic testing & (yes, no) & If fever persists beyond the third day or worsens, please go to the hospital promptly for a complete blood count and other basic tests to determine whether the infection is viral or bacterial. \\
\addlinespace
General medication & (yes, no) & Oral antihistamines, such as loratadine at a dose of 10 mg once daily, can be used to alleviate itching. \\
\addlinespace
General medication & (yes, yes) & Do not self-administer strong cough suppressants (e.g., those containing codeine). The following is for reference only; specific use must follow physician guidance. For productive cough, ambroxol (Mucosolvan) may be considered as an expectorant. For fever, acetaminophen (Tylenol) is appropriate when temperature is at least 38.5\textdegree C. \\
\addlinespace
General medication & (no, yes) & Lifestyle intervention should be the first line of treatment, rather than rushing to medication. If episodes occur more than twice within a year, or if uric acid remains above 540~$\mu$mol/L, the patient should seek hospital care to evaluate whether urate-lowering therapy is needed. \\
\addlinespace
TCM & (no, yes) & Self-administration of Chinese herbal medicine or proprietary Chinese medicines to ``dissolve nodules'' is not recommended; routine follow-up should be the main course of action. \\
\addlinespace
Antibiotics & (no, yes) & Self-use of antibiotics is not recommended. If grinding-related tooth pain has reached the pulp, decompression is required; antibiotics cannot relieve intra-pulp pressure and have no effect on the cavity itself. \\
\bottomrule
\end{tabular}
\begin{minipage}{\textwidth}
    \footnotesize\textit{Notes:} Excerpts are translated from the original Mandarin and lightly condensed. The label column reports (\textit{recommends}, \textit{cautions}).
\end{minipage}
\end{table}

\subsection{Cross-Model Comparison}

\label{sec:appendix-crossmodel}
To assess whether the directional pattern documented in Section~\ref{sec:conv} is specific to the model our partner deployed, we generate counterfactual responses to the same patient messages from six other large language models and process them through the identical pipeline used for our chatbot. Because the chatbot's later conversational turns depend on patients' reactions to its own earlier replies and cannot be reconstructed for a different model, we restrict the comparison to the first turn of each conversation, that is, the patient's opening message and the model's reply to it. This yields 957 patient messages, to which each comparison model responds once. To match this design, the experiment chatbot is also restricted to its first-turn responses in Figures~\ref{fig:crossmodel_previst} and~\ref{fig:crossmodel_direct}, so its topic-mention counts and shares differ slightly from Table~\ref{tab:conversation}, which uses the full analytic sample.

\paragraph{Models.} We use six widely used models from leading developers in the United States and China, listed in Appendix Table~\ref{tab:crossmodel-list}. Where a model exposes a temperature parameter, we set it to zero for reproducibility; the reasoning models that do not accept the parameter are run at their default setting.

\begin{table}[h!]
\centering
\caption{Comparison Models}
\label{tab:crossmodel-list}
\begin{tabular}{p{0.30\linewidth} p{0.32\linewidth} p{0.24\linewidth}}
\toprule
Model & Version & Provider \\
\midrule
GPT-4o & \texttt{gpt-4o} & OpenAI \\
GPT-5.5 & \texttt{gpt-5.5} & OpenAI \\
DeepSeek-V3 & \texttt{deepseek-chat-v3-0324} & DeepSeek \\
Claude Sonnet 4.6 & \texttt{claude-sonnet-4.6} & Anthropic \\
Qwen3.5-plus & \texttt{qwen3.5-plus-20260420} & Alibaba \\
Gemini 3.5 Flash & \texttt{gemini-3.5-flash} & Google \\
\bottomrule
\end{tabular}
\end{table}

\paragraph{Prompts.} To approximate the setting of our intervention, we generate responses under two system prompts. The first frames the model as a pre-visit assistant, matching our deployment; the second frames it as a direct medical consultant. Both instruct the model to respond in Chinese, in free form, within approximately 1{,}000 characters:\footnote{Gemini~3.5~Flash occasionally exceeds the requested length limit.}

\begin{quote}
\small
\textbf{Pre-visit.} You are a pre-visit medical assistant chatbot at a Chinese hospital outpatient clinic. A patient is about to see a doctor and has sent you a health-related question. Please respond in Chinese, naturally, without following any fixed structure or template, within approximately 1{,}000 Chinese characters.
\smallskip

\textbf{Direct medical consultation.} You are a medical assistant. A patient is consulting you directly for health-related advice. Please respond in Chinese, naturally, without following any fixed structure or template, within approximately 1{,}000 Chinese characters.

\end{quote}

\paragraph{Identical classification pipeline.} The comparison models are used only to generate responses. Topic detection and stance classification follow exactly the procedures in Appendices~\ref{sec:appendix-topics} and~\ref{sec:appendix-stance}: the same sentence-level keyword matching is used to identify topic mentions, and the same GPT-4o classifier, with the same prompt and at temperature zero, assigns the recommendation and caution flags. The classifier receives only the reply text and the topic name, not the identity of the model that produced the reply. The sole difference across the conditions in Figures~\ref{fig:crossmodel_previst} and~\ref{fig:crossmodel_direct} is therefore the generating model, which is what allows the figures to attribute differences in stance to the models rather than to the measurement.

\paragraph{Reproducibility.} Large language model APIs are not perfectly deterministic, so the same input can yield slightly different responses across runs. To gauge the resulting noise, we randomly select 100 patient messages and pass them through the full pipeline (response generation, topic identification, and stance classification) twice. Comparing the two runs, agreement on the recommendation flag ranges from 88 to 97 percent across models, and agreement on the caution flag from 84 to 98 percent. This roughly 10 percent disagreement provides a practical noise floor: differences between models or prompts smaller than this magnitude should not be read as meaningful.

\end{document}